\documentclass[10pt]{iopart}

\usepackage{graphicx}
\usepackage{url}
\usepackage{esint}

\usepackage[OT2,T1]{fontenc}
\usepackage[russian,ngerman,english]{babel}

\newcommand{\muas}[0]{\hbox{\rm $\mu$as}}

\newcommand{\ve}[1]{\mbox{\boldmath$#1$}}

\let\oldbibitem\bibitem
\renewcommand\bibitem[2][]{\oldbibitem{#2}}

\begin{document}

\title[Light propagation in 2PN approximation in the field of one moving monopole]
{Light propagation in 2PN approximation in the field of one moving monopole $\;$ I. Initial value problem}

\author{Sven Zschocke}

\address{Institute of Planetary Geodesy - Lohrmann Observatory, Dresden Technical University,
Helmholtzstrasse 10, D-01069 Dresden, Germany}

\ead{sven.zschocke@tu-dresden.de}

\begin{abstract}
In  this investigation the light propagation in the gravitational field of one arbitrarily moving body with monopole structure is considered
in the second post-Newtonian approximation. It is found that the light trajectory depends on the acceleration of
the body. Some of these acceleration terms are important in order to get well-defined logarithmic functions with dimensionless arguments,
while all the other acceleration terms are negligible on the pico-second level of accuracy in time-delay measurements.   
The expressions of the observables total light deflection and time delay are determined.
\end{abstract}

\pacs{95.10.Jk, 95.10.Ce, 95.30.Sf, 04.25.Nx, 04.80.Cc}


\section{Introduction}

The precision of astrometric observations has made impressive progress during the recent decades. Especially two space-based  
astrometry missions of the European Space Agency (ESA) exemplify the rapid evolution in the accuracy of astrometric measurements.  
The first ESA astrometry mission has been the Hipparcos mission, launched on 8 August 1989, which has provided astrometric data with  
milli-arcsecond (mas) accuracy in angular observations \cite{Hipparcos,Hipparcos1,Hipparcos2}. The second ESA astrometry mission is the  
Gaia mission, launched on 19 December 2013, which is aiming at the micro-arcsecond (\muas) level of accuracy in angular observations \cite{GAIA}.  
The Gaia mission is an all-sky survey in astrometry, photometry and spectroscopy, which will be complete up to stellar magnitude of $V=20$
of altogether about $10^9$ celestial light sources, mainly stars of our galaxy, but also quasars, exoplanets, and Solar System objects.
Preliminary results of first Gaia Data Release (DR1) have recently been published within a series of articles  
\cite{GAIA1,GAIA2,GAIA3,GAIA4,GAIA5,GAIA6,GAIA7,GAIA133}, which are already more precise than those in all existing star catalogues. The astrometric  
precision will further be improved substantially by the second (DR2) and third (DR3) Gaia Data Release announced for April 2018 and in the fall of 2020, 
while the final release for the nominal mission is expected at the end of 2022 which will provide the  
full astrometric, photometric, and radial-velocity catalogues.  
In view of these extensive advancements it is clear that in foreseeable future the astrometric precision will arrive at the domain of sub-micro-arcsecond  
(sub-\muas) and perhaps even the nano-arcsecond (nas) level of accuracy. In this respect the medium-sized missions Theia \cite{Theia},  
NEAT \cite{NEAT1,NEAT2} and Gaia-NIR \cite{Gaia_NIR} are mentioned, which have been proposed to ESA and aim to operate at such fields of 
ultra-high-precision astrometry.  

The primary information in astrometry is carried by light signals emitted by the celestial objects and which propagate from the light source 
towards the observer. So the progress in high-precision astrometry depends directly on improvements in the theory of light propagation  
and based on these advancements all subsequent issues of relativistic astrometry can be investigated.  
Consequently, the precise description of the trajectory of a light signal in the near-zone of the curved space-time of 
the Solar System is a fundamental aspect in the science of relativistic astrometry and triggered by this reason substantial advancements 
in the theory of light propagation have been achieved in recent decades. For a survey on the present status in the theory of 
light propagation we take the liberty to refer to the text books \cite{Brumberg1991,Kopeikin_Efroimsky_Kaplan} as well as to 
the references \cite{KlionerKopeikin1992,Kopeikin1997,KopeikinSchaefer1999,KopeikinSchaefer1999_Gwinn_Eubanks,Klioner2003a,Klioner2003b,KopeikinMashhoon2002} and  
the recent reviews in \cite{Zschocke1,Zschocke2}. Particularly with regard to astrometry on the sub-micro-arcsecond or nano-arcsecond level,  
it is absolutely necessary to account for second post-Newtonian (2PN) effects in the theory of light propagation  
\cite{Xu_Wu,Xu_Gong_Wu_Soffel_Klioner,Minazzoli1,Deng_Xie,2PN_Light_PropagationA,Deng_2015,Xie_Huang,Minazzoli2,Conference_Cambridge}. 
Several post-post-Newtonian effects on light deflection in the gravitational field of one monopole at rest have been determined a long time ago  
\cite{EpsteinShapiro,FischbachFreeman,RichterMatzner1,RichterMatzner2,RichterMatzner3,Cowling,BodennerWill2003}. An exceptional progress has been made in 
\cite{Brumberg1991,Brumberg1987}, where the coordinate velocity and trajectory as function of coordinate time have been determined in 2PN approximation 
for light signals propagating  
in the Schwarzschild field. Generalizations of that solution for the case of the parametrized post-post-Newtonian metric have been given  
in \cite{Article_Zschocke1}, where the numerical magnitude of each individual term has been determined.  
Especially, the occurrence of so-called enhanced post-post-Newtonian terms have been ascertained and it was clarified that they are caused by   
a physically inadequate choice of the parametrization of the light rays (namely the use of coordinate dependent impact vectors)  
which, however, is inevitable in real astrometric data reduction.  
Two alternative approaches to the calculation of propagation time and direction of light rays
have been formulated recently. Both approaches allow one to avoid explicit integration of the geodesic equations for light rays.   
The first approach \cite{LePoncinLafitteLinetTeyssandier2004,TeyssandierLePoncinLafitte2008,Teyssandier,Hees_Bertone_Poncin_Lafitte_2014b}  
is based on the use of Synge's world function.
Another approach based on the eikonal concept has been developed in \cite{AshbyBertotti2010} in order to investigate
the light propagation in the field of a spherically symmetric body.

Thus far, all these investigations in 2PN approximation have been focussed on the problem of light propagation in the field of one monopole at rest,  
while in reality the Solar System bodies are in motion. In order to account for the motion of the bodies one may take the analytical solution  
for the light trajectory in the field of one monopole at rest and afterwards one may implement the body's retarded position, an idea which is   
induced by some kind of educated guess. In fact, such an approach would be sufficient for \muas-astrometry.  
Though, such procedure does not allow to account for those terms which are proportional to the speed of the massive body, and it turns out that  
for the sub-\muas-astrometry it is really necessary to consider light trajectories in the field of moving bodies  
in order to account for the just mentioned terms \cite{Zschocke1,Zschocke2}.  
However, investigations about light propagation in 2PN approximation in the field of moving bodies  
are extremely rare. So far, to the best of our knowledge, there are only three investigations dealing with moving deflecting bodies in the 
post-post-Newtonian approximation. (i) In \cite{Bruegmann2005} the light propagation in 2PN approximation in the field of two moving 
point-like bodies has been determined but with approximations which are of interest in case of studying light propagation in the field of 
binary pulsars rather than Solar System objects. 
(ii) In \cite{Moving_Kerr_Black_Hole1,Moving_Kerr_Black_Hole2} the problem of time-delay of a light signal in the field of a Kerr-Newman black hole 
in uniform motion has been determined, but this study does also not aim at astrometry in the Solar System.  
(iii) In \cite{Zschocke3} the light trajectory in the field of one arbitrarily moving 
point-like body in the 2PN approximation has been determined, but the problem of ill-defined logarithms has not been addressed and 
the fact that the light trajectory depends on the acceleration of the body has not been recognized. 
It has not escaped our attention that the logarithms were not well-defined in \cite{Zschocke3}, but neither the importance of this circumstance  
nor its relation to the acceleration terms has been recovered.  
More explicitly, the reinvestigation of the problem of light propagation in 2PN approximation in the field of one moving monopole has  
revealed two peculiarities: First, it has been found that some logarithmic terms are improperly defined because their arguments have the
dimension of a length which spoils a clear mathematical meaning of these functions. Second, it has been recognized that the solution for the light  
trajectory depends on the acceleration of the moving body and these terms are related to the problem of incorrectly defined logarithmic terms.

The manuscript is organized as follows: A compendium of the exact field equations and exact geodesic equation is given in Section \ref{Section1}.  
The metric tensor and geodesic equation in the 2PN approximation is provided in Section \ref{Section2}. 
In Sections \ref{Section3a} and \ref{Section3b} the integration procedure and the consideration of retardation is described. 
The coordinate velocity of a light signal in 2PN approximation in given in Section \ref{Section4}.  
Section \ref{Section5} contains the light trajectory in 2PN approximation and shows the relation between logarithmic terms 
and acceleration terms which allows to rearrange the 2PN solution for the light trajectory in such a way that it contains only
well-defined logarithmic functions.    
Expressions for total light deflection and time delay are derived in Section \ref{Section6}. 
A summary is given in Section \ref{Section7}. The notation, parameters, impact vectors, relations of the partial integration, and the light propagation in  
post-Newtonian (1PN and 1.5PN) and first post-Minkowskian (1PM) approximation are presented in the appendices.

\section{The metric tensor and geodesic equation}\label{Section1}  

\subsection{The exact field equations} 

According to the general theory of relativity, the geometry of curved space-time is determined by the field equations of gravity  
which relate the metric tensor $g_{\mu\nu}$ of curved space-time to the stress-energy tensor of matter $T_{\mu\nu}\,$  
and which can be written in the following form \cite{Kopeikin_Efroimsky_Kaplan,MTW,Landau_Lifschitz,Book_Fock,Einstein1,Einstein2}  
(e.g. Sec. 17.1 in \cite{MTW}),  
\begin{eqnarray}
R_{\mu\nu} - \frac{1}{2}\,g_{\mu\nu}\,R &=& \frac{8\,\pi\,G}{c^4}\,T_{\mu\nu}\,,  
\label{Field_Equations_5}
\end{eqnarray}

\noindent
where $R_{\mu \nu} = \Gamma^{\rho}_{\mu\nu,\rho} - \Gamma^{\rho}_{\mu\rho,\nu} + \Gamma^{\rho}_{\sigma\rho}\,\Gamma^{\sigma}_{\mu\nu}
- \Gamma^{\rho}_{\sigma\nu}\,\Gamma^{\sigma}_{\mu\rho}\,$ is the Ricci tensor (cf. Eq.~(8.47) in \cite{MTW})  
\footnote{In view of $\Gamma^{\rho}_{\mu\rho,\nu} = \left(\ln \sqrt{-g}\right)_{,\mu\nu}$ (cf. Eq.~(8.51a) in \cite{MTW} or Eq.~(29.a) in \cite{Einstein2}) the  
symmetry of Ricci tensor $R_{\mu\nu} = R_{\nu\mu}$ is evident due to $\Gamma^{\rho}_{\mu\nu} = \Gamma^{\rho}_{\nu\mu}$.}, 
the Christoffel symbols are  
\begin{eqnarray}
\Gamma^{\alpha}_{\mu\nu} &=& \frac{1}{2}\,g^{\alpha\beta}
\left(g_{\beta\mu,\nu} + g_{\beta\nu,\mu} - g_{\mu\nu,\beta}\right),   
\label{Christoffel_Symbols}
\end{eqnarray}

\noindent 
$R = g^{\mu\nu} R_{\mu \nu}\,$ is the Ricci scalar (cf. Eq.~(8.48) in \cite{MTW}), and the signature of metric tensor is $\left(-,+,+,+\right)$.  
Frequently, the form of the field equations (\ref{Field_Equations_5}) is not convenient for practical calculations in celestial mechanics and 
relativistic astrometry. A more feasible form is  
arrived by the Landau-Lifschitz formulation of the field equations \cite{Kopeikin_Efroimsky_Kaplan,MTW,Landau_Lifschitz,Poisson_Lecture_Notes}, 
where instead of the metric $g_{\alpha\beta}$ the inverse gothic metric {\foreignlanguage{italian}{g}}$^{\alpha\beta} = \sqrt{-g}\,g^{\alpha\beta}$  
is used, where $g = {\rm det} \left(g_{\alpha\beta}\right)$ is the determinant of the metric tensor. The factor $\sqrt{-g}$ implies that the  
gothic metric is actually not a tensor but a tensor density. From the gothic metric one can obtain the metric tensor uniquely at any stage of the calculations 
where the fact that ${\rm det} \left(g_{\alpha\beta}\right) = {\rm det}(${\foreignlanguage{italian}{g}}$^{\alpha\beta})$ becomes useful.  
In this formulation the contravariant components of the gothic metric are decomposed as follows (cf. Eq.~(5.1) in \cite{Thorne}),  
\begin{eqnarray}
\sqrt{-g}\,g^{\alpha\beta} &=& \eta^{\alpha \beta} - \overline{h}^{\alpha \beta}\,,
\label{metric_20}
\end{eqnarray}

\noindent
which is especially useful in case of an asymptotically flat space-time. 
Here, $\overline{h}^{\alpha \beta}$ is the trace-reversed metric perturbation which describes the deviation of the metric tensor of curved space-time 
from the metric tensor of Minkowskian space-time.  
In line with the resolutions of International Astronomical Union (IAU) \cite{IAU_Resolution1} we adopt harmonic coordinates
$\left(ct,\ve{x}\right)$, a class of coordinate systems which obey the harmonic gauge condition \cite{Brumberg1991,Kopeikin_Efroimsky_Kaplan}
\begin{eqnarray}
\frac{\partial \sqrt{-g}\,g^{\alpha\beta}}{\partial x^{\alpha}} &=& 0\,,
\label{Harmonic_Gauge}
\end{eqnarray}

\noindent
usually called de Donder gauge \cite{De_Donder} which has also been introduced independently in \cite{Lanczos}.
The harmonic coordinates $\left(ct,\ve{x}\right)$ are curvilinear but they can be treated as though they were Cartesian coordinates \cite{Thorne}.   
The exact field equations (\ref{Field_Equations_5}) in terms of harmonic coordinates can be written as follows 
(cf. Eq.~(36.37) in \cite{MTW} or Eq.~(5.2b) in \cite{Thorne}):
\begin{eqnarray}
\opensquare\; 
\overline{h}^{\alpha \beta} &=& - \frac{16\,\pi\,G}{c^4}\,\left(\tau^{\alpha \beta} + t^{\alpha \beta}\right),
\label{Field_Equations_10}
\end{eqnarray}

\noindent
where  
$\opensquare = \eta_{\mu\nu}\,\partial_{\mu}\,\partial_{\nu}$ is the (flat) d'Alembert operator and   
\begin{eqnarray}
\tau^{\alpha \beta} &=& \left( - g\right)\,T^{\alpha \beta}\,,
\label{metric_35}
\\
\nonumber\\
t^{\alpha \beta} &=& \left( - g\right)\,t_{\rm LL}^{\alpha \beta} + \frac{c^4}{16\,\pi\,G}\;
\left(\overline{h}^{\alpha\mu}_{\;\;\;,\;\nu}\;\overline{h}^{\beta \nu}_{\;\;\;,\;\mu}
- \overline{h}^{\alpha\beta}_{\;\;\;,\;\mu\nu}\;\overline{h}^{\mu \nu}\right),
\label{metric_40}
\end{eqnarray}

\noindent
where $t_{\rm LL}^{\alpha \beta}$ is the Landau-Lifschitz pseudotensor of gravitational field \cite{Landau_Lifschitz}, which is symmetric 
in its indices and in explicit form given by Eq.~(20.22) in \cite{MTW} or by Eqs.~(3.503) - (3.505) in \cite{Kopeikin_Efroimsky_Kaplan}.  
We shall assume that the gravitational system is isolated (Fock-Sommerfeld boundary conditions), that means flatness of the metric at spatial
infinity and the constraint of no-incoming gravitational radiation is imposed at past null infinity ${\cal J}^{-}$ 
(cf. notation in Section 34 in \cite{MTW} and Figure 34.2. in \cite{MTW}).
In terms of trace-reversed metric perturbation these conditions read as follows
\cite{Kopeikin_Efroimsky_Kaplan,KlionerKopeikin1992,Book_Fock,IAU_Resolution1,Radiation_Condition,Zschocke_Multipole_Expansion},
\begin{eqnarray}
&& \hspace{-0.5cm} \lim_{r \rightarrow \infty \atop t  + \frac{r}{c} = {\rm const}}\,\overline{h}^{\mu \nu}\left(t,\ve{x}\right) \!=\! 0\,,
\label{Asymptotic_1}
\\
\nonumber\\
&& \hspace{-0.5cm} \lim_{r \rightarrow \infty \atop t + \frac{r}{c} = {\rm const}}
\left(\frac{\partial}{\partial r} r\,\overline{h}^{\mu \nu}\left(t,\ve{x}\right)
+ \frac{\partial}{\partial ct} \,r\,\overline{h}^{\mu \nu}\left(t,\ve{x}\right)\right) = 0\,,
\label{Asymptotic_2}
\end{eqnarray}

\noindent
where $r = \left|\ve{x}\right|$.
In addition, $r\,\partial_{\alpha}\,\overline{h}^{\mu \nu}$ should be bounded at spatial infinity \cite{Book_Fock,Radiation_Condition}.
Then, a formal solution of (\ref{Field_Equations_10}) is given by  
\begin{eqnarray}
\overline{h}^{\alpha \beta} \left(t,\ve{x}\right) &=& \frac{4\,G}{c^4}\,
\int d^3 x^{\prime}\,
\frac{\tau^{\alpha\beta}\left(u, \ve{x}^{\prime}\right) + t^{\alpha\beta}\left(u, \ve{x}^{\prime}\right)}{\left| \ve{x} - \ve{x}^{\prime} \right|}\,,
\label{Introduction_2}
\end{eqnarray}

\noindent
where the integral runs over three-dimensional volume of the entire space-time. This solution represents an implicit integro-differential equation 
because the metric perturbations appear on both sides of this equation, which can be solved by iteration \cite{Poisson_Lecture_Notes,Thorne,Blanchet_Damour1};  
see also Eqs.~(3.530a) - (3.530d) in \cite{Kopeikin_Efroimsky_Kaplan}. What is of primary importance here is  
the fact that the retarded time $\displaystyle u = t - \frac{\displaystyle \left| \ve{x} - \ve{x}^{\prime} \right|}{\displaystyle c}$ naturally appears 
in the formal solution of the exact field equations (\ref{Introduction_2}), which states that a space-time point $\left(u,\ve{x}^{\prime}\right)$ (e.g. located  
inside the matter distribution) is in causal contact with a space-time point $\left(t,\ve{x}\right)$ (e.g. located outside the matter source).

\subsection{The exact geodesic equation} 

In general relativity light signals in curved space-time propagate along null geodesics which are governed by the geodesic equation and 
isotropic condition \cite{Brumberg1991,Kopeikin_Efroimsky_Kaplan,MTW},  
\begin{eqnarray}
\frac{d^2 x^{\alpha}\left(\lambda\right)}{d \lambda^2}
+ \Gamma^{\alpha}_{\mu\nu}\,\frac{d x^{\mu}\left(\lambda\right)}{d \lambda}\,
\frac{d x^{\nu}\left(\lambda\right)}{d \lambda} &=& 0\,,  
\label{Geodetic_Equation1}
\\
\nonumber\\ 
g_{\alpha\beta}\,\frac{d x^{\alpha}\left(\lambda\right)}{d \lambda}\,\frac{d x^{\beta}\left(\lambda\right)}{d \lambda} = 0\,, 
\label{Null_Condition1}
\end{eqnarray}

\noindent
where $x^{\alpha}\left(\lambda\right)$ is the four-coordinate of the light signal (photons) which depends on the affine curve parameter $\lambda$.   
The geodesic equation    
(\ref{Geodetic_Equation1}) is valid for any free falling particle in space-time and can be defined by the requirement that a tangent vector  
along the geodesics remains tangent when parallel transported along it. The null condition (\ref{Null_Condition1}) 
is only valid for massless particles (photons) and asserts that the length of the tangent vector along 
the light trajectory remains zero.  
The geodesic equation (\ref{Geodetic_Equation1}) and isotropic condition (\ref{Null_Condition1}) are valid in any coordinate system. They can  
be rewritten in terms of coordinate time $x^0=x^0\left(\lambda\right)$ (recalling that $x^0=ct$) \cite{Brumberg1991,Kopeikin_Efroimsky_Kaplan,MTW} 
(e.g. Eqs.~(3.220) - (3.224) in \cite{Kopeikin_Efroimsky_Kaplan}),  
\begin{eqnarray}
\frac{\ddot{x}^{i}\left(t\right)}{c^2} + \Gamma^{i}_{\mu\nu} \frac{\dot{x}^{\mu}\left(t\right)}{c} \frac{\dot{x}^{\nu}\left(t\right)}{c}
- \Gamma^{0}_{\mu\nu} \frac{\dot{x}^{\mu}\left(t\right)}{c} \frac{\dot{x}^{\nu}\left(t\right)}{c} \frac{\dot{x}^{i}\left(t\right)}{c} = 0\;, 
\label{Geodetic_Equation2}
\\
\nonumber\\
g_{\alpha\beta}\,\frac{\dot{x}^{\alpha}\left(t\right)}{c}\,\frac{\dot{x}^{\beta}\left(t\right)}{c} = 0\,,
\label{Null_Condition2}
\end{eqnarray}

\noindent
where a dot denotes total derivative with respect to coordinate time and $\dot{x}^{i}\left(t\right)$ are the three-components 
of the coordinate velocity of the photon which absolute value differs from the speed of light in flat space $\left|\dot{\ve x}\right| \neq c$ .  
For a unique solution of (\ref{Geodetic_Equation2}) initial-boundary conditions are required  
\cite{Brumberg1991,KlionerKopeikin1992,Kopeikin1997,KopeikinSchaefer1999_Gwinn_Eubanks,Zschocke1,Zschocke2},  
\begin{eqnarray}
\ve{x}_0 &=& \ve{x}\left(t\right)\bigg|_{t=t_0} \quad {\rm and} \quad \ve{\sigma} = \frac{\dot{\ve{x}}\left(t\right)}{c}\bigg|_{t = - \infty}\,,
\label{Introduction_6}
\end{eqnarray}

\noindent
with $\ve{x}_0$ being the position of the light source at the moment $t_0$ of emission of the light-signal and $\ve{\sigma}$
being the unit-direction ($\ve{\sigma} \cdot \ve{\sigma} = 1$) of the light ray at past null infinity.  
The first integration of geodesic equation (\ref{Geodetic_Equation2}) yields the coordinate velocity of the light signal,  
\begin{eqnarray}
\frac{\dot{\ve{x}}\left(t\right)}{c} &=& \int\limits_{- \infty}^{t} d c {\rm t} \,\frac{\ddot{\ve{x}}\left({\rm t}\right)}{c^2} 
= \ve{\sigma} + \frac{\Delta \dot{\ve{x}}\left(t\right)}{c}\,,  
\label{Integral_1}
\end{eqnarray}

\noindent
where the integration variable ${\rm t}$ (roman style) runs from lower limit of integration $ - \infty$ to the upper limit of integration $t$ and 
where $\Delta \dot{\ve{x}}\left(t\right)/c$ denotes the correction to the unit-direction $\ve{\sigma}$ of the light ray at past null infinity.  
The second integration of geodesic equation yields the trajectory of the light signal,
\begin{eqnarray}
\ve{x}\left(t\right) &=& \int\limits_{t_0}^{t} d c {\rm t} \,\frac{\dot{\ve{x}}\left({\rm t}\right)}{c} 
= \ve{x}_0 + c \left(t - t_0\right) \ve{\sigma} + \Delta \ve{x}\left(t,t_0\right)\,, 
\label{Integral_2}
\end{eqnarray}

\noindent 
where the integration variable ${\rm t}$ runs from lower limit of integration $t_0$ to the upper limit of integration $t$  
and where $\Delta \ve{x}\left(t,t_0\right)$ denotes the corrections to the unperturbed light trajectory 
\begin{eqnarray}
\ve{x}_{\rm N}\left(t\right) &=& \ve{x}_0 + c \left(t - t_0\right) \ve{\sigma}\,.  
\label{Unperturbed_Lightray} 
\end{eqnarray} 

\noindent
The metric in the Christoffel symbols (\ref{Christoffel_Symbols}) are functions of the space-time coordinates $(t,\ve{x})$, while if one inserts  
these symbols in the geodesic equation (\ref{Geodetic_Equation2}) then the metric becomes relevant  
at the coordinates of the photon $\ve{x}\left(t\right)$. Consequently, the derivatives of the metric tensor  
contained in the Christoffel symbols of the geodesic equation (\ref{Geodetic_Equation2}) must be taken along the light trajectory,  
\begin{eqnarray}
g_{\alpha \beta, \mu} &=& \frac{\partial g_{\alpha \beta}\left(t,\ve{x}\right)}{\partial x^{\mu}}
\Bigg|_{\ve{x}=\ve{x}\mbox{$\left(t\right)$}}\,, 
\label{Derivatives_1}
\end{eqnarray}

\noindent
where $\ve{x}\left(t\right)$ is the exact light trajectory. The equation (\ref{Derivatives_1}) means that                                  
the differentiations of the metric in (\ref{Geodetic_Equation2}) have to be performed with respect to the space-time coordinates,
and afterwards the light trajectory has to be substituted. The metric $g_{\alpha\beta}$ of the Solar System, however,  
and therefore also the trajectories $\ve{x}\left(t\right)$ of light signals propagating in the Solar System, 
can be determined only approximately, which will be the topic of the next Section.  

\section{The metric tensor and geodesic equation in 2PN approximation}\label{Section2}

The Solar System consists of $N$ bodies which move under the influence of their mutual gravitational interaction around 
the barycenter of the Solar System. Because the field equations (\ref{Introduction_2}) for the case of $N$ moving bodies 
cannot be determined in their exact form one has to resort on approximation schemes. There are two important approximations:   
the post-Minkowskian (weak-field) approximation and the post-Newtonian (weak-field slow-motion) approximation.  
The post-Minkowskian approximation represents a series expansion of the metric tensor in powers of the gravitational constant 
and is briefly viewed in \ref{Appendix3}, while the post-Newtonian approximation represents  
a series expansion of the metric tensor in inverse powers of the speed of light and will be considered now. 

\subsection{The metric tensor in the second post-Newtonian approximation} 

Because the gravitational fields in the Solar System are weak ($m_A \ll P_A$, where $m_A$ and $P_A$ are Schwarzschild radius and equatorial radius of body $A$) 
and the velocities of the bodies are slow ($v_A \ll c$, where $v_A$ is the orbital velocity of body $A$ and $c$ is the speed of light) the utilization  
of the post-Newtonian approximation is well justified  
\cite{Brumberg1991,Kopeikin_Efroimsky_Kaplan,MTW,Poisson_Lecture_Notes,Book_Clifford_Will,Poisson_Will,Book_PN,Will_PN}, 
which admits an expansion of the metric tensor  
in inverse powers of the speed of light (weak-field slow-motion approximation),  
\begin{eqnarray}
g_{\alpha\beta}\left(t,\ve{x}\right) &=& \eta_{\alpha \beta} + \sum\limits_{n=2}^{\infty} h_{\alpha\beta}^{\left(n\right)}\left(t,\ve{x}\right),  
\label{Reply_1}
\end{eqnarray}

\noindent
where $h^{(n)}_{\alpha\beta} \sim {\cal O} \left(c^{-n}\right)$ are tiny perturbations to the flat Minkowskian metric.  
The post-Newtonian expansion (\ref{Reply_1}) actually represents a non-analytic series because besides simple inverse powers in the speed of light  
$\left(1/c\right)^n$ it also involves powers of logarithmic terms $\left(1/c\right)^n \left(\ln c\right)^m$  
\cite{Blanchet_Damour1,Book_PN,Poujade_Blanchet,Blanchet_Damour3}.  
Such non-analytic terms are associated with the problem of wave tails (see also appendix in \cite{Thorne})  
caused by the non-linear structure of the gravitational field equations 
(\ref{Field_Equations_10}) but emerge at the order ${\cal O}\left(c^{-8}\right)$;  
cf. Eq.~(5.5) in \cite{Blanchet_Damour1} or Eq.~(5.3) in \cite{Blanchet_Damour3}.  
One has furthermore to keep in mind that the post-Newtonian expansion (\ref{Reply_1})  
is only valid inside the near-zone of the gravitational system where the retardations are assumed to be small  
\cite{Kopeikin_Efroimsky_Kaplan,MTW,Poisson_Lecture_Notes,Poisson_Will,Expansion_2PN}. The near-zone can be thought of as three-dimensional sphere 
around the gravitating system with radius (cf. Figure 7.7 in \cite{Kopeikin_Efroimsky_Kaplan} or Figure 36.3 in \cite{MTW})  
\begin{eqnarray}
\left|\ve{x}\right| &\ll& \lambda_{\rm gr}\,,  
\label{near-zone_1}
\end{eqnarray}

\noindent 
where $\lambda_{\rm gr}$ is a characteristic wavelength of gravitational waves emitted by the gravitational system. To have an idea about the 
magnitude for the Solar System one may consider the wavelength of gravitational radiation emitted by Jupiter during its revolution around the Sun 
which amounts to be about $\lambda_{\rm gr} \sim 10^{17}\,{\rm m}$ \cite{Kopeikin_Efroimsky_Kaplan,Zschocke2,MTW}.  
The condition in (\ref{near-zone_1}) is only a rather rough estimate which refers to the term near-zone of the Solar System, while a more 
restrictive condition will be provided later in (\ref{near-zone_2}) subject to relation (\ref{constraint_acceleration}).
In 2PN approximation, the post-Newtonian expansion (\ref{Reply_1}) for the covariant and contravariant components of the metric tensor reads 
\cite{Brumberg1991,Kopeikin_Efroimsky_Kaplan,KlionerKopeikin1992,Poisson_Lecture_Notes,Poisson_Will},
\begin{eqnarray}
\fl \hspace{1.5cm} g_{\alpha \beta}\left(t,\ve{x}\right) = \eta_{\alpha \beta} + h^{(2)}_{\alpha\beta}\left(t,\ve{x}\right)
+ h^{(3)}_{\alpha\beta}\left(t,\ve{x}\right) + h^{(4)}_{\alpha\beta}\left(t,\ve{x}\right) + {\cal O} \left(c^{-5}\right),
\label{post_Newtonian_metric_B}
\\
\nonumber\\
\fl \hspace{1.5cm} g^{\alpha \beta}\left(t,\ve{x}\right) = \eta^{\alpha \beta} - h_{(2)}^{\alpha\beta}\left(t,\ve{x}\right)
- h_{(3)}^{\alpha\beta}\left(t,\ve{x}\right) - h_{(4)}^{\alpha\beta}\left(t,\ve{x}\right) + {\cal O}\left(c^{-5}\right). 
\label{post_Newtonian_metric_C}
\end{eqnarray}

\noindent 
The metric perturbations in 2PN approximation for the case of one monopole in slow but otherwise arbitrary motion read 
as follows \cite{Deng_Xie,Zschocke3,Blanchet_Faye_Ponsot} (cf. Eqs.~(C7) - (C12) and (C14) - (C16) in \cite{Zschocke3}),  
\begin{eqnarray}
\fl h^{(2)}_{00}\left(t,\ve{x}\right) = + \frac{2\,m_A}{r_A\left(t\right)}\,, \quad
h^{(2)}_{ij}\left(t,\ve{x}\right) = + \frac{2\,m_A}{r_A\left(t\right)}\,\delta_{ij}\,, \quad
h^{(3)}_{0i}\left(t,\ve{x}\right) = - \frac{4\,m_A}{r_A\left(t\right)}\,\frac{v_A^i\left(t\right)}{c}\,,
\label{Metric_1}
\\
\nonumber\\
\fl h^{(4)}_{0i}\left(t,\ve{x}\right) = + 4\,m_A\,\frac{a_A^i\left(t\right)}{c^2}\,,
\label{Metric_4}
\\
\nonumber\\
\fl h^{(4)}_{00}\left(t,\ve{x}\right) = + \frac{4\,m_A}{r_A\left(t\right)}\,\frac{v^2_A\left(t\right)}{c^2}
- \frac{m_A}{r_A\left(t\right)}\,\frac{\left(\ve{n}_A\left(t\right) \cdot \ve{v}_A\left(t\right)\right)^2}{c^2}
- m_A\,\frac{\left(\ve{n}_A\left(t\right) \cdot \ve{a}_A\left(t\right)\right)}{c^2}
- \frac{2\,m^2_A}{r^2_A\left(t\right)}\,,
\nonumber\\
\label{Metric_5}
\\
\nonumber\\
\fl h^{(4)}_{ij}\left(t,\ve{x}\right) =
- \frac{m_A}{r_A\left(t\right)}\,\frac{\left(\ve{n}_A\left(t\right) \cdot \ve{v}_A\left(t\right)\right)^2}{c^2}\,\delta_{ij}
+ \frac{4\,m_A}{r_A\left(t\right)}\,\frac{v_A^i\left(t\right)}{c}\,\frac{v_A^j\left(t\right)}{c}
\nonumber\\
\nonumber\\
\fl \hspace{1.9cm} - m_A\,\frac{\left(\ve{n}_A\left(t\right) \cdot \ve{a}_A\left(t\right)\right)}{c^2}\,\delta_{ij}
 + \frac{m^2_A}{r^2_A\left(t\right)}\,\delta_{ij}
+ \frac{m^2_A}{r^2_A\left(t\right)}\,n_A^i\left(t\right)\,n_A^j\left(t\right),
\label{Metric_6}
\end{eqnarray}

\noindent
while $h_{0i}^{(2)}=h_{00}^{(3)}=h_{ij}^{(3)}=0$ \cite{Brumberg1991,Zschocke3,MTW,IAU_Resolution1,Book_Clifford_Will,Poujade_Blanchet,DSX1,DSX2}.  
Thereby, the three-vector pointing from the position of the monopole towards the field point (e.g. Eq.~(B.4) in \cite{KlionerPeip2003}),  
\begin{eqnarray}
\ve{r}_A\left(t\right) &=& \ve{x} - \ve{x}_A\left(t\right),
\label{vector}
\end{eqnarray}
 
\noindent 
and its unit-vector $\ve{n}_A\left(t\right) = \ve{r}_A\left(t\right)/r_A\left(t\right)$ have been introduced.  
The metric (\ref{Metric_1}) - (\ref{Metric_6}) is valid for one monopole $A$ in arbitrary motion where the physical origin of the motion is not specified;  
for instance one might imagine just rockets tied to that body 
(cf. comment in the text below Eq.~(C.22) in \cite{KlionerPeip2003} and in the text below Eq.~(C5) in \cite{Zschocke3}).  
In addition the following asymptotic conditions are implicitly involved,
\begin{eqnarray}
\lim_{t \rightarrow - \infty} \ve{a}_A\left(t\right) &=& 0\,,
\label{Condition_Acceleration}
\\
\lim_{t \rightarrow - \infty} \ve{v}_A\left(t\right) &=& 0\,.
\label{Condition_Velocity}
\end{eqnarray}

\noindent
The dynamical asymptotic condition (\ref{Condition_Acceleration}) means that the body is asymptotically free at past timelike infinity $I^{-}$ 
(cf. notation in Section 34 in \cite{MTW} and Figure 34.2. in \cite{MTW}).
The kinematical asymptotic condition (\ref{Condition_Velocity}) follows from (\ref{Condition_Acceleration})
and the requirement that the orbital motion of the body is bounded by the near-zone of the gravitational system.  
Besides these conditions and the requirement of slow-motion the worldline $\ve{x}_A\left(t\right)$ of body $A$ is supposed to be arbitrary.

\subsection{The geodesic equation in the second post-Newtonian approximation} 

Inserting the post-Newtonian expansion (\ref{post_Newtonian_metric_B}) - (\ref{post_Newtonian_metric_C}) into (\ref{Geodetic_Equation2}) one obtains 
the geodesic equation in the second post-Newtonian approximation \cite{Brumberg1991,KopeikinSchaefer1999_Gwinn_Eubanks,Bruegmann2005,Zschocke3},  
\begin{eqnarray}
\fl \hspace{0.3cm} \frac{\ddot{x}^i}{c^2} = + \frac{1}{2}\,h_{00,i}^{(2)}
- h_{00,j}^{(2)} \frac{\dot{x}^i}{c}\frac{\dot{x}^j}{c}
- h_{ij,k}^{(2)}\,\frac{\dot{x}^j}{c}\frac{\dot{x}^k}{c}
+ \frac{1}{2}\,h_{jk,i}^{(2)}\,\frac{\dot{x}^j}{c}\frac{\dot{x}^k}{c}
- h_{ij,0}^{(2)} \frac{\dot{x}^j}{c}
\nonumber\\
\nonumber\\
\fl \hspace{1.2cm} + \frac{1}{2}\,h_{jk,0}^{(2)} \frac{\dot{x}^i}{c}
\frac{\dot{x}^j}{c}\frac{\dot{x}^k}{c}
- \frac{1}{2}\,h_{00,0}^{(2)}\,\frac{\dot{x}^i}{c}
- h_{0i,j}^{(3)} \frac{\dot{x}^j}{c}
+ h_{0j,i}^{(3)} \frac{\dot{x}^j}{c}
- h_{0j,k}^{(3)}\frac{\dot{x}^i}{c}\frac{\dot{x}^j}{c}\frac{\dot{x}^k}{c}
\nonumber\\
\nonumber\\
\fl \hspace{1.2cm} - h_{0i,0}^{(3)} - \frac{1}{2}\,h_{ij}^{(2)}\,h_{00,j}^{(2)}
- h_{00}^{(2)}\,h_{00,j}^{(2)}\,\frac{\dot{x}^i}{c}\,\frac{\dot{x}^j}{c}
+ h_{is}^{(2)}\,h_{sj,k}^{(2)}\,\frac{\dot{x}^j}{c}\,\frac{\dot{x}^k}{c}
- \frac{1}{2}\,h_{is}^{(2)}\,h_{jk,s}^{(2)}\,\frac{\dot{x}^j}{c}\,\frac{\dot{x}^k}{c}
\nonumber\\
\nonumber\\
\fl \hspace{1.2cm} + \frac{1}{2}\,h_{00,i}^{(4)}\,
- h_{00,j}^{(4)}\,\frac{\dot{x}^i}{c}\,\frac{\dot{x}^j}{c}
- h_{ij,k}^{(4)}\,\frac{\dot{x}^j}{c}\,\frac{\dot{x}^k}{c}
+ \frac{1}{2}\,h_{jk,i}^{(4)}\,\frac{\dot{x}^j}{c}\,\frac{\dot{x}^k}{c}
\nonumber\\
\nonumber\\
\fl \hspace{1.2cm} + h_{0j,i}^{(4)}\,\frac{\dot{x}^j}{c} - h_{0i,j}^{(4)}\,\frac{\dot{x}^j}{c}
- h_{0j,k}^{(4)}\,\frac{\dot{x}^i}{c}\,\frac{\dot{x}^j}{c}\,\frac{\dot{x}^k}{c}
- h_{0i,0}^{(4)} + {\cal O}\left(c^{-5}\right), 
\label{Geodesic_Equation3}
\end{eqnarray}

\noindent
where the time-argument has been omitted so that $\ddot{x}^i = \ddot{x}^i\left(t\right)$ and $\dot{x}^i = \dot{x}^i\left(t\right)$.  
The last term in (\ref{Geodesic_Equation3}) is seemingly of
the order ${\cal O}\left(c^{-5}\right)$ but contributes to order ${\cal O}\left(c^{-4}\right)$ to the light trajectory \cite{Zschocke3}.
In order to obtain (\ref{Geodesic_Equation3}) the following relations among the covariant and contravariant components of the metric tensor have been used,
\begin{eqnarray}
h_{00}^{(2)} &=& h^{00}_{(2)}\,,\; h_{ij}^{(2)} = h^{ij}_{(2)}\,,
\nonumber\\
h_{0i}^{(3)} &=& - h^{0i}_{(3)}\,,\; h_{0i}^{(4)} = - h^{0i}_{(4)}\,,
\nonumber\\
h_{00}^{(4)} &=& h^{00}_{(4)} - h^{00}_{(2)}\,h^{00}_{(2)}\,,\;
h_{ij}^{(4)} = h^{ij}_{(4)} + h^{ik}_{(2)}\,h^{kj}_{(2)}\,,
\end{eqnarray}

\noindent
which result from $g_{\alpha\mu}\,g^{\mu \beta} = \delta_{\alpha}^{\beta} = {\rm diag}\left(+1,+1,+1,+1\right)$.  
The formal solution for the first and second integration of the geodesic equation (\ref{Geodesic_Equation3}) is written as follows,  
\begin{eqnarray}
\fl \frac{\dot{\ve{x}}\left(t\right)}{c} = \ve{\sigma} +
\frac{\Delta \dot{\ve{x}}_{\rm 1PN}\left(t\right)}{c} + \frac{\Delta \dot{\ve{x}}_{\rm 1.5PN}\left(t\right)}{c} 
+ \frac{\Delta \dot{\ve{x}}_{\rm 2PN}\left(t\right)}{c} + {\cal O}\left(c^{-5}\right),
\label{Expansion_Coordinate_Velocity}
\\
\nonumber\\
\fl \ve{x}\left(t\right) = \ve{x}_0 + c \left(t-t_0\right) \ve{\sigma} + \Delta \ve{x}_{\rm 1PN}\left(t,t_0\right)
+ \Delta \ve{x}_{\rm 1.5PN}\left(t,t_0\right) + \Delta \ve{x}_{\rm 2PN}\left(t,t_0\right) + {\cal O}\left(c^{-5}\right),
\nonumber\\ 
\label{Expansion_Coordinate_Trajectory}
\end{eqnarray}

\noindent
where $\Delta \ve{x}_{\rm 1PN} = {\cal O}\left(c^{-2}\right)$, $\Delta \ve{x}_{\rm 1.5PN} = {\cal O}\left(c^{-3}\right)$, and
$\Delta \ve{x}_{\rm 2PN} = {\cal O}\left(c^{-4}\right)$. 
Let us recall that the metric perturbations in (\ref{Metric_1}) - (\ref{Metric_6}) are functions of the field-points $(t,\ve{x})$, while  
in the geodesic equation (\ref{Geodesic_Equation3}) the metric perturbations must be taken at the coordinates of the photon $\ve{x}\left(t\right)$, so that   
according to Eq.~(\ref{Derivatives_1}) we have  
\begin{eqnarray}
h_{\alpha \beta, \mu}^{(n)} &=& \frac{\partial h_{\alpha \beta}^{(n)}\left(t,\ve{x}\right)}{\partial x^{\mu}}
\Bigg|_{\ve{x}=\ve{x}\mbox{\normalsize $\left(t\right)$}}\,.  
\label{geodesic_equation_3}
\end{eqnarray}

\noindent
The equation (\ref{geodesic_equation_3}) states that first of all the differentiations in (\ref{Geodesic_Equation3}) have to be performed with respect to  
the space-time coordinates, and afterwards the light trajectory has to be substituted in the appropriate approximation, as  
it will be enlightened within the next Section.  

\section{Basic procedure of the integration of geodesic equation}\label{Section3a} 

Faced with the fact that the worldline of the body is unknown, we will naturally be obliged to integrate the geodesic equation (\ref{Geodesic_Equation3}) by 
parts. Integration by parts is a technique to integrate the product of two functions, when one of the functions forming the product is given as the  
total derivative of another function,  
\begin{eqnarray}
\fl \hspace{1.5cm} \int\limits d c{\rm t}\,f\left({\rm t}\right) \left[\frac{d}{d c {\rm t}}\,g\left({\rm t}\right)\right] 
= f\left({\rm t}\right)\,g\left({\rm t}\right)\,  
- \, \int\limits d c{\rm t}\,\left[\frac{d}{d c {\rm t}}\,f\left({\rm t}\right)\right]\,g\left({\rm t}\right).  
\label{Integration_by_Part}
\end{eqnarray}

\noindent
The result in (\ref{Integration_by_Part}) still involves an integral, but the new integral will either be of simpler structure or it turns out to be of higher  
order and can be neglected. It has to be emphasized that the first and second integration of geodesic equation (\ref{Geodesic_Equation3}) implicates  
an appreciable amount of algebraic effort. In particular, the complete list incorporates a few hundred integrals.  
Therefore, the description of the integration procedure is restricted to a very few but typical integrals in favor of a clear description  
of the basic ideas of the entire approach. The exemplifying considerations are fairly detailed and will provide a comprehensive  
understanding about how to integrate the geodesic equation.

\subsection{First example of the first integration}\label{First_Integration_First_Example}  

The first integration of geodesic equation yields the coordinate velocity of the light signal and is defined by Eq.~(\ref{Integral_1}) where the  
integrand is given by (\ref{Geodesic_Equation3}). 

We consider the first integration of the first term on the fourth line in (\ref{Geodesic_Equation3}).  
According to (\ref{geodesic_equation_3}) after spatial differentiation we substitute the light ray and obtain,  
\begin{eqnarray}
\fl \ve{{\cal I}}_A\left(t\right) = \frac{1}{2} \int\limits_{-\infty}^t d c{\rm t}\,
\frac{\partial h_{00}^{(4)}\left({\rm t},\ve{x}\right)}{\partial \ve{x}}\Bigg|_{\ve{x}=\ve{x}\mbox{\normalsize $\left({\rm t}\right)$}} 
\; = \; \ve{{\cal I}}_{A_1}\left(t\right) + \ve{{\cal I}}_{A_2}\left(t\right) + \ve{{\cal I}}_{A_3}\left(t\right) + \ve{{\cal I}}_{A_4}\left(t\right),   
\label{Example_1}
\end{eqnarray}

\noindent  
which separates into four integrals, each of which corresponds to one of the four summands of the metric coefficient $h^{(4)}_{00}$ in  
(\ref{Metric_5}). Due to $\left(r_A^m\right)_{,i} = m\,r_A^i \left(r_A\right)^{m-2}$,  
where $m = ..., -2, -1, \pm 0, +1, +2, ...$ denotes integer powers and $i = 1, 2, 3$ labels the vectorial components,  
we get the following integrals,  
\begin{eqnarray}
\fl \ve{{\cal I}}_{A_1}\left(t\right) = - 2\,m_A \int\limits_{-\infty}^t d c{\rm t}\,
\frac{\ve{r}_A\left({\rm t}\right)}{r_A^3\left({\rm t}\right)} 
\,\frac{v^2_A\left({\rm t}\right)}{c^2}\,,  
\label{Example_Integral_1}
\\
\nonumber\\
\fl \ve{{\cal I}}_{A_2}\left(t\right) = - \,\frac{m_A}{c^2} \int\limits_{-\infty}^t d c{\rm t}\,
\frac{\ve{r}_A\left({\rm t}\right) \cdot \ve{v}_A\left({\rm t}\right)}{r_A^3\left({\rm t}\right)}\,
\ve{v}_A\left({\rm t}\right)
+ \frac{3}{2}\,\frac{m_A}{c^2} \int\limits_{-\infty}^t d c{\rm t}\,
\frac{\ve{r}_A\left({\rm t}\right)}{r_A^5\left({\rm t}\right)}\,
\left(\ve{r}_A\left({\rm t}\right) \cdot \ve{v}_A\left({\rm t}\right)\right)^2 \,, 
\nonumber\\
\label{Example_Integral_2} 
\\
\fl \ve{{\cal I}}_{A_3}\left(t\right) = - \frac{1}{2}\,\frac{m_A}{c^2} \int\limits_{-\infty}^t d c{\rm t}\, 
\frac{\ve{a}_A\left({\rm t}\right)}{r_A\left({\rm t}\right)}  
+ \frac{1}{2}\,\frac{m_A}{c^2} \int\limits_{-\infty}^t d c{\rm t}\,
\frac{\ve{r}_A\left({\rm t}\right) \cdot \ve{a}_A\left({\rm t}\right)}{r_A^3\left({\rm t}\right)}\,
\ve{r}_A\left({\rm t}\right),  
\label{Example_Integral_3} 
\\
\nonumber\\
\fl \ve{{\cal I}}_{A_4}\left(t\right) = + 2\,m_A^2 \int\limits_{-\infty}^t d c{\rm t}\,
\frac{\ve{r}_A\left({\rm t}\right)}{r_A^4\left({\rm t}\right)}\,,  
\label{Example_Integral_4} 
\end{eqnarray}

\noindent 
where the three-vector \footnote{We shall avoid to introduce a further notation which distinguishes among (\ref{vector_B})
which points from the position of the body $\ve{x}_A\left(t\right)$ towards the exact position of the light signal $\ve{x}\left(t\right)$
and (\ref{vector}) which points from the position of the body $\ve{x}_A\left(t\right)$ towards the field point $\ve{x}$.}  
\begin{eqnarray}
\ve{r}_A\left(t\right) &=& \ve{x}\left(t\right) - \ve{x}_A\left(t\right)
\label{vector_B}
\end{eqnarray}

\noindent 
points from the instantaneous position of the body towards the spatial position of the light signal;  
we notice that the same three-vector has been introduced by Eq.~(6.8) in \cite{KlionerKopeikin1992} or by Eq.~(B.22) in \cite{KlionerPeip2003}.  
In line with the geodesic equation in  
2PN approximation (\ref{Geodesic_Equation3}) these integrals have to be determined up to terms of the order ${\cal O}\left(c^{-5}\right)$. 
The evaluation of each of the integrals in (\ref{Example_Integral_1}) - (\ref{Example_Integral_4}) proceeds in very similar way. 
As typical example let us consider the integral in (\ref{Example_Integral_1}). In view of the prefactor $m_A/c^2 \sim {\cal O}\left(c^{-4}\right)$, 
we may approximate the three-vector $\ve{r}_A\left({\rm t}\right)$ in (\ref{Example_Integral_1}) by it's Newtonian approximation,  
\begin{eqnarray}
\ve{r}^{\rm N}_A\left(t\right) &=& \ve{x}_0 + c \left(t - t_0\right) \ve{\sigma} - \ve{x}_A\left(t\right),
\label{Example_20}
\end{eqnarray}

\noindent
due to $\ve{r}_A\left(t\right) = \ve{r}^{\rm N}_A\left(t\right) + {\cal O}\left(c^{-2}\right)$. 
Then, the integral (\ref{Example_Integral_1}) reads 
\begin{eqnarray}
\ve{{\cal I}}_{A_1}\left(t\right) = - 2\,m_A \int\limits_{-\infty}^t d c{\rm t}\,
\frac{\ve{r}^{\rm N}_A\left({\rm t}\right)}{\left(r^{\rm N}_A\left({\rm t}\right)\right)^3}
\,\frac{v^2_A\left({\rm t}\right)}{c^2} + {\cal O}\left(c^{-5}\right).
\label{Example_Integral_1b}
\end{eqnarray}

\noindent  
First of all, the numerator $\ve{r}_A^{\rm N}$ is rewritten  
in the form $\ve{r}_A^{\rm N} = \ve{d}_A^{\rm N} + \left(\ve{\sigma} \cdot \ve{r}_A^{\rm N}\right)\,\ve{\sigma}$  
which follows from the definition (\ref{Impact_Vector_1}), so we have  
\begin{eqnarray}
\fl \ve{{\cal I}}_{A_1}\left(t\right) = - 2\,m_A \int\limits_{-\infty}^t d c{\rm t}\,
\frac{\ve{d}^{\rm N}_A\left({\rm t}\right)}{\left(r^{\rm N}_A\left({\rm t}\right)\right)^3}
\,\frac{v^2_A\left({\rm t}\right)}{c^2} 
- 2\,m_A\,\ve{\sigma} \int\limits_{-\infty}^t d c{\rm t}\,
\frac{\ve{\sigma} \cdot \ve{r}^{\rm N}_A\left({\rm t}\right)}{\left(r^{\rm N}_A\left({\rm t}\right)\right)^3}
\,\frac{v^2_A\left({\rm t}\right)}{c^2} + {\cal O}\left(c^{-5}\right).  
\nonumber\\ 
\label{Example_Integral_1_A}
\end{eqnarray}

\noindent
The only unknown in (\ref{Example_Integral_1_A}) is the worldline of the body, $\ve{x}_A\left(t\right)$, which necessitates integration by parts.  
Using the relations (\ref{Appendix_Time_Derivative_3}) and (\ref{Appendix_Time_Derivative_7}) in \ref{Appendix_Integration_by_Parts} we conclude that  
\begin{eqnarray}
\ve{{\cal I}}_{A_1}\left(t\right) &=& - 2\,m_A \! \int\limits_{-\infty}^t \!\! d c{\rm t}\,
\left[\frac{d}{d c {\rm t}}\frac{1}{r^{\rm N}_A\left({\rm t}\right)}
\frac{1}{r^{\rm N}_A\left({\rm t}\right) - \ve{\sigma} \cdot \ve{r}^{\rm N}_A\left({\rm t}\right)}\right]
\ve{d}_A^{\rm N}\left({\rm t}\right)\,\frac{v^2_A\left({\rm t}\right)}{c^2} 
\nonumber\\ 
\nonumber\\ 
&& + 2\,m_A\,\ve{\sigma} \!\! \int\limits_{-\infty}^t \! d c{\rm t}
\left[\frac{d}{d c {\rm t}}\frac{1}{r^{\rm N}_A\left({\rm t}\right)}\right]\,\frac{v^2_A\left({\rm t}\right)}{c^2} 
+ {\cal O}\left(c^{-5}\right).  
\label{Example_Integral_1_B}
\end{eqnarray}

\noindent
Now we are in the position to integrate by parts. With the aid of (\ref{Integration_by_Part}) and recalling relation (\ref{Time_Derivative_Impact_Vector_1})  
we obtain for the integral (\ref{Example_Integral_1}) the solution  
\begin{eqnarray}
\fl \ve{{\cal I}}_{A_1}\left(t\right) = - 2\,m_A\,\frac{\ve{d}^{\rm N}_A\left(t\right)}{r^{\rm N}_A\left(t\right)}\,
\frac{1}{r^{\rm N}_A\left(t\right) - \ve{\sigma} \cdot \ve{r}^{\rm N}_A\left(t\right)}\,\frac{v^2_A\left(t\right)}{c^2}  
+ 2\,m_A\,\frac{\ve{\sigma}}{r^{\rm N}_A\left(t\right)}\,\frac{v^2_A\left(t\right)}{c^2} + {\cal O}\left(c^{-5}\right).   
\label{Solution_Example_1}
\end{eqnarray}

\noindent
Let us also consider the integral in (\ref{Example_Integral_4}). The prefactor $m_A^2 \sim {\cal O}\left(c^{-4}\right)$ allows to 
approximate the three-vector $\ve{r}_A\left({\rm t}\right)$ in (\ref{Example_Integral_4}) by it's Newtonian approximation (\ref{Example_20}) 
due to $\ve{r}_A\left({\rm t}\right) = \ve{r}^{\rm N}_A\left({\rm t}\right) + {\cal O}\left(c^{-2}\right)$, so we get 
\begin{eqnarray}
\ve{{\cal I}}_{A_4}\left(t\right) &=& + 2\,m_A^2 \int\limits_{-\infty}^t d c{\rm t}\,
\frac{\ve{r}^{\rm N}_A\left({\rm t}\right)}{\left(r^{\rm N}_A\left({\rm t}\right)\right)^4} + {\cal O}\left(c^{-5}\right)\,. 
\label{Example_Integral_4_A}
\end{eqnarray}
 
\noindent 
Rewriting the numerator of that integral in the same way as in the previous example we obtain  
\begin{eqnarray}
\fl \ve{{\cal I}}_{A_4}\left(t\right) = + 2\,m_A^2 \int\limits_{-\infty}^t d c{\rm t}\,  
\frac{\ve{d}^{\rm N}_A\left({\rm t}\right)}{\left(r^{\rm N}_A\left({\rm t}\right)\right)^4} 
+ 2\,m_A^2 \ve{\sigma} \int\limits_{-\infty}^t d c{\rm t}\, 
\frac{\ve{\sigma} \cdot \ve{r}^{\rm N}_A\left({\rm t}\right)}{\left(r^{\rm N}_A\left({\rm t}\right)\right)^4} + {\cal O}\left(c^{-5}\right).  
\label{Example_Integral_4_B}
\end{eqnarray}

\noindent
By means of the relations (\ref{Appendix_Time_Derivative_4}) and (\ref{Appendix_Time_Derivative_7}) in \ref{Appendix_Integration_by_Parts} 
we conclude that  
\begin{eqnarray}
\fl \ve{{\cal I}}_{A_4}\left(t\right) = + m_A^2 \!\int\limits_{-\infty}^t \! d c{\rm t}
\left[\frac{d}{d c {\rm t}}\left(\!\frac{\ve{\sigma}\cdot\ve{r}_A^{\rm N}\left({\rm t}\right)}{\left(d_A^{\rm N}\left({\rm t}\right)\right)^2
\left(r_A^{\rm N}\left({\rm t}\right)\right)^2}
+ \frac{1}{\left(d_A^{\rm N}\left({\rm t}\right)\right)^3}
\arctan \frac{\ve{\sigma}\cdot\ve{r}_A^{\rm N}\left({\rm t}\right)}{d_A^{\rm N}\left({\rm t}\right)}\right)\right]
\ve{d}^{\rm N}_A\left({\rm t}\right)  
\nonumber\\
\nonumber\\
\fl \hspace{1.7cm} - m_A^2\, \ve{\sigma} \int\limits_{-\infty}^t d c{\rm t}\,
\left[\frac{d}{d c {\rm t}}\,\frac{1}{\left(r^{\rm N}_A\left({\rm t}\right)\right)^2}\right] + {\cal O}\left(c^{-5}\right).  
\label{Example_Integral_4_C}
\end{eqnarray}

\noindent 
Then, integration by parts (\ref{Integration_by_Part}) and recalling relation (\ref{Time_Derivative_Impact_Vector_1}) we obtain 
for the integral (\ref{Example_Integral_4}) the solution  
\begin{eqnarray}
\ve{{\cal I}}_{A_4}\left(t\right) &=& + m_A^2 \frac{\ve{d}^{\rm N}_A\left(t\right)}{\left(d^{\rm N}_A\left(t\right)\right)^2}
\frac{\ve{\sigma} \cdot \ve{r}^{\rm N}_A\left(t\right)}{\left(r^{\rm N}_A\left(t\right)\right)^2} 
+ m_A^2 \frac{\ve{d}^{\rm N}_A\left(t\right)}{\left(d^{\rm N}_A\left(t\right)\right)^3} 
\arctan \frac{\ve{\sigma} \cdot \ve{r}^{\rm N}_A\left(t\right)}{d^{\rm N}_A\left(t\right)} 
\nonumber\\
\nonumber\\
&& - m_A^2 \frac{\ve{\sigma}}{\left(r^{\rm N}_A\left(t\right)\right)^2} 
+ \frac{\pi}{2} m_A^2\frac{\ve{d}^{\rm N}_A\left( - \infty\right)}{\left(d^{\rm N}_A\left( - \infty\right)\right)^3} 
+ {\cal O}\left(c^{-5}\right).  
\label{Solution_Example_4}
\end{eqnarray}

\noindent
The last term in (\ref{Solution_Example_4}) originates from the lower limit of integration and is finite  
because the orbital motion of the body is supposed to be inside the domain of the near-zone of the Solar System.  
On the other side, such terms cannot be determined 
exactly because there is no exact statement for the spatial position of the body at past timelike infinity $\ve{x}_A\left( - \infty\right)$, hence  
$\ve{d}^{\rm N}_A\left( - \infty\right)$ remains an unknown parameter. It is, therefore, a remarkable and providential feature that all integration constants 
cancel against each other in the final result for the coordinate velocity of the light signal.

\subsection{Second example of the first integration}\label{First_Integration_Second_Example}  

Now let us look at the first integration (\ref{Integral_1}) of terms in the first line of geodesic equation (\ref{Geodesic_Equation3}).  
For instance, let us consider the integration of the first term on the first line in (\ref{Geodesic_Equation3}),  
\begin{eqnarray}
\ve{{\cal I}}_B\left(t\right) = \frac{1}{2} \int\limits_{-\infty}^t d c{\rm t}\,
\frac{\partial h_{00}^{(2)}\left({\rm t},\ve{x}\right)}{\partial \ve{x}}\Bigg|_{\ve{x}=\ve{x}\mbox{\normalsize $\left({\rm t}\right)$}}
= - m_A \int\limits_{-\infty}^t d c{\rm t}\,
\frac{\ve{r}_A\left({\rm t}\right)}{r^3_A\left({\rm t}\right)}\,.  
\label{Example_Integral_5}
\end{eqnarray}

\noindent
In view of the prefactor $m_A \sim {\cal O}\left(c^{-2}\right)$ the three-vector $\ve{r}_A\left({\rm t}\right)$ in (\ref{Example_Integral_5}) can be approximated 
by it's first post-Newtonian approximation,  
\begin{eqnarray}
\ve{r}^{\rm 1PN}_A\left(t\right) &=& \ve{x}_0 + c \left(t - t_0\right) \ve{\sigma}
+ \Delta \ve{x}_{\rm 1PN}\left(t,t_0\right) - \ve{x}_A\left(t\right),
\label{Example_55}
\end{eqnarray}

\noindent
because of $\ve{r}_A\left({\rm t}\right) = \ve{r}^{\rm 1PN}_A\left({\rm t}\right) + {\cal O}\left(c^{-3}\right)$, and where  
the perturbation $\Delta \ve{x}_{\rm 1PN}\left(t,t_0\right)$ has been given by Eqs.~(\ref{Delta_t1}) and (\ref{Light_Trajectory_M_1}). Hence 
\begin{eqnarray}
\ve{{\cal I}}_B\left(t\right) &=& - m_A \int\limits_{-\infty}^t d c{\rm t}\,
\frac{\ve{r}^{\rm 1PN}_A\left({\rm t}\right)}{\left(r^{\rm 1PN}_A\left({\rm t}\right)\right)^3} + {\cal O}\left(c^{-5}\right)\,.
\label{Example_Integral_5_A}
\end{eqnarray}

\noindent 
We write the three-vector (\ref{Example_55}) and its absolute value in the following form 
\begin{eqnarray}
\ve{r}^{\rm 1PN}_A\left(t\right) &=& \ve{r}^{\rm N}_A\left(t\right) + \Delta \ve{x}_{\rm 1PN}\left(t,t_0\right),  
\label{Example_Integral_5_B}
\\
\nonumber\\
r^{\rm 1PN}_A\left(t\right) &=& r^{\rm N}_A\left(t\right)
+ \frac{\ve{r}^{\rm N}_A\left(t\right) \cdot \Delta \ve{x}_{\rm 1PN}\left(t,t_0\right)}{r^{\rm N}_A\left(t\right)} + {\cal O}\left(c^{-3}\right). 
\label{Example_Integral_5_C}
\end{eqnarray}

\noindent
Then, inserting (\ref{Example_Integral_5_B}) and (\ref{Example_Integral_5_C}) into (\ref{Example_Integral_5_A}) and series expansion yields   
\begin{eqnarray}
\fl \ve{{\cal I}}_B\left(t\right) = \ve{{\cal I}}_{B_1}\left(t\right) + \ve{{\cal I}}_{B_2}\left(t\right) + \ve{{\cal I}}_{B_3}\left(t\right),  
\label{Example_Integral_5_D}
\\
\nonumber\\
\fl \ve{{\cal I}}_{B_1}\left(t\right) = 
- m_A \int\limits_{-\infty}^t d c{\rm t}\,\frac{\ve{r}^{\rm N}_A\left({\rm t}\right)}{\left(r^{\rm N}_A\left({\rm t}\right)\right)^3} 
+ {\cal O}\left(c^{-5}\right),  
\label{Example_Integral_5_E}
\\
\nonumber\\
\fl \ve{{\cal I}}_{B_2}\left(t\right) =  
+ 3\,m_A \int\limits_{-\infty}^t d c{\rm t}\,
\frac{\ve{r}^{\rm N}_A\left({\rm t}\right)}{\left(r^{\rm N}_A\left({\rm t}\right)\right)^5}
\left(\ve{r}^{\rm N}_A\left({\rm t}\right) \cdot \Delta \ve{x}_{\rm 1PN}\left({\rm t},t_0\right)\right) 
+ {\cal O}\left(c^{-5}\right),  
\label{Example_Integral_5_F}
\\
\nonumber\\
\fl \ve{{\cal I}}_{B_3}\left(t\right) = 
- m_A \int\limits_{-\infty}^t d c{\rm t}\, 
\frac{\Delta \ve{x}_{\rm 1PN}\left({\rm t},t_0\right)}{\left(r^{\rm N}_A\left({\rm t}\right)\right)^3} + {\cal O}\left(c^{-5}\right).  
\label{Example_Integral_5_G}
\end{eqnarray}

\noindent
The integrands of these integrals depend on the unknown worldline of the body $\ve{x}_A\left(t\right)$ and can be solved by integration by parts.  
For the determination of the integral (\ref{Example_Integral_5_E}) one has to rewrite the numerator in the form  
$\ve{r}_A^{\rm N} = \ve{d}_A^{\rm N} + \left(\ve{\sigma} \cdot \ve{r}_A^{\rm N}\right)\,\ve{\sigma}$, while for the denominator 
one has to insert the exact relation (\ref{Exact_Time_Derivative_3}). Moreover, a second integration by parts is necessary 
which also yields terms proportional to $m_A\,v_A^2$.     
On the other side, the integral in (\ref{Example_Integral_5_F}) can be solved by one 
integration by parts. Here we will restrict our exemplifying considerations on the evaluation of the integral in (\ref{Example_Integral_5_G}). 
By inserting the expressions (\ref{Delta_t1}) and (\ref{Light_Trajectory_M_1}) into (\ref{Example_Integral_5_G}) one obtains  
\begin{eqnarray} 
\fl \ve{{\cal I}}_{B_3}\left(t\right) =
+ 2\,m_A^2 \int\limits_{-\infty}^t d c{\rm t}\,
\frac{1}{\left(r^{\rm N}_A\left({\rm t}\right)\right)^3}\,
\frac{\ve{d}_A^{\rm N}\left({\rm t}\right)}{r^{\rm N}_A\left({\rm t}\right) - \ve{\sigma} \cdot \ve{r}^{\rm N}_A\left({\rm t}\right)} 
\nonumber\\ 
\nonumber\\ 
\fl - 2\,m_A^2 \frac{\ve{d}_A^{\rm N}\left(t_0\right)}{r^{\rm N}_A\left(t_0\right) - \ve{\sigma} \cdot \ve{r}^{\rm N}_A\left(t_0\right)}
\int\limits_{-\infty}^t d c{\rm t}\,
\frac{1}{\left(r^{\rm N}_A\left({\rm t}\right)\right)^3}\,
 - 2\,m_A^2\,\ve{\sigma} \! \! \int\limits_{-\infty}^t d c{\rm t}\,
\frac{\ln \left(r^{\rm N}_A\left({\rm t}\right) - \ve{\sigma} \cdot \ve{r}^{\rm N}_A\left({\rm t}\right)\right)}{\left(r^{\rm N}_A\left({\rm t}\right)\right)^3} 
\nonumber\\ 
\nonumber\\ 
\fl + 2\,m_A^2 \ln \left(r^{\rm N}_A\left(t_0\right) - \ve{\sigma} \cdot \ve{r}^{\rm N}_A\left(t_0\right)\right) \ve{\sigma} 
\int\limits_{-\infty}^t d c{\rm t}\,\frac{1}{\left(r^{\rm N}_A\left({\rm t}\right)\right)^3}  
+ {\cal O}\left(c^{-5}\right).
\label{Example_Integral_5_F_1}
\end{eqnarray}

\noindent
The integrand of the first integral in (\ref{Example_Integral_5_F_1}) is rewritten by means of  
\begin{eqnarray}
\frac{1}{r^{\rm N}_A\left({\rm t}\right) - \ve{\sigma} \cdot \ve{r}^{\rm N}_A\left({\rm t}\right)}
&=& \frac{r^{\rm N}_A\left({\rm t}\right) + \ve{\sigma} \cdot \ve{r}^{\rm N}_A\left({\rm t}\right)}{\left(d^{\rm N}_A\left({\rm t}\right)\right)^2}\,.  
\label{Example_Integral_5_F_2}
\end{eqnarray}

\noindent 
Then, using relations (\ref{Appendix_Time_Derivative_2}), (\ref{Appendix_Time_Derivative_3}) and (\ref{Appendix_Time_Derivative_7}) 
in \ref{Appendix_Integration_by_Parts} we may rewrite the integral in (\ref{Example_Integral_5_F_1}) as follows, 
\begin{eqnarray}
\fl \ve{{\cal I}}_{B_3}\left(t\right) =
+ 2\,m_A^2 \int\limits_{-\infty}^t d c{\rm t}\,
\left[\frac{d}{d c {\rm t}} 
\left(\frac{1}{d^{\rm N}_A\left({\rm t}\right)}\,\arctan \frac{\ve{\sigma} \cdot \ve{r}^{\rm N}_A\left({\rm t}\right)}{d^{\rm N}_A\left({\rm t}\right)}  
- \frac{1}{r^{\rm N}_A\left({\rm t}\right)}\right) \right] \frac{\ve{d}^{\rm N}_A\left({\rm t}\right)}{\left(d^{\rm N}_A\left({\rm t}\right)\right)^2}  
\nonumber\\
\nonumber\\
\fl - 2\,m_A^2 \frac{\ve{d}_A^{\rm N}\left(t_0\right)}{r^{\rm N}_A\left(t_0\right) - \ve{\sigma} \cdot \ve{r}^{\rm N}_A\left(t_0\right)}
\int\limits_{-\infty}^t d c{\rm t}\,
\left[\frac{d}{d c {\rm t}}\,\frac{1}{r^{\rm N}_A\left({\rm t}\right)}\,
\frac{1}{r^{\rm N}_A\left({\rm t}\right) - \ve{\sigma} \cdot \ve{r}^{\rm N}_A\left({\rm t}\right)} \right]  
\nonumber\\
\nonumber\\
\fl - 2\,m_A^2\,\ve{\sigma} \! \int\limits_{-\infty}^t d c{\rm t}\,
\ln \left(r^{\rm N}_A\left({\rm t}\right) - \ve{\sigma} \cdot \ve{r}^{\rm N}_A\left({\rm t}\right)\right)
\left[\frac{d}{d c {\rm t}}\,\frac{1}{r^{\rm N}_A\left({\rm t}\right)}\,
\frac{1}{r^{\rm N}_A\left({\rm t}\right) - \ve{\sigma} \cdot \ve{r}^{\rm N}_A\left({\rm t}\right)} \right]
\nonumber\\
\nonumber\\
\fl + 2\,m_A^2 \ln \left(r^{\rm N}_A\left(t_0\right) - \ve{\sigma} \cdot \ve{r}^{\rm N}_A\left(t_0\right)\right) \ve{\sigma} 
\! \! \! \int\limits_{-\infty}^t d c{\rm t} 
\left[\frac{d}{d c {\rm t}}\,\frac{1}{r^{\rm N}_A\left({\rm t}\right)}
\frac{1}{r^{\rm N}_A\left({\rm t}\right) - \ve{\sigma} \cdot \ve{r}^{\rm N}_A\left({\rm t}\right)} \right] 
+ {\cal O}\left(c^{-5}\right)\!.
\label{Example_Integral_5_F_3}
\end{eqnarray}

\noindent
Now we can perform the integration by parts according to relation (\ref{Integration_by_Part}). While the integrals in the first, second, and fourth line 
are straightforward, the evaluation of the integral in the third line implies a second integration by parts and runs as follows: 
\begin{eqnarray}
\fl - 2\,m_A^2\,\ve{\sigma} \! \int\limits_{-\infty}^t d c{\rm t}\,
\ln \left(r^{\rm N}_A\left({\rm t}\right) - \ve{\sigma} \cdot \ve{r}^{\rm N}_A\left({\rm t}\right)\right)
\left[\frac{d}{d c {\rm t}}\,\frac{1}{r^{\rm N}_A\left({\rm t}\right)}\,
\frac{1}{r^{\rm N}_A\left({\rm t}\right) - \ve{\sigma} \cdot \ve{r}^{\rm N}_A\left({\rm t}\right)} \right]
\nonumber\\ 
\nonumber\\ 
\fl = - 2\,m_A^2\,
\frac{\ve{\sigma}}{r^{\rm N}_A\left(t\right)}\,
\frac{\ln \left(r^{\rm N}_A\left(t\right) - \ve{\sigma} \cdot \ve{r}^{\rm N}_A\left(t\right)\right)}
{r^{\rm N}_A\left(t\right) - \ve{\sigma} \cdot \ve{r}^{\rm N}_A\left(t\right)} 
- 2\,m_A^2\,\ve{\sigma} \! \int\limits_{-\infty}^t d c{\rm t}\,
\frac{1}{\left(r^{\rm N}_A\left({\rm t}\right)\right)^2}\,\frac{1}{r^{\rm N}_A\left({\rm t}\right) - \ve{\sigma} \cdot \ve{r}^{\rm N}_A\left({\rm t}\right)}\,,  
\nonumber\\ 
\label{Example_Integral_5_F_4}
\end{eqnarray}

\noindent
where we have used relation (\ref{Appendix_Time_Derivative_1}) in \ref{Appendix_Integration_by_Parts}. The remaining integral in the second line  
of (\ref{Example_Integral_5_F_4}) is treated by means of relation (\ref{Example_Integral_5_F_2}) as well as relations  
(\ref{Appendix_Time_Derivative_6B}) and (\ref{Appendix_Time_Derivative_1}) in \ref{Appendix_Integration_by_Parts}. Altogether, we obtain for 
the integral in (\ref{Example_Integral_5_F_3}) the following solution,  
\begin{eqnarray}
\fl \ve{{\cal I}}_{B_3}\left(t\right) =
+ 2\,m_A^2 
\left(\frac{1}{d^{\rm N}_A\left(t\right)}\,\arctan \frac{\ve{\sigma} \cdot \ve{r}^{\rm N}_A\left(t\right)}{d^{\rm N}_A\left(t\right)}
- \frac{1}{r^{\rm N}_A\left(t\right)}\right) \frac{\ve{d}^{\rm N}_A\left(t\right)}{\left(d^{\rm N}_A\left(t\right)\right)^2}
\nonumber\\
\nonumber\\
\fl - 2\,\frac{m_A^2}{r^{\rm N}_A\left(t\right)}\,\frac{1}{r^{\rm N}_A\left(t\right) - \ve{\sigma} \cdot \ve{r}^{\rm N}_A\left(t\right)}  
 \frac{\ve{d}_A^{\rm N}\left(t_0\right)}{r^{\rm N}_A\left(t_0\right) - \ve{\sigma} \cdot \ve{r}^{\rm N}_A\left(t_0\right)}
+ 2\,\frac{m_A^2}{\left(d^{\rm N}_A\left(t\right)\right)^2} \,\ve{\sigma}\,
\ln \frac{r^{\rm N}_A\left(t\right) - \ve{\sigma} \cdot \ve{r}^{\rm N}_A\left(t\right)}{r^{\rm N}_A\left(t\right)} 
\nonumber\\
\nonumber\\
\fl - 2\,\frac{m_A^2}{r^{\rm N}_A\left(t\right)}\,\frac{\ve{\sigma}}{r^{\rm N}_A\left(t\right) - \ve{\sigma} \cdot \ve{r}^{\rm N}_A\left(t\right)}\, 
\ln \frac{r^{\rm N}_A\left(t\right) - \ve{\sigma} \cdot \ve{r}^{\rm N}_A\left(t\right)}
{r^{\rm N}_A\left(t_0\right) - \ve{\sigma} \cdot \ve{r}^{\rm N}_A\left(t_0\right)} 
\nonumber\\
\nonumber\\
\fl + \pi\, m_A^2\frac{\ve{d}^{\rm N}_A\left( - \infty\right)}{\left(d^{\rm N}_A\left( - \infty\right)\right)^3}
- 2\,\frac{m_A^2}{\left(d^{\rm N}_A\left(-\infty\right)\right)^2}\,\ve{\sigma}\,\ln \left(2\right)  
+ {\cal O}\left(c^{-5}\right), 
\label{Example_Integral_5_F_5}
\end{eqnarray} 

\noindent
where the last term originates from 
$\displaystyle \lim_{t \rightarrow - \infty} \ln \frac{r^{\rm N}_A - \ve{\sigma} \cdot \ve{r}^{\rm N}_A}{r^{\rm N}_A} = \ln \left(2\right)$.  
As mentioned above, the terms in the last line of (\ref{Example_Integral_5_F_5}) represent finite constants and do cancel each other  
in the final solution of the first integration of geodesic equation. The complete cancelation of all these integration constants 
is used as an important check for the correctness of the entire integration procedure.  
Furthermore, we notice that all the logarithms of the structure $\ln r^{\rm N}_A\left(t\right)$ as appearing in the last term of the
second line in (\ref{Example_Integral_5_F_5}) do also cancel each other in the final solution for the first integration of geodesic equation.

\subsection{The intermediate result of the first integration}\label{First_Integration_Intermediate_Result}  

In general, the first integration (\ref{Integral_1}) of geodesic equation (\ref{Geodesic_Equation3}) is performed along this procedure 
which has been elucidated by the examples in (\ref{Example_1}) and (\ref{Example_Integral_5}). In the \ref{Appendix_Integration_by_Parts}  
a list of all relevant relations are presented which are required in order to execute each integral by means of integration by parts. 
While the evaluation of all these integrals is actually not as much tricky, the integration procedure becomes rather cumbersome in view of the considerable
amount of integrals and individual terms so that the algebraic expressions become very lengthy. Therefore, the computer algebra system Maple \cite{Maple}  
has extensively been used in order to facilitate and to assist the entire integration procedure. 
Basically, there are two primary steps:  
(i) Each individual integral is calculated analytically and cross-checked by differentiation.
(ii) Summarize and simplify all those three-scalars which have one and the same vectorial prefactor           
(there are only four linearly independent vectorial prefactors: $\ve{\sigma}, \ve{v}_A, \ve{a}_A$ and either $\ve{d}^{\rm N}_A$ or $\ve{r}^{\rm N}_A$).
It is important to realize that  
the final solution of that procedure yields the coordinate velocity as function of $\ve{r}^{\rm N}_A\left(t\right)$. Afterwards,  
by means of Eqs.~(\ref{Example_Integral_5_B}) and (\ref{Example_Integral_5_C}), one has to rewrite this solution in terms of  
$\ve{r}^{\rm 1PN}_A\left(t\right)$ and $\ve{r}^{\rm N}_A\left(t\right)$. Then, for the formal solution one arrives at the following structure:  
\begin{eqnarray}
\fl \frac{\dot{\ve{x}}_{\rm 2PN}\left(t\right)}{c} = \ve{\sigma} + m_A\,\ve{F}_1\left(\ve{r}^{\rm 1PN}_A\left(t\right)\right)  
+ m_A\,\ve{F}_2\left(\ve{r}^{\rm N}_A\left(t\right),\ve{v}_A\left(t\right),\ve{a}_A\left(t\right)\right) 
+ m_A^2\,\ve{F}_3\left(\ve{r}^{\rm N}_A\left(t\right)\right),  
\nonumber\\ 
\label{First_Integration_Preliminary}
\end{eqnarray}

\noindent 
where $\ve{r}^{\rm N}_A\left(t\right)$ and $\ve{r}^{\rm 1PN}_A\left(t\right)$ are given by Eqs.~(\ref{Example_20}) and (\ref{Example_55}), respectively.  
That means, the solution for the coordinate velocity of the light signal (\ref{First_Integration_Preliminary}) is given in terms of the  
instantaneous position $\ve{x}_A\left(t\right)$ of the massive body. As it has been enlightened by the examples and as mentioned above, 
the solution in (\ref{First_Integration_Preliminary}), that means the vectorial functions $\ve{F}_1$ and $\ve{F}_2$ and $\ve{F}_3$,  
are of rather involved structure and as an intermediate result will not be presented here, but they  
can be retrieved from the final result for the coordinate velocity (\ref{First_Integration}) by series expansion in inverse powers of the speed of light.  
As emphasized above, the effect of retardation is hidden in the intermediate solution (\ref{First_Integration_Preliminary}). This fact   
allows us to rewrite (\ref{First_Integration_Preliminary}) in terms of the retarded position of the body $\ve{x}_A\left(s\right)$, which will 
lead us to the considerably simpler expression in (\ref{First_Integration}) for the coordinate velocity of the light signal. This fact will be the subject of 
Section \ref{Section3b}. We stress that both the expressions, Eq.~(\ref{First_Integration_Preliminary}) and Eq.~(\ref{First_Integration}),  
are identical up to terms of the order ${\cal O}\left(c^{-5}\right)$.

\subsection{First example of the second integration}\label{Second_Integration_First_Example}  

The second integration of geodesic equation yields the trajectory of the light signal and is defined by Eq.~(\ref{Integral_2}) where the  
integrand is given by (\ref{First_Integration_Preliminary}). 

In Section \ref{First_Integration_First_Example} we have considered the first integration of the first term on the fourth line  
in (\ref{Geodesic_Equation3}). Now we will consider the second integration of this term, that means the following integral
\begin{eqnarray}
\ve{{\cal J}}_A\left(t,t_0\right) = \int\limits_{t_0}^t d c{\rm t}\,\ve{{\cal I}}_{A}\left({\rm t}\right),  
\label{Example_A1}
\end{eqnarray}

\noindent
where the integrand is given by Eq.~(\ref{Example_1}). Accordingly, we have to consider  
\begin{eqnarray}
\ve{{\cal J}}_A\left(t,t_0\right) = \ve{{\cal J}}_{A_1}\left(t,t_0\right) + \ve{{\cal J}}_{A_2}\left(t,t_0\right) + \ve{{\cal J}}_{A_3}\left(t,t_0\right) + \ve{{\cal J}}_{A_4}\left(t,t_0\right),  
\label{Example_A2}
\end{eqnarray}

\noindent 
where 
\begin{eqnarray}
\ve{{\cal J}}_{A_n}\left(t,t_0\right) &=& \int\limits_{t_0}^t d c{\rm t}\,\ve{{\cal I}}_{A_n}\left({\rm t}\right) \quad {\rm with} \quad n=1,2,3,4\,,
\label{Example_IntegralA_n}
\end{eqnarray}
 
\noindent
and the functions $\ve{{\cal I}}_{A_n}\left({\rm t}\right)$ are given by (\ref{Example_Integral_1}) - (\ref{Example_Integral_4}). 
Explicit solutions were presented for $\ve{{\cal I}}_{A_1}$ in Eq.~(\ref{Solution_Example_1}) and $\ve{{\cal I}}_{A_4}$ in Eq.~(\ref{Solution_Example_4}).  
As typical example let us consider $\ve{\cal J}_{A_1}$. By inserting the function $\ve{{\cal I}}_{A_1}$ in (\ref{Solution_Example_1}) 
into (\ref{Example_IntegralA_n}) we get the integral  
\begin{eqnarray}
\fl \ve{\cal J}_{A_1}\left(t,t_0\right) = - \, 2\,m_A \int\limits_{t_0}^t d c{\rm t}\, 
\bigg[\frac{\ve{d}^{\rm N}_A\left({\rm t}\right)}{r^{\rm N}_A\left({\rm t}\right)}\,
\frac{1}{r^{\rm N}_A\left({\rm t}\right) - \ve{\sigma} \cdot \ve{r}^{\rm N}_A\left({\rm t}\right)}\,\frac{v^2_A\left({\rm t}\right)}{c^2} 
- \ve{\sigma}\,\frac{1}{r^{\rm N}_A\left({\rm t}\right)}\,\frac{v^2_A\left({\rm t}\right)}{c^2}\bigg].   
\label{Example_B} 
\end{eqnarray}

\noindent
By means of relations (\ref{Appendix_Time_Derivative_1}) and (\ref{Appendix_Time_Derivative_14}) in \ref{Appendix_Integration_by_Parts} we  
may rewrite that integral as follows,  
\begin{eqnarray}
\fl \ve{\cal J}_{A_1}\left(t,t_0\right) = - \, 2\,m_A \int\limits_{t_0}^t d c{\rm t}\,
\bigg[\frac{d}{d c {\rm t}}\,\frac{1}{r^{\rm N}_A\left({\rm t}\right) - \ve{\sigma} \cdot \ve{r}^{\rm N}_A\left({\rm t}\right)}\bigg]  
\ve{d}^{\rm N}_A\left({\rm t}\right)\frac{v^2_A\left({\rm t}\right)}{c^2}
\nonumber\\
\nonumber\\
\fl \hspace{2.0cm} - \, 2\,m_A \,\ve{\sigma} \int\limits_{t_0}^t d c{\rm t}\, \bigg[\frac{d}{d c {\rm t}}\, 
\ln \left(r^{\rm N}_A\left({\rm t}\right) - \ve{\sigma} \cdot \ve{r}^{\rm N}_A\left({\rm t}\right)\right)\bigg] 
\frac{v^2_A\left({\rm t}\right)}{c^2} + {\cal O}\left(c^{-5}\right).  
\label{Example_B1}
\end{eqnarray}

\noindent  
Now we may integrate by parts according to (\ref{Integration_by_Part}) and obtain the following solution for the integral,  
\begin{eqnarray}
\fl \ve{\cal J}_{A_1}\left(t,t_0\right) = 
- 2\,m_A\left(\frac{\ve{d}^{\rm N}_A\left(t\right)}{r^{\rm N}_A\left(t\right) - \ve{\sigma} \cdot \ve{r}^{\rm N}_A\left(t\right)}\,\frac{v^2_A\left(t\right)}{c^2} 
- \frac{\ve{d}^{\rm N}_A\left(t_0\right)}{r^{\rm N}_A\left(t_0\right) - \ve{\sigma} \cdot \ve{r}^{\rm N}_A\left(t_0\right)}\,\frac{v^2_A\left(t_0\right)}{c^2}\right)
\nonumber\\
\nonumber\\
\fl - 2 m_A \ve{\sigma}  
\left( \frac{v^2_A\left(t\right)}{c^2} \ln \left(r^{\rm N}_A\left(t\right) - \ve{\sigma} \cdot \ve{r}^{\rm N}_A\left(t\right)\right) 
- \frac{v^2_A\left(t_0\right)}{c^2} \ln \left(r^{\rm N}_A\left(t_0\right) - \ve{\sigma} \cdot \ve{r}^{\rm N}_A\left(t_0\right)\right) \right) 
+ {\cal O}\left(c^{-5}\right)\!.   
\nonumber\\ 
\label{Example_D}
\end{eqnarray}

\noindent 
As further example we will consider $\ve{\cal J}_{A_4}$. By inserting the function $\ve{{\cal I}}_{A_4}$ in (\ref{Solution_Example_4}) 
into (\ref{Example_IntegralA_n}) we get the integral  
\begin{eqnarray}
\fl \ve{\cal J}_{A_4}\left(t,t_0\right) = m_A^2 \! \int\limits_{t_0}^t \! d c{\rm t}
\bigg[\frac{\ve{d}^{\rm N}_A\left({\rm t}\right)}{\left(d^{\rm N}_A\left({\rm t}\right)\right)^2}
\frac{\ve{\sigma} \cdot \ve{r}^{\rm N}_A\left({\rm t}\right)}{\left(r^{\rm N}_A\left({\rm t}\right)\right)^2}
+ \frac{\ve{d}^{\rm N}_A\left({\rm t}\right)}{\left(d^{\rm N}_A\left({\rm t}\right)\right)^3}
\arctan \frac{\ve{\sigma} \cdot \ve{r}^{\rm N}_A\left({\rm t}\right)}{d^{\rm N}_A\left({\rm t}\right)}
- \frac{\ve{\sigma}}{\left(r^{\rm N}_A\left({\rm t}\right)\right)^2} \bigg], 
\nonumber\\
\label{Example_C1}
\end{eqnarray}

\noindent 
where the integration constant (i.e. the last term in the second line of (\ref{Solution_Example_4})) is omitted because all integration constants 
cancel each other. Subject to relations (\ref{Appendix_Time_Derivative_2}), (\ref{Appendix_Time_Derivative_6B}),  
and (\ref{Appendix_Time_Derivative_17}) in \ref{Appendix_Integration_by_Parts} we may rewrite the integral as follows,  
\begin{eqnarray}
\fl \ve{\cal J}_{A_4}\left(t,t_0\right) = + m_A^2 \int\limits_{t_0}^t d c{\rm t}
\bigg[\frac{d}{d c {\rm t}} \ve{\sigma} \cdot \ve{r}^{\rm N}_A\left({\rm t}\right)\, 
\arctan \frac{\ve{\sigma} \cdot \ve{r}^{\rm N}_A\left({\rm t}\right)}{d^{\rm N}_A\left({\rm t}\right)}\bigg] 
\frac{\ve{d}^{\rm N}_A\left({\rm t}\right)}{\left(d^{\rm N}_A\left({\rm t}\right)\right)^3}  
\nonumber\\
\nonumber\\
\fl \hspace{2.0cm} - m_A^2\,\ve{\sigma} \int\limits_{t_0}^t d c{\rm t} 
\bigg[\frac{d}{d c {\rm t}} \,\frac{1}{d^{\rm N}_A\left({\rm t}\right)} 
\arctan \frac{\ve{\sigma} \cdot \ve{r}^{\rm N}_A\left({\rm t}\right)}{d^{\rm N}_A\left({\rm t}\right)}\bigg] 
+ {\cal O}\left(c^{-5}\right),  
\label{Example_C2}
\end{eqnarray}

\noindent
where the logarithms have cancelled each other. Integration by parts according to (\ref{Integration_by_Part}) yields  
the following solution for that integral:    
\begin{eqnarray}
\fl \ve{\cal J}_{A_4}\left(t,t_0\right) = + m_A^2\;  
\ve{\sigma} \cdot \ve{r}^{\rm N}_A\left(t\right)\; \arctan \frac{\ve{\sigma} \cdot \ve{r}^{\rm N}_A\left(t\right)}{d^{\rm N}_A\left(t\right)} 
\frac{\ve{d}^{\rm N}_A\left(t\right)}{\left(d^{\rm N}_A\left(t\right)\right)^3}  
\nonumber\\ 
\nonumber\\ 
\fl \hspace{1.5cm} - m_A^2\;
\ve{\sigma} \cdot \ve{r}^{\rm N}_A\left(t_0\right)\; \arctan \frac{\ve{\sigma} \cdot \ve{r}^{\rm N}_A\left(t_0\right)}{d^{\rm N}_A\left(t_0\right)}
\frac{\ve{d}^{\rm N}_A\left(t_0\right)}{\left(d^{\rm N}_A\left(t_0\right)\right)^3} 
\nonumber\\ 
\nonumber\\ 
\fl \hspace{1.5cm} - m_A^2 \frac{\ve{\sigma}}{d^{\rm N}_A\left(t\right)}
\arctan \frac{\ve{\sigma} \cdot \ve{r}^{\rm N}_A\left(t\right)}{d^{\rm N}_A\left(t\right)}
+ m_A^2 \frac{\ve{\sigma}}{d^{\rm N}_A\left(t_0\right)}
\arctan \frac{\ve{\sigma} \cdot \ve{r}^{\rm N}_A\left(t_0\right)}{d^{\rm N}_A\left(t_0\right)} 
+ {\cal O}\left(c^{-5}\right). 
\label{Example_E}
\end{eqnarray}

\subsection{Second example of the second integration}\label{Second_Integration_Second_Example}  

In Section \ref{First_Integration_Second_Example} we have considered the first integration of the first term on the first line
in (\ref{Geodesic_Equation3}). Now we will consider the second integration of this term, that means the following integral
\begin{eqnarray}
\ve{{\cal J}}_B\left(t,t_0\right) = \int\limits_{t_0}^t d c{\rm t}\,\ve{{\cal I}}_{B}\left({\rm t}\right),  
\label{Example_H1}
\end{eqnarray}

\noindent
where the integrand is given by Eq.~(\ref{Example_Integral_5}). According to (\ref{Example_Integral_5_D}), we have to consider  
\begin{eqnarray}
\ve{{\cal J}}_B\left(t,t_0\right) = \ve{{\cal J}}_{B_1}\left(t,t_0\right) + \ve{{\cal J}}_{B_2}\left(t,t_0\right) + \ve{{\cal J}}_{B_3}\left(t,t_0\right),  
\label{Example_H2}
\end{eqnarray}

\noindent
where
\begin{eqnarray}
\ve{{\cal J}}_{B_n}\left(t,t_0\right) &=& \int\limits_{t_0}^t d c{\rm t}\,\ve{{\cal I}}_{B_n}\left({\rm t}\right) \quad {\rm with} \quad n=1,2,3\,,
\label{Example_IntegralB_n}
\end{eqnarray}

\noindent
and the functions $\ve{{\cal I}}_{B_n}\left({\rm t}\right)$ are given by (\ref{Example_Integral_5_E}) - (\ref{Example_Integral_5_G}). 
An explicit solution has been presented for $\ve{{\cal I}}_{B_3}$ in Eq.~(\ref{Example_Integral_5_F_5}). By inserting the  
solution (\ref{Example_Integral_5_F_5}) into (\ref{Example_IntegralB_n}) we get  
\begin{eqnarray}
\fl \ve{{\cal J}}_{B_3}\left(t,t_0\right) = 2\,m_A^2 \int\limits_{t_0}^t d c{\rm t}
\bigg[\left(\frac{1}{d^{\rm N}_A\left({\rm t}\right)}\,\arctan \frac{\ve{\sigma} \cdot \ve{r}^{\rm N}_A\left({\rm t}\right)}{d^{\rm N}_A\left({\rm t}\right)}
- \frac{1}{r^{\rm N}_A\left({\rm t}\right)}\right) \frac{\ve{d}^{\rm N}_A\left({\rm t}\right)}{\left(d^{\rm N}_A\left({\rm t}\right)\right)^2}
\nonumber\\
\nonumber\\
\fl - \,\frac{\ve{d}_A^{\rm N}\left(t_0\right)}{r^{\rm N}_A\left(t_0\right) - \ve{\sigma} \cdot \ve{r}^{\rm N}_A\left(t_0\right)}
\frac{1}{r^{\rm N}_A\left({\rm t}\right)}\,
\frac{1}{r^{\rm N}_A\left({\rm t}\right) - \ve{\sigma} \cdot \ve{r}^{\rm N}_A\left({\rm t}\right)}
+ \frac{\ve{\sigma}}{\left(d^{\rm N}_A\left({\rm t}\right)\right)^2}\,
\ln \left(r^{\rm N}_A\left({\rm t}\right) - \ve{\sigma} \cdot \ve{r}^{\rm N}_A\left({\rm t}\right)\right) 
\nonumber\\
\nonumber\\
\fl - \frac{\ve{\sigma}}{r^{\rm N}_A\left({\rm t}\right)}\,
\frac{1}{r^{\rm N}_A\left({\rm t}\right) - \ve{\sigma} \cdot \ve{r}^{\rm N}_A\left({\rm t}\right)}\,
\ln \frac{r^{\rm N}_A\left({\rm t}\right) - \ve{\sigma} \cdot \ve{r}^{\rm N}_A\left({\rm t}\right)}
{r^{\rm N}_A\left(t_0\right) - \ve{\sigma} \cdot \ve{r}^{\rm N}_A\left(t_0\right)}\bigg],  
\label{Example_Integral_H_1}
\end{eqnarray}

\noindent
where we have taken into account that all the logarithms having the structure $\ln r^{\rm N}_A\left(t\right)$ do cancel against 
each other in the final result for the coordinate velocity of the light signal (cf. text below Eq.~(\ref{Example_Integral_5_F_5})).   
Also the integration constants, i.e. the terms in the last line of (\ref{Example_Integral_5_F_5}), have been omitted because all these constants  
cancel each other in the final solution for the coordinate velocity of the light signal.  
By making use of relations (\ref{Appendix_Time_Derivative_1}), (\ref{Appendix_Time_Derivative_14}), (\ref{Appendix_Time_Derivative_17}), 
(\ref{Appendix_Time_Derivative_18}) in \ref{Appendix_Integration_by_Parts} we may rewrite the integral as follows, 
\begin{eqnarray}
\fl \ve{{\cal J}}_{B_3}\left(t,t_0\right) = 2 m_A^2 \! \int\limits_{t_0}^t \! d c{\rm t}
\bigg[\frac{d}{d c {\rm t}} \left(\ve{\sigma} \cdot \ve{r}^{\rm N}_A\left({\rm t}\right) 
\arctan \frac{\ve{\sigma} \cdot \ve{r}^{\rm N}_A\left({\rm t}\right)}{d^{\rm N}_A\left({\rm t}\right)} 
- d^{\rm N}_A\left({\rm t}\right) \ln r^{\rm N}_A\left({\rm t}\right) \right) \bigg]  
\frac{\ve{d}^{\rm N}_A\left({\rm t}\right)}{\left(d^{\rm N}_A\left({\rm t}\right)\right)^3}  
\nonumber\\
\nonumber\\
\fl + \,2 m_A^2 \int\limits_{t_0}^t d c{\rm t} 
\bigg[\frac{d}{d c {\rm t}} \,\ln \left(r^{\rm N}_A\left({\rm t}\right) - \ve{\sigma} \cdot \ve{r}^{\rm N}_A\left({\rm t}\right)\right) \bigg] 
\frac{\ve{d}^{\rm N}_A\left({\rm t}\right)}{\left(d^{\rm N}_A\left({\rm t}\right)\right)^2}  
\nonumber\\
\nonumber\\
\fl - \,2 m_A^2 \frac{\ve{d}_A^{\rm N}\left(t_0\right)}{r^{\rm N}_A\left(t_0\right) - \ve{\sigma} \cdot \ve{r}^{\rm N}_A\left(t_0\right)}
\int\limits_{t_0}^t d c{\rm t}
\bigg[\frac{d}{d c {\rm t}}\,\frac{1}{r^{\rm N}_A\left({\rm t}\right) - \ve{\sigma} \cdot \ve{r}^{\rm N}_A\left({\rm t}\right)}\bigg] 
\nonumber\\
\nonumber\\
\fl + \,2 m_A^2 \ve{\sigma} \int\limits_{t_0}^t d c {\rm t}
\bigg[\frac{d}{d c {\rm t}} \bigg(r^{\rm N}_A\left({\rm t}\right) 
+ \ve{\sigma} \cdot \ve{r}^{\rm N}_A\left({\rm t}\right)\,
\ln \left(r^{\rm N}_A\left({\rm t}\right) - \ve{\sigma} \cdot \ve{r}^{\rm N}_A\left({\rm t}\right)\right)\bigg)\bigg] 
\frac{1}{\left(d^{\rm N}_A\left({\rm t}\right)\right)^2}  
\nonumber\\
\nonumber\\
\fl - \,2 m_A^2 \ve{\sigma} \int\limits_{t_0}^t d c{\rm t} 
\bigg[\frac{d}{d c {\rm t}} \,\frac{1}{r^{\rm N}_A\left({\rm t}\right) - \ve{\sigma} \cdot \ve{r}^{\rm N}_A\left({\rm t}\right)}\bigg] 
\ln \frac{r^{\rm N}_A\left({\rm t}\right) - \ve{\sigma} \cdot \ve{r}^{\rm N}_A\left({\rm t}\right)}
{r^{\rm N}_A\left(t_0\right) - \ve{\sigma} \cdot \ve{r}^{\rm N}_A\left(t_0\right)}  
+ {\cal O}\left(c^{-5}\right). 
\label{Example_Integral_H_2}
\end{eqnarray}

\noindent 
Then, by integration by parts (\ref{Integration_by_Part}) and keeping in mind relation (\ref{Time_Derivative_Impact_Vector_1}), we obtain  
\begin{eqnarray}
\fl \ve{{\cal J}}_{B_3}\left(t,t_0\right) = + 2 m_A^2 
\left(\ve{\sigma} \cdot \ve{r}^{\rm N}_A\left(t\right)
\arctan \frac{\ve{\sigma} \cdot \ve{r}^{\rm N}_A\left(t\right)}{d^{\rm N}_A\left(t\right)}
- d^{\rm N}_A\left(t\right) \ln r^{\rm N}_A\left(t\right) \right) 
\frac{\ve{d}^{\rm N}_A\left(t\right)}{\left(d^{\rm N}_A\left(t\right)\right)^3}
\nonumber\\
\nonumber\\
\fl - 2 m_A^2
\left(\ve{\sigma} \cdot \ve{r}^{\rm N}_A\left(t_0\right)
\arctan \frac{\ve{\sigma} \cdot \ve{r}^{\rm N}_A\left(t_0\right)}{d^{\rm N}_A\left(t_0\right)}
- d^{\rm N}_A\left(t_0\right) \ln r^{\rm N}_A\left(t_0\right) \right)
\frac{\ve{d}^{\rm N}_A\left(t_0\right)}{\left(d^{\rm N}_A\left(t_0\right)\right)^3}
\nonumber\\
\nonumber\\
\fl + \,2 m_A^2 
\ln \left(r^{\rm N}_A\left(t\right) - \ve{\sigma} \cdot \ve{r}^{\rm N}_A\left(t\right)\right) 
\frac{\ve{d}^{\rm N}_A\left(t\right)}{\left(d^{\rm N}_A\left(t\right)\right)^2}
- \,2 m_A^2
\ln \left(r^{\rm N}_A\left(t_0\right) - \ve{\sigma} \cdot \ve{r}^{\rm N}_A\left(t_0\right)\right)
\frac{\ve{d}^{\rm N}_A\left(t_0\right)}{\left(d^{\rm N}_A\left(t_0\right)\right)^2}
\nonumber\\
\nonumber\\
\fl - \,2 m_A^2 \frac{\ve{d}_A^{\rm N}\left(t_0\right)}{r^{\rm N}_A\left(t_0\right) - \ve{\sigma} \cdot \ve{r}^{\rm N}_A\left(t_0\right)}
\left(\frac{1}{r^{\rm N}_A\left(t\right) - \ve{\sigma} \cdot \ve{r}^{\rm N}_A\left(t\right)}
- \frac{1}{r^{\rm N}_A\left(t_0\right) - \ve{\sigma} \cdot \ve{r}^{\rm N}_A\left(t_0\right)}\right) 
\nonumber\\
\nonumber\\
\fl + \,2 m_A^2 \ve{\sigma} 
\left(\frac{\ve{\sigma} \cdot \ve{r}^{\rm N}_A\left(t\right)}{\left(d^{\rm N}_A\left(t\right)\right)^2}\,
\ln \left(r^{\rm N}_A\left(t\right) - \ve{\sigma} \cdot \ve{r}^{\rm N}_A\left(t\right)\right)
- \frac{\ve{\sigma} \cdot \ve{r}^{\rm N}_A\left(t_0\right)}{\left(d^{\rm N}_A\left(t_0\right)\right)^2}\,
\ln \left(r^{\rm N}_A\left(t_0\right) - \ve{\sigma} \cdot \ve{r}^{\rm N}_A\left(t_0\right)\right)\right)
\nonumber\\
\nonumber\\
\fl + \,2 m_A^2 \ve{\sigma} \left(\frac{r^{\rm N}_A\left(t\right)}{\left(d^{\rm N}_A\left(t\right)\right)^2}
 - \frac{r^{\rm N}_A\left(t_0\right)}{\left(d^{\rm N}_A\left(t_0\right)\right)^2}\right)  
- \,2 m_A^2 \ve{\sigma} 
\,\frac{1}{r^{\rm N}_A\left(t\right) - \ve{\sigma} \cdot \ve{r}^{\rm N}_A\left(t\right)}
\ln \frac{r^{\rm N}_A\left(t\right) - \ve{\sigma} \cdot \ve{r}^{\rm N}_A\left(t\right)}
{r^{\rm N}_A\left(t_0\right) - \ve{\sigma} \cdot \ve{r}^{\rm N}_A\left(t_0\right)}
\nonumber\\
\nonumber\\
\fl - \,2 m_A^2 \ve{\sigma}
\left(\frac{1}{r^{\rm N}_A\left(t\right) - \ve{\sigma} \cdot \ve{r}^{\rm N}_A\left(t\right)} 
- \frac{1}{r^{\rm N}_A\left(t_0\right) - \ve{\sigma} \cdot \ve{r}^{\rm N}_A\left(t_0\right)}\right)  
+ {\cal O}\left(c^{-5}\right), 
\label{Example_Integral_H_3}
\end{eqnarray}

\noindent
where for the last integral in (\ref{Example_Integral_H_2}) two integration by parts were required.

\subsection{The intermediate result of the second integration}

The entire algorithm of second integration proceeds in exactly the way enlightened by the examples. All these integrals are treated by integration by parts 
with the aid of the relations listed in \ref{Appendix_Integration_by_Parts}. The examples have also shown that the integrals of second integration turn 
out to be of similar structure to those one encounters in the first integration of geodesic equation. Nevertheless, as elucidated by 
the example in (\ref{Example_Integral_H_3}), the expressions are of remarkable algebraic structure.  
Furthermore, in view of the considerable amount of integrals, the integration procedure is rather cumbersome.  
The computer algebra system Maple \cite{Maple} has extensively been used to run the entire integration procedure of the second integration 
of geodesic equation. That means, like in the first integration, there are two basic steps:
(i) Calculation of each integral analytically and cross-check by differentiation. 
(ii) Summarize and simplify the entire three-scalars which have the same vectorial prefactor           
(there are only four linearly independent vectorial prefactors: $\ve{\sigma}, \ve{v}_A, \ve{a}_A$ and either $\ve{d}^{\rm N}_A$ or $\ve{r}^{\rm N}_A$).
The final solution of that procedure yields the trajectory  
of a light signal in terms of $\ve{r}^{\rm N}_A\left(t\right)$. Afterwards,  
by means of Eqs.~(\ref{Example_Integral_5_B}) and (\ref{Example_Integral_5_C}), one may rewrite this solution in terms of       
$\ve{r}^{\rm 1PN}_A\left(t\right)$ and $\ve{r}^{\rm N}_A\left(t\right)$. Then, one arrives at the following structure for the formal solution:
\begin{eqnarray}
\fl \hspace{0.5cm} \ve{x}_{\rm 2PN}\left(t\right) = \ve{x}_0 + \left(t - t_0\right) \ve{\sigma}  
\nonumber\\ 
\fl \hspace{1.5cm} + \, m_A\, \ve{G}_1\left(\ve{r}^{\rm 1PN}_A\left(t\right)\right)  
+ m_A\, \ve{G}_2\left(\ve{r}^{\rm N}_A\left(t\right),\ve{v}_A\left(t\right),\ve{a}_A\left(t\right)\right) 
+ m_A^2\, \ve{G}_3\left(\ve{r}^{\rm N}_A\left(t\right)\right),    
\label{Second_Integration_Preliminary}
\end{eqnarray}

\noindent 
where the position of the
light ray and the position of the massive body $\ve{x}_A\left(t\right)$ are taken at the very same instant of time. 
This solution, like the solution in (\ref{First_Integration_Preliminary}), is of pretty involved algebraic structure and as an intermediate result 
the vectorial functions $\ve{G}_1$ and $\ve{G}_2$ and $\ve{G}_3$ will not be given here, but they can be retrieved from the final 
result for the light trajectory (\ref{Second_Integration}) by series expansion in inverse powers of the speed of light. 
As a matter of fact, the effect of retardation is hidden in the intermediate solution (\ref{Second_Integration_Preliminary}), which allows  
us to rewrite (\ref{Second_Integration_Preliminary}) in terms of the retarded position of the body $\ve{x}_A\left(s\right)$ and   
results in the considerably simpler expression in (\ref{Second_Integration}) for the trajectory of the light signal, which will be the subject of 
Section \ref{Section3b}. We stress that both the expressions, Eq.~(\ref{Second_Integration_Preliminary}) and Eq.~(\ref{Second_Integration}), are identical  
up to terms of the order ${\cal O}\left(c^{-5}\right)$.

\section{Transformation in terms of retarded time}\label{Section3b}

The solutions for coordinate velocity in Eq.~(\ref{First_Integration_Preliminary}) and trajectory in Eq.~(\ref{Second_Integration_Preliminary}) 
of a light signal depend on the position of the body $\ve{x}_A\left(t\right)$ where $t$ is the coordinate time of the global reference system. 
These expressions are of rather involved structure, which can drastically be simplified if they are expressed in terms of the retarded 
position of the body $\ve{x}_A\left(s\right)$ where $s$ is the retarded time defined  
by Eq.~(\ref{Retarded_Time_1}). This fact has been used for the first time in \cite{Klioner2003a} in order to simplify the mathematical 
expressions for the photon's coordinate velocity and trajectory; cf. Eqs.~(31) - (33) and (35) - (37) in \cite{Klioner2003a}.  
Though, the advantage to express the solutions (\ref{First_Integration_Preliminary}) and (\ref{Second_Integration_Preliminary}) 
in terms of the retarded position of the body is threefold: 
\begin{enumerate} 
\item[(i)] 
The mathematical expressions in (\ref{First_Integration_Preliminary}) and (\ref{Second_Integration_Preliminary}) which are  
functions of $\ve{x}_A\left(t\right)$ become considerably simpler when expressed in terms of the retarded position $\ve{x}_A\left(s\right)$.
\item[(ii)] The results (\ref{First_Integration_Preliminary}) and (\ref{Second_Integration_Preliminary}) can directly be compared 
with the post-Minkowskian approach if they are given in terms of $\ve{x}_A\left(s\right)$.  
\item[(iii)] If (\ref{First_Integration_Preliminary}) and (\ref{Second_Integration_Preliminary}) are expressed in terms of the 
retarded position $\ve{x}_A\left(s\right)$ then the effect of retardation is evident.  
\end{enumerate} 

\noindent 
The items (i) and (ii) are mathematical arguments hence of rather formal nature, 
while item (iii) is triggered by physical reasons which will be considered in more detail in what follows.  

It is a fundamental consequence of the exact field equations of gravity (\ref{Field_Equations_10}) that gravitational action travels with the 
finite speed of light. In fact, the first detection of gravitational waves from the inspiral of a binary neutron star and the determination of the location  
of the source by subsequent observations in the electromagnetic spectrum \cite{Ligo1,Ligo2} has been used to constrain the difference between the speed of gravity 
and the speed of light to be between $ - 3 \times 10^{- 15}$ and $ + 7 \times 10^{- 16}$ times the speed of light \cite{Ligo3}. The effect of retardation  
of gravity can explicitly be read off from the general solution (\ref{Introduction_2}) of the field equations, where the retardation is evident.  
Consequently, in case of one pointlike body the gravitational field at $\left(t,\ve{x}\right)$ is generated by the body at its retarded position 
$\ve{x}_A\left(s\right)$, where the retarded time $s$ is given by the following implicit relation (e.g.. Eq.~(7.13) in \cite{Kopeikin_Efroimsky_Kaplan}),
\begin{eqnarray}
s &=& t - \frac{r_A\left(s\right)}{c}\,, 
\label{Retarded_Time_1}
\end{eqnarray}

\noindent
with 
\begin{eqnarray}
\ve{r}_A\left(s\right) &=& \ve{x} - \ve{x}_A\left(s\right). 
\label{vector_A}
\end{eqnarray}
 
\noindent 
This fundamental statement is not only valid for the far-zone of a gravitational system, where the gravitational field
decouples from the matter and propagates as a gravitational wave, but also inside the near-zone of the gravitational system where the gravitational fields
and the matter sources are still entangled with each other.  

\begin{figure}[!ht]
\begin{indented}
\item[]
\includegraphics[scale=0.14]{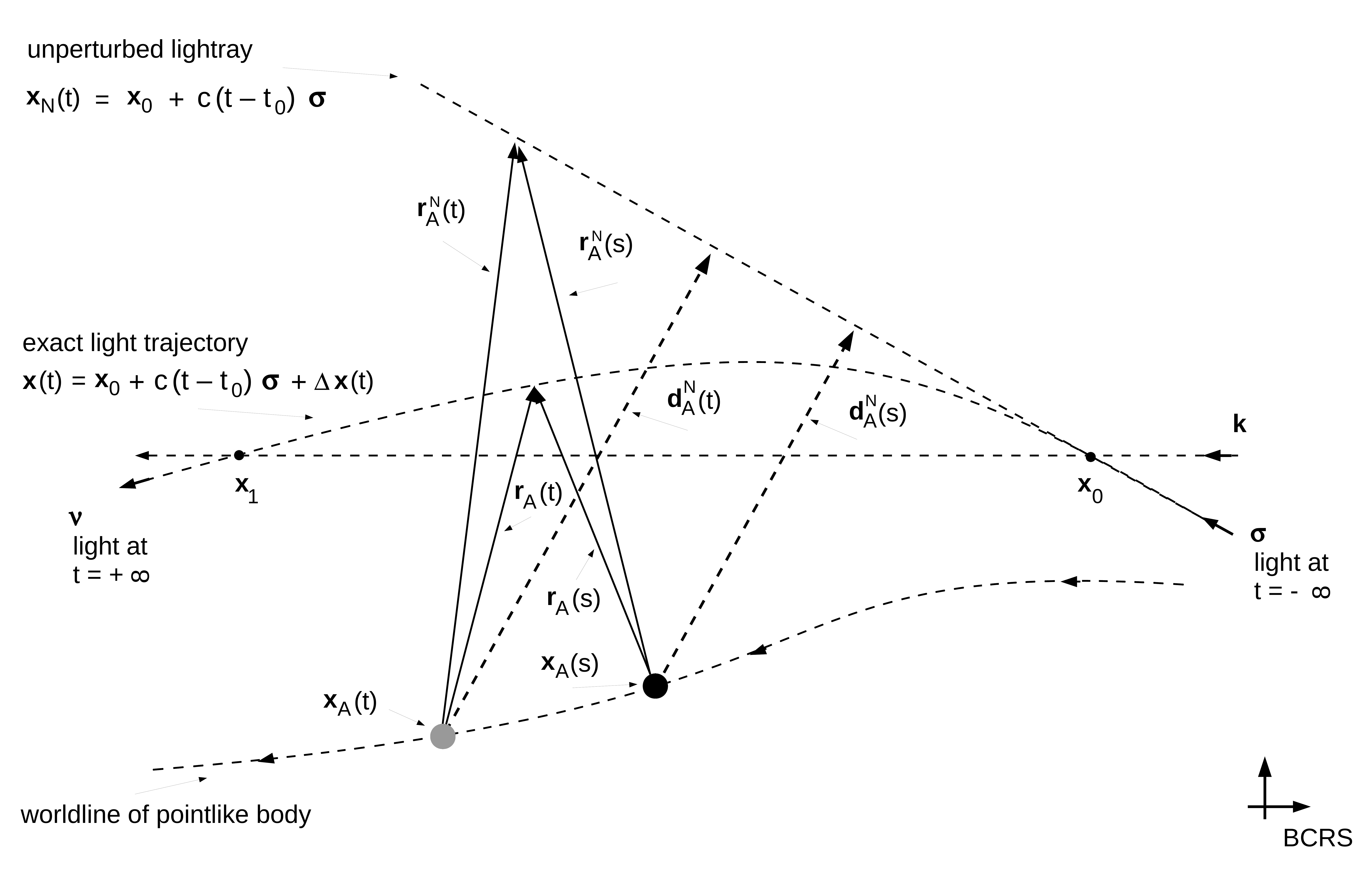}
\end{indented}
\vspace{10pt}
\caption{\label{Diagram} A geometrical representation of light propagation through the gravitational field
of one pointlike massive body $A$ moving along its worldline $\ve{x}_A\left(t\right)$. The unperturbed light ray $\ve{x}_{\rm N}\left(t\right)$
is defined by Eq.~(\ref{Unperturbed_Lightray}) and propagates along a straight line in direction $\ve{\sigma}$.
The exact light ray $\ve{x}\left(t\right)$ is defined by Eq.~(\ref{Integral_2}) and propagates along a curved trajectory.
The three-vector $\ve{r}_A\left(t\right)$ is defined by Eq.~(\ref{vector_B}) and points from the massive body $A$ at its  
instantaneous position $\ve{x}_A\left(t\right)$ (gray sphere) toward the exact photon's position $\ve{x}\left(t\right)$ at instant $t$.  
Because of the fact that gravitational action travels with the finite speed of light, the light ray at $\ve{x}\left(t\right)$ is
influenced by the gravitational field generated by the body at its retarded position $\ve{x}_A\left(s\right)$
(black sphere). The spatial vector $\ve{r}_A\left(s\right)$ is defined by Eq.~(\ref{vector_C}) and points from the massive body $A$ at its retarded  
position $\ve{x}_A\left(s\right)$ toward the exact photon's position $\ve{x}\left(t\right)$ at instant $t$. The impact vectors $\ve{d}_A^{\rm N}\left(t\right)$
and $\ve{d}_A^{\rm N}\left(s\right)$ are defined by Eqs.~(\ref{Impact_Vector_1}) and (\ref{Impact_Vector_2}), while the impact vectors
$\ve{d}_A\left(t\right)$ and $\ve{d}_A\left(s\right)$ are defined by Eqs.~(\ref{Impact_Vector_5}) and (\ref{Impact_Vector_6}) but they are not illustrated here.}
\end{figure}

The post-Newtonian expansion (\ref{Reply_1}) is based on the assumption that the retardation (\ref{Retarded_Time_1}) is small. Especially, this assumption  
implies that a series expansion of the retarded position of the massive body becomes meaningful \cite{Kopeikin_Efroimsky_Kaplan,MTW,Poisson_Lecture_Notes} 
\begin{eqnarray}
\fl \hspace{2.0cm} \ve{x}_A\left(s\right) = \ve{x}_A\left(t\right) + \frac{1}{1!}\,\dot{\ve{x}}_A\left(t\right)\,\left(s - t\right)
+ \frac{1}{2!}\,\ddot{\ve{x}}_A\left(t\right)\,\left(s - t\right)^2 + {\cal O}\left(c^{-3}\right).   
\label{retarded_position}
\end{eqnarray}

\noindent
Accordingly, the retarded time can also be expanded in terms of inverse powers of the speed of light,
\begin{eqnarray}
s &=& t - \frac{r_A\left(t\right)}{c} - \frac{\ve{r}_A\left(t\right) \cdot \ve{v}_A\left(t\right)}{c^2}
+ {\cal O}\left(c^{-3}\right)\,.
\label{retarded_time_A}
\end{eqnarray}

\noindent
In particular, using (\ref{retarded_position}) and (\ref{retarded_time_A}) one obtains the following important relations \cite{Zschocke2,Zschocke_Soffel},  
\begin{eqnarray}
\fl \ve{r}_A\left(s\right) = \ve{r}_A\left(t\right) + \frac{\ve{v}_A\left(t\right)}{c}\,r_A\left(t\right)
+ \frac{\ve{v}_A\left(t\right)}{c}\,\frac{\ve{r}_A\left(t\right) \cdot \ve{v}_A\left(t\right)}{c}
- \frac{1}{2}\,\frac{\ve{a}_A\left(t\right)}{c}\,\frac{r^2_A\left(t\right)}{c} + {\cal O}\left(c^{-3}\right),
\nonumber\\
\label{retarded_time_B}
\\
\fl \frac{\ve{v}_A\left(s\right)}{c} = \frac{\ve{v}_A\left(t\right)}{c} - \frac{\ve{a}_A\left(t\right)}{c}\,\frac{r_A\left(t\right)}{c}
+ {\cal O}\left(c^{-3}\right),
\label{retarded_time_C}
\\
\fl \frac{\ve{a}_A\left(s\right)}{c} = \frac{\ve{a}_A\left(t\right)}{c} + {\cal O}\left(\dot{\ve{a}}_A\right)  
+ {\cal O}\left(c^{-3}\right),
\label{retarded_time_D}
\end{eqnarray}

\noindent
where $\ve{v}_A\left(t\right) = \dot{\ve{x}}_A\left(t\right)$ and $\ve{a}_A\left(t\right) = \ddot{\ve{x}}_A\left(t\right)$ is the velocity
and acceleration of the body. These equations relate the expressions $\ve{r}_A\left(t\right)$, $\ve{v}_A\left(t\right)$, $\ve{a}_A\left(t\right)$ 
at coordinate time with the expressions $\ve{r}_A\left(s\right)$, $\ve{v}_A\left(s\right)$, $\ve{a}_A\left(s\right)$ at retarded time; 
such relations were also used in \cite{Klioner2003a,KlionerPeip2003}.  
Because the metric perturbations in 2PN approximation, Eqs.~(\ref{Metric_1}) - (\ref{Metric_6}), are functions of these expressions, 
they can be either expressed in terms of coordinate time $t$ or in terms of retarded time $s$, where it becomes obvious that the retardation effects  
are present in the post-Newtonian expansion of the metric. In other words, the relations (\ref{retarded_time_B}) - (\ref{retarded_time_D})  
allow one to demonstrate that the post-post-Newtonian approximation of the metric tensor (\ref{post_Newtonian_metric_B}) when expressed in terms 
of retarded time $s$ would coincide with the post-post-Minkowskian approximation of the metric tensor (\ref{metric_perturbation_1PM_5})  
up to terms of the order ${\cal O} \left(c^{-5}\right)$.  

Like the 2PN metric perturbations (\ref{Metric_1}) - (\ref{Metric_6}), also the 2PN solutions for the coordinate velocity and trajectory of the light signal 
(\ref{First_Integration_Preliminary}) and (\ref{Second_Integration_Preliminary}) are functions of $\ve{r}_A\left(t\right)$, $\ve{v}_A\left(t\right)$, 
$\ve{a}_A\left(t\right)$. Hence, using the relations (\ref{retarded_time_B}) - (\ref{retarded_time_D}) one may express these solutions either  
in terms of coordinate time or in terms of retarded time. No additional assumptions are imposed. As mentioned above, if we do so then the advantage is  
that the mathematical expressions for (\ref{First_Integration_Preliminary}) and (\ref{Second_Integration_Preliminary}) become drastically simplified  
(a fact which has also been recognized in \cite{Klioner2003a}) and    
the retardation which is somehow hidden in (\ref{First_Integration_Preliminary}) and (\ref{Second_Integration_Preliminary}) becomes more explicit.  
Let us consider that issue further.  

The equations (\ref{Retarded_Time_1}) - (\ref{retarded_time_D}) are valid for any field point $\ve{x}$ which is arbitrary, and  
in particular the series expansions (\ref{retarded_time_A}) - (\ref{retarded_time_D}) are based on the series expansion of the retarded position 
of the body (\ref{retarded_position}) without reference to the field point $\ve{x}$.  
So if we replace the spatial field point $\ve{x}$ in Eq.~(\ref{vector_A}) by the spatial position of the light signal $\ve{x}\left(t\right)$, and 
introduce the three-vector \footnote{We shall avoid to introduce a further notation which distinguishes among  (\ref{vector_C})
which points from the retarded position of the body $\ve{x}_A\left(s\right)$ towards the exact position of the light signal $\ve{x}\left(t\right)$
and (\ref{vector_A}) which points from the retarded position of the body $\ve{x}_A\left(s\right)$ towards the field point $\ve{x}$.} 
\begin{eqnarray}
\ve{r}_A\left(s\right) &=& \ve{x}\left(t\right) - \ve{x}_A\left(s\right),
\label{vector_C}
\end{eqnarray}

\noindent 
then all these equations (\ref{Retarded_Time_1}) - (\ref{retarded_time_D}) will keep their validity.  
For later purposes we also introduce the three-vector  
\begin{eqnarray}
\ve{r}_A\left(s_0\right) &=& \ve{x}\left(t_0\right) - \ve{x}_A\left(s_0\right),
\label{vector_D}
\end{eqnarray}

\noindent
where the retarded time $s_0$ is a specific case of (\ref{Retarded_Time_1}) and related to the time of emission of the light signal
(cf. Eqs.~(7.75) and (7.361) in \cite{Kopeikin_Efroimsky_Kaplan} or Eq.~(46) in \cite{KopeikinSchaefer1999})  
\begin{eqnarray}
s_0 &=& t_0 - \frac{r_A\left(s_0\right)}{c}\,.
\label{Retarded_Time_0}
\end{eqnarray}

\noindent 
The solutions of first and second integration of geodesic equation, formally given by (\ref{First_Integration_Preliminary}) and  
(\ref{Second_Integration_Preliminary}), depend on the three-vectors $\ve{r}^{\rm N}_A\left(t\right)$ and $\ve{r}^{\rm 1PN}_A\left(t\right)$ as given by  
Eqs.~(\ref{Example_20}) and (\ref{Example_55}). As mentioned, the solutions for coordinate velocity and trajectory adopt the most simple mathematical structure  
if they are presented in terms of retarded time $s$. Accordingly, we introduce the following three-vectors 
(the three-vector defined in Eq.~(\ref{Distance_Vector_N}) has also been used in Eq.~(B.2) in \cite{Klioner2003a}),  
\begin{eqnarray}
\ve{r}^{\rm N}_A\left(s\right) = \ve{x}_0 + c \left(t-t_0\right) \ve{\sigma} - \ve{x}_A\left(s\right),
\label{Distance_Vector_N}
\\
\nonumber\\
\ve{r}^{\rm 1PN}_A\left(s\right) = \ve{x}_0 + c \left(t - t_0\right) \ve{\sigma} + \Delta \ve{x}_{\rm 1PN}\left(s,s_0\right) - \ve{x}_A\left(s\right),  
\label{Distance_Vector_1PN}
\end{eqnarray}

\noindent
where $\Delta \ve{x}_{\rm 1PN}\left(s,s_0\right)$ is given by Eqs.~(\ref{Delta_s1}) and (\ref{Retarded_Light_Trajectory_M_1}). Then,  
according to (\ref{retarded_time_B}), the three-vectors at retarded time  
in Eqs.~(\ref{Distance_Vector_N}) and (\ref{Distance_Vector_1PN}) and the three-vectors at coordinate time in Eqs.~(\ref{Example_20}) and (\ref{Example_55})  
are related to each other as follows:  
\begin{eqnarray}
\fl \ve{r}^{\rm N}_A\left(s\right) = \ve{r}^{\rm N}_A\left(t\right) + \frac{\ve{v}_A\left(t\right)}{c} r^{\rm N}_A\left(t\right)
+ \frac{\ve{v}_A\left(t\right)}{c} \frac{\ve{r}^{\rm N}_A\left(t\right) \cdot \ve{v}_A\left(t\right)}{c}
- \frac{1}{2} \frac{\ve{a}_A\left(t\right)}{c} \frac{\left(r^{\rm N}_A\left(t\right)\right)^2}{c} 
\nonumber\\
\fl \hspace{1.7cm} + {\cal O}\left(c^{-3}\right),
\label{retarded_time_E}
\\
\nonumber\\
\fl \ve{r}^{\rm 1PN}_A\left(s\right) = \ve{r}^{\rm 1PN}_A\left(t\right) + \frac{\ve{v}_A\left(t\right)}{c} r^{\rm 1PN}_A\left(t\right)
+ \frac{\ve{v}_A\left(t\right)}{c} \frac{\ve{r}^{\rm 1PN}_A\left(t\right) \cdot \ve{v}_A\left(t\right)}{c}
- \frac{1}{2} \frac{\ve{a}_A\left(t\right)}{c} \frac{\left(r^{\rm 1PN}_A\left(t\right)\right)^2}{c} 
\nonumber\\
\fl \hspace{1.7cm} + \,{\cal O}\left(c^{-3}\right). 
\label{retarded_time_F}
\end{eqnarray}

\noindent 
By means of relations (\ref{retarded_time_C}) - (\ref{retarded_time_D}) and (\ref{retarded_time_E}) - (\ref{retarded_time_F}) the solutions  
of first integration in (\ref{First_Integration_Preliminary}) and second integration in (\ref{Second_Integration_Preliminary}) of geodesic equation,  
first of all given in terms of the instantaneous position of the massive body, can be rewritten in terms of  
retarded position of the body, which will be the topic of the next Section.  

Before we proceed further, another comment should be in order about the near-zone. The series-expansions in (\ref{retarded_time_A}) - (\ref{retarded_time_D})
and (\ref{retarded_time_E}) - (\ref{retarded_time_F}) are useful as long as the retardations are small, a requirement
which is well-justified in the near-zone of the Solar System \cite{Kopeikin_Efroimsky_Kaplan,MTW,Poisson_Lecture_Notes,Poisson_Will,Book_PN}.
It especially implies that for the acceleration of the body we have the restriction that
\begin{eqnarray}
\frac{a_A\left(t\right)\,r_A\left(t\right)}{c^2} &\ll& \frac{v_A\left(t\right)}{c} \ll 1\;,
\label{constraint_acceleration}
\end{eqnarray}

\noindent
for any moment of time. This relation can be used in order to set limits on the spatial radius of the near-zone of the Solar System,
\begin{eqnarray}
\left|\ve{x}\right| &\le& 10^{14}\,{\rm m}\,,
\label{near-zone_2}
\end{eqnarray}

\noindent
which is about $0.01$ light-year.

\section{The coordinate velocity of a light signal in 2PN approximation}\label{Section4} 

According to Eq.~(\ref{Integral_1}), the first integration of geodesic equation (\ref{Geodesic_Equation3}) yields the coordinate velocity  
of a light signal in 2PN approximation. As described in Section \ref{Section3a} the integration is performed by iteration and by means 
of integration by parts. Such a solution is, first of all, given in terms of the instantaneous position of the body $\ve{x}_A\left(t\right)$ 
as formally given by Eq.~(\ref{First_Integration_Preliminary}). According to Section \ref{Section3b},  
the solution for the coordinate velocity can be reexpressed in terms of the retarded position of the body $\ve{x}_A\left(s\right)$ using relations  
(\ref{retarded_time_C}) - (\ref{retarded_time_D}) and (\ref{retarded_time_E}) - (\ref{retarded_time_F}).  
Then, the first integration of geodesic equation in 2PN approximation reads as follows:  
\begin{eqnarray}
\fl \frac{\dot{\ve{x}}_{\rm 2PN}\left(t\right)}{c} = \ve{\sigma}
+ m_A\,\ve{A}_1\left(\ve{r}^{\rm 1PN}_A\left(s\right)\right)
+ m_A\,\ve{A}_2\left(\ve{r}^{\rm N}_A\left(s\right),\ve{v}_A\left(s\right)\right)
+ m_A^2\,\ve{A}_3\left(\ve{r}^{\rm N}_A\left(s\right)\right)  
\nonumber\\
\nonumber\\
\fl \hspace{1.7cm} + \, m_A\,\ve{\epsilon}_1\left(\ve{r}^{\rm N}_A\left(s\right),\ve{v}_A\left(s\right)\right),    
\label{First_Integration}
\end{eqnarray}

\noindent
where the arguments are given by Eqs.~(\ref{Distance_Vector_N}) and (\ref{Distance_Vector_1PN}). 
Observe that even though the velocity of the photon (\ref{First_Integration_Preliminary}), which is expressed in terms of $\ve{x}_A\left(t\right)$,
depends on the acceleration of the body, the velocity of the photon (\ref{First_Integration}), which is expressed in terms of $\ve{x}_A\left(s\right)$,
does not depend on the acceleration of the body. Furthermore, we notice that in  
Eq.~(\ref{First_Integration}) one may replace the approximate arguments by the exact expression in (\ref{vector_C}) because 
of $\ve{r}_A\left(s\right) = \ve{r}^{\rm N}_A\left(s\right) + {\cal O}\left(c^{-2}\right)$ and 
$\ve{r}_A\left(s\right) = \ve{r}^{\rm 1PN}_A\left(s\right) + {\cal O}\left(c^{-3}\right)$
such a replacement would cause an error of the order ${\cal O}\left(c^{-5}\right)$ which is beyond 2PN approximation.  
The vectorial functions are given as follows: 
\begin{eqnarray}
\fl \ve{A}_1\left(\ve{x}\right) =
- 2\,\left(\frac{\ve{\sigma} \times \left(\ve{x} \times \ve{\sigma}\right)}
{x \left(x - \ve{\sigma} \cdot \ve{x}\right)}
+ \frac{\ve{\sigma}}{x} \right),
\label{Vectorial_Function_A1}
\\
\nonumber\\
\fl \ve{A}_2\left(\ve{x},\ve{v}\right) = + \,2\,\frac{\ve{\sigma} \times \left(\ve{x} \times \ve{\sigma}\right)}
{x \left(x - \ve{\sigma} \cdot \ve{x}\right)}\,
\frac{\ve{\sigma} \cdot \ve{v}}{c}
+ \frac{4}{x}\,\frac{\ve{v}}{c}
+ 2\,\frac{\ve{\sigma} \times \left(\ve{x} \times \ve{\sigma}\right)}
{x^2}\,\frac{\ve{\sigma} \cdot \ve{v}}{c}
- 2\,\frac{\ve{\sigma}}{x^2}\,
\frac{\ve{x} \cdot \ve{v}}{c}
\nonumber\\
\nonumber\\
\fl \hspace{1.8cm} -\,2\,\frac{\ve{\sigma} \times \left(\ve{x} \times \ve{\sigma}\right)}
{x^2\,\left(x - \ve{\sigma} \cdot \ve{x}\right)}\,
\frac{\left(\ve{\sigma} \times \left(\ve{x} \times \ve{\sigma}\right) \right) \cdot \ve{v}}{c}\,,  
\label{Vectorial_Function_A2}
\\
\nonumber\\
\fl \ve{A}_3\left(\ve{x}\right) = - \frac{1}{2}\,\frac{\ve{\sigma} \cdot \ve{x}}{x^4}\,\ve{x}
+ 8\,\frac{\ve{\sigma} \times \left(\ve{x} \times \ve{\sigma}\right)}{x^2\,\left(x - \ve{\sigma} \cdot\ve{x} \right)}\,
+ 4\,\frac{\ve{\sigma} \times \left(\ve{x} \times \ve{\sigma}\right)}{x\,\left(x - \ve{\sigma} \cdot \ve{x} \right)^2}\,
- 4\,\frac{\ve{\sigma}}{x\,\left(x - \ve{\sigma} \cdot \ve{x} \right)}
+ \frac{9}{2}\,\frac{\ve{\sigma}}{x^2}
\nonumber\\
\nonumber\\
\fl \hspace{1.6cm} -\,\frac{15}{4}\,\left(\ve{\sigma} \cdot \ve{x}\right)\,
\frac{\ve{\sigma} \times \left(\ve{x} \times \ve{\sigma}\right)}{x^2\,\left|\ve{\sigma} \times \ve{x}\right|^2}
- \frac{15}{4}\,\frac{\ve{\sigma} \times \left(\ve{x} \times \ve{\sigma}\right)}{\left|\ve{\sigma} \times \ve{x}\right|^3}
\left(\arctan \frac{\ve{\sigma} \cdot \ve{x}}{\left|\ve{\sigma} \times \ve{x}\right|} + \frac{\pi}{2}\right). 
\label{Vectorial_Function_A3}
\end{eqnarray}

\noindent
The result in Eq.~(\ref{First_Integration}) agrees with Eq.~(35) in a preceding investigation \cite{Zschocke3}, while the vectorial functions  
in Eqs.~(\ref{Vectorial_Function_A1}) - (\ref{Vectorial_Function_A3}) coincide with the vectorial functions in 
Eqs.~(36) - (38) in \cite{Zschocke3}; notice that the vectorial function $\ve{\epsilon}_1$ was contained in $\ve{A}_2$ in \cite{Zschocke3}. 
The explicit form of the vectorial function $\ve{\epsilon}_1$  
has already been determined in \cite{Zschocke3} but was not presented there explicitly, only an upper limit has been given by Eq.~(39) in \cite{Zschocke3}.  
The vectorial function $\ve{\epsilon}_1$ was given at the first time in \cite{Zschocke_Proceeding_2} and reads explicitly   
\begin{eqnarray}
\fl \ve{\epsilon}_1\left(\ve{x},\ve{v}\right) =
- \frac{v^2}{c^2}\,\frac{\ve{\sigma} \times \left(\ve{x} \times \ve{\sigma}\right)}{x - \ve{\sigma} \cdot \ve{x}}\,\frac{1}{x}
- 2\,\left(\frac{\ve{v} \cdot \ve{x}}{c\,x}\right)^2\,
\frac{\ve{\sigma} \times \left(\ve{x} \times \ve{\sigma}\right)}{x - \ve{\sigma} \cdot \ve{x}}\,\frac{1}{x}
\nonumber\\
\fl \hspace{0.35cm} -\, 2\,
\left(\frac{\ve{\sigma} \cdot \ve{v}}{c}\right)^2\,\frac{\ve{\sigma} \times \left(\ve{x} \times \ve{\sigma}\right)}{x - \ve{\sigma} \cdot \ve{x}}\,\frac{1}{x}
+\, 4\,\left(\frac{\ve{\sigma} \cdot \ve{v}}{c}\right) \, \left(\frac{\ve{v} \cdot \ve{x}}{c\,x}\right) \,
\frac{\ve{\sigma} \times \left(\ve{x} \times \ve{\sigma}\right)}{x - \ve{\sigma} \cdot \ve{x}}\,\frac{1}{x}
\nonumber\\
\fl  \hspace{0.35cm} +\, 4\,\frac{\ve{v}}{c}\,\left(\frac{\ve{v} \cdot \ve{x}}{c\,x}\right) \,\frac{1}{x}
- 4\,\frac{\ve{v}}{c}\,\left(\frac{\ve{\sigma} \cdot \ve{v}}{c}\right)\,\frac{1}{x}
 - \, \frac{v^2}{c^2}\,\frac{\ve{\sigma}}{x}
- 2\, \left(\frac{\ve{v} \cdot \ve{x}}{c\,x}\right)^2\,\frac{\ve{\sigma}}{x}
+ 2\, \left(\frac{\ve{\sigma} \cdot \ve{v}}{c}\right)^2\,\frac{\ve{\sigma}}{x}\,. 
\label{epsilon_1}
\end{eqnarray}

\noindent
For the determination of the upper limit of $\left|\ve{\epsilon}_1\right|$ it is useful to introduce the angles
\begin{eqnarray}
\left(\frac{\ve{v} \cdot \ve{x}}{c\,x}\right) &=& \frac{v}{c}\,\cos \alpha \,,
\label{angle_alpha}
\\
\nonumber\\
\left(\frac{\ve{\sigma} \cdot \ve{v}}{c}\right) &=& \frac{v}{c}\,\cos \beta \,. 
\label{angle_beta}
\end{eqnarray}

\noindent
Then, one obtains for the upper limit of the absolute value of $\ve{\epsilon}_1$   
\begin{eqnarray}
\left|\ve{\epsilon}_1\left(\ve{x},\ve{v}\right) \right| &\le& \frac{18}{\left|\ve{\sigma} \times \ve{x}\right|}\,\frac{v^2}{c^2} 
+ \frac{9}{x}\,\frac{v^2}{c^2}\,,  
\label{estimate_epsilon_1}
\end{eqnarray}

\noindent
which is a marginal correction of the estimate given by Eq.~(39) in \cite{Zschocke3}. Using the parameters in Table~\ref{Table1}, 
numerical values of the upper limit of this expression  
in (\ref{estimate_epsilon_1}) are given in Table~\ref{Table2} which indicate the negligibility of this term 
for light deflection measurements on nas-level of accuracy within the Solar System.  
It should be emphasized that the expression in Eq.~(\ref{First_Integration}) is identical to the expression in Eq.~(\ref{First_Integration_Preliminary})  
up to terms of the order ${\cal O}\left(c^{-5}\right)$.

\section{The trajectory of a light signal in 2PN approximation}\label{Section5}

\subsection{The preliminary solution for the light trajectory} 

According to Eq.~(\ref{Integral_2}), the second integration of geodesic equation (\ref{Geodesic_Equation3}) yields the trajectory  
of the light signal in 2PN approximation. The integration proceeds similar to the first integration, that means 
by iteration and by means of integration by parts, see Section \ref{Section3a}. Such a solution is, first of all, given in terms 
of the instantaneous position of the body $\ve{x}_A\left(t\right)$ as formally given by Eq.~(\ref{Second_Integration_Preliminary}).   
Afterwards, this solution of the light trajectory is reexpressed in terms of the retarded position $\ve{x}_A\left(s\right)$ of the body, 
see Section \ref{Section3b} and  
especially the relations (\ref{retarded_time_C}) - (\ref{retarded_time_D}) and (\ref{retarded_time_E}) - (\ref{retarded_time_F}).  
Then, the second integration of geodesic equation in 2PN approximation reads as follows:    
\begin{eqnarray}
\fl \ve{x}_{\rm 2PN}\left(t\right) = \ve{x}_0 + c \left(t - t_0\right) \ve{\sigma} 
+ \,m_A\,\bigg(\ve{\tilde B}_1\left(\ve{r}^{\rm 1PN}_A\left(s\right)\right)
- \ve{\tilde B}_1\left(\ve{r}^{\rm 1PN}_A\left(s_0\right)\right) \bigg)
\nonumber\\
\nonumber\\
\fl \hspace{1.7cm} + \,m_A\,\bigg(\ve{\tilde B}_2\left(\ve{r}^{\rm N}_A\left(s\right),\ve{v}_A\left(s\right)\right)
- \ve{\tilde B}_2\left(\ve{r}^{\rm N}_A\left(s_0\right),\ve{v}_A\left(s_0\right)\right) \bigg)
\nonumber\\
\nonumber\\
\fl \hspace{1.7cm} +\,m_A^2\,\bigg(\ve{\tilde B}_3\left(\ve{r}^{\rm N}_A\left(s\right)\right) - \ve{\tilde B}_3\left(\ve{r}^{\rm N}_A\left(s_0\right)\right)\bigg)  
\nonumber\\
\nonumber\\
\fl \hspace{1.7cm} +\, \,m_A\,\bigg(\ve{\tilde \epsilon}_2\left(\ve{r}^{\rm N}_A\left(s\right),\ve{v}_A\left(s\right),\ve{a}_A\left(s\right)\right)
- \ve{\tilde \epsilon}_2\left(\ve{r}^{\rm N}_A\left(s_0\right),\ve{v}_A\left(s_0\right),\ve{a}_A\left(s_0\right)\right) \bigg), 
\label{Second_Integration}
\end{eqnarray} 

\noindent 
where the arguments have been given by Eqs.~(\ref{Distance_Vector_N}) and (\ref{Distance_Vector_1PN}). Like in (\ref{First_Integration}),  
one may replace the approximate arguments in Eq.~(\ref{Second_Integration}) by the exact expression in (\ref{vector_C}) because
of $\ve{r}_A\left(s\right) = \ve{r}^{\rm N}_A\left(s\right) + {\cal O}\left(c^{-2}\right)$ and
$\ve{r}_A\left(s\right) = \ve{r}^{\rm 1PN}_A\left(s\right) + {\cal O}\left(c^{-3}\right)$
such a replacement would cause an error of the order ${\cal O}\left(c^{-5}\right)$ which is beyond 2PN approximation.  
The vectorial functions are given as follows:  
\begin{eqnarray}  
\fl \ve{\tilde B}_1\left(\ve{x}\right) = -\,2\, \frac{\ve{\sigma} \times \left(\ve{x} \times \ve{\sigma}\right)}{x - \ve{\sigma} \cdot \ve{x}}
+\,2\,\ve{\sigma}\,\ln \left(x - \ve{\sigma} \cdot \ve{x}\right),
\label{Vectorial_Function_B1}
\\
\nonumber\\
\fl \ve{\tilde B}_2\left(\ve{x},\ve{v}\right) =
+ 2\,\frac{\ve{\sigma} \times \left(\ve{x} \times \ve{\sigma}\right)}{x - \ve{\sigma} \cdot \ve{x}}\,\frac{\ve{\sigma} \cdot \ve{v}}{c}
-\,2\,\frac{\ve{v}}{c}\,\ln \left(x - \ve{\sigma} \cdot \ve{x}\right) +\,2\,\frac{\ve{v}}{c}\,,  
\label{Vectorial_Function_B2}
\\
\nonumber\\
\fl \ve{\tilde B}_3\left(\ve{x}\right) = + 4\,\frac{\ve{\sigma}}{x - \ve{\sigma} \cdot \ve{x}}
+\,4\,\frac{\ve{\sigma} \times \left(\ve{x} \times \ve{\sigma}\right)}{\left(x - \ve{\sigma} \cdot \ve{x}\right)^2}
+\,\frac{1}{4}\,\frac{\ve{x}}{x^2}
-\,\frac{15}{4}\,\frac{\ve{\sigma}}{\left|\ve{\sigma} \times \ve{x}\right|} \,
\arctan \frac{\ve{\sigma} \cdot \ve{x}}{\left|\ve{\sigma} \times \ve{x}\right|}
\nonumber\\
\nonumber\\
\fl  \hspace{1.6cm} -\,\frac{15}{4}\,\left(\ve{\sigma} \cdot \ve{x}\right) \frac{\ve{\sigma} \times \left(\ve{x} \times \ve{\sigma}\right)}
{\left|\ve{\sigma} \times \ve{x}\right|^3} \left(\arctan \frac{\ve{\sigma} \cdot \ve{x}}{\left|\ve{\sigma} \times \ve{x}\right|} 
+ \frac{\pi}{2}\right),   
\label{Vectorial_Function_B3}
\end{eqnarray}

\noindent 
where we notice that the last term in the vectorial function $\ve{\tilde B}_2$ vanishes in case of a closed system of $N$ bodies subject to 
relation (\ref{Retarded_Light_Trajectory_M_3}).
The result in Eq.~(\ref{Second_Integration}) agrees with Eq.~(41) in a preceding investigation \cite{Zschocke3}, while the vectorial functions
in Eqs.~(\ref{Vectorial_Function_B1}) - (\ref{Vectorial_Function_B3}) coincide with the vectorial functions in
Eqs.~(42) - (44) in \cite{Zschocke3}; notice that the vectorial function $\ve{\tilde \epsilon}_2$ was contained in $\ve{B}_2$ in \cite{Zschocke3}.
The expression for the vectorial function $\ve{\tilde \epsilon}_2$ was given at the first time in \cite{Zschocke_Proceeding_2}, which   
is separated into $\ve{\tilde \epsilon}^A_2$ proportional to the velocity of the body plus a term $\ve{\tilde \epsilon}^B_2$ proportional to the  
acceleration of the body,  
\begin{eqnarray}
\fl \ve{\tilde \epsilon}_2\left(\ve{x},\ve{v},\ve{a}\right) 
= \ve{\tilde \epsilon}^A_2\left(\ve{x},\ve{v}\right) + \ve{\tilde \epsilon}^B_2\left(\ve{x},\ve{a}\right), 
\label{epsilon_2}
\\
\nonumber\\  
\fl \ve{\tilde \epsilon}^A_2\left(\ve{x},\ve{v}\right) 
= - \frac{v^2}{c^2}\,\frac{\ve{\sigma} \times \left(\ve{x} \times \ve{\sigma}\right)}{x - \ve{\sigma} \cdot \ve{x}}
+ \frac{v^2}{c^2}\, \ve{\sigma}\,\ln \left(x - \ve{\sigma} \cdot \ve{x}\right), 
\label{epsilon_2_v} 
\\
\nonumber\\
\fl \ve{\tilde \epsilon}^B_2\left(\ve{x},\ve{a}\right)  
= + 2 \frac{\ve{a}}{c^2} \left(x - \ve{\sigma} \cdot \ve{x} \right) \left( 1 - \ln\left(x - \ve{\sigma} \cdot \ve{x}\right)\right) 
+ 2 \frac{\ve{\sigma} \cdot \ve{a}}{c^2}\,\left(\ve{\sigma} \times \left(\ve{x} \times \ve{\sigma}\right)\right) \ln\left(x - \ve{\sigma} \cdot \ve{x}\right).  
\nonumber\\
\label{epsilon_2_a} 
\end{eqnarray}
 
\noindent 
The vectorial function $\ve{\tilde \epsilon}^A_2$ has also been determined in  
\cite{Zschocke3} but was not presented there explicitly. Its upper limit reads 
\begin{eqnarray}
\left|\ve{\tilde \epsilon}^A_2\left(\ve{x},\ve{v}\right)\right| \le \frac{v^2}{c^2} 
\sqrt{\left(\frac{x + \ve{\sigma} \cdot \ve{x}}{\left|\ve{\sigma} \times \ve{x} \right|}\right)^2 
+ \left(\ln\left(x - \ve{\sigma} \cdot \ve{x}\right)\right)^2}\;, 
\label{upper_limit_epsilon_2_v}
\end{eqnarray}

\noindent 
which confirms Eq.~(45) in \cite{Zschocke3}. 
The vectorial function $\ve{\tilde \epsilon}^B_2$ was not determined in \cite{Zschocke3}, where   
it was asserted that the light trajectory is supposed to contain no terms proportional to the acceleration of the body,  
see text below Eq.~(45) in \cite{Zschocke3}. But according to Eq.~(\ref{epsilon_2_a}), this assertion  
is incorrect because there are terms in the vectorial function $\ve{\tilde \epsilon}_2$ which are proportional to the acceleration of the body.  
In the Appendix we show that the terms proportional to the acceleration of the body are in exact agreement with the results  
in \cite{KopeikinSchaefer1999} up to the order ${\cal O}\left(c^{-5}\right)$; cf. Eq.~(\ref{Remnant_Integrals_70}). For the sake of completeness  
an upper limit of the vectorial function $\ve{\tilde \epsilon}^B_2$ is also given, 
\begin{eqnarray}
\fl \left|\ve{\tilde \epsilon}^B_2\left(\ve{x},\ve{a}\right)\right| \le 2\,\frac{a}{c^2}
\bigg(\left(x - \ve{\sigma} \cdot \ve{x}\right) \left(1 + \left| \ln \left(x - \ve{\sigma} \cdot \ve{x}\right) \right| \right) 
+ \left|\ve{\sigma} \times \ve{x}\right| \left| \ln \left(x - \ve{\sigma} \cdot \ve{x}\right) \right| \bigg).  
\label{upper_limit_epsilon_2_a}
\end{eqnarray}
 
\noindent 
Before we proceed further, it should be noticed that the expression in Eq.~(\ref{Second_Integration}) is identical to the expression  
in Eq.~(\ref{Second_Integration_Preliminary}) up to terms of the order ${\cal O}\left(c^{-5}\right)$. However,    
the arguments of the logarithms in Eqs.~(\ref{Vectorial_Function_B1}) - (\ref{Vectorial_Function_B2}) and Eqs.~(\ref{epsilon_2_v}) - (\ref{epsilon_2_a}) 
as well as in (\ref{upper_limit_epsilon_2_v}) - (\ref{upper_limit_epsilon_2_a})  
have the physical dimension of a length and for this reason they are not well-defined mathematical objects.  
This problem will be issued in the next Section.

\subsection{Cancellation of improperly defined logarithmic terms}  

Let us consider the logarithmic functions in the solution for the light trajectory (\ref{Second_Integration}), as there are   
\begin{eqnarray} 
\fl \ve{S}_1 = +\,2\,m_A\,\ve{\sigma}\,
\ln \frac{r^{\rm 1PN}_A\left(s\right) - \ve{\sigma}\cdot\ve{r}^{\rm 1PN}_A\left(s\right)}
{r^{\rm 1PN}_A\left(s_0\right)-\ve{\sigma}\cdot\ve{r}^{\rm 1PN}_A\left(s_0\right)}\,, 
\label{logarithmic_term_1} 
\\
\nonumber\\ 
\fl \ve{S}_2 = -\,2\,\frac{m_A}{c}\,
\bigg[\ve{v}_A\left(s\right)\,\ln \left(r^{\rm N}_A\left(s\right) - \ve{\sigma} \cdot \ve{r}^{\rm N}_A\left(s\right)\right)  
- \left(s \leftrightarrow s_0\right) \bigg],  
\label{logarithmic_term_2}
\\
\nonumber\\ 
\fl \ve{S}_3 = + \frac{m_A}{c^2}\,\ve{\sigma} 
\bigg[v_A^2\left(s\right)\,\ln \left(r^{\rm N}_A\left(s\right) - \ve{\sigma} \cdot \ve{r}^{\rm N}_A\left(s\right)\right) 
- \left(s \leftrightarrow s_0\right) \bigg],  
\label{logarithmic_term_3}
\\
\nonumber\\
\fl \ve{S}_4 = - 2\,\frac{m_A}{c^2}\bigg[\ve{a}_A\left(s\right)
\left(r^{\rm N}_A\left(s\right) - \ve{\sigma} \cdot \ve{r}^{\rm N}_A\left(s\right) \right)
\ln \left(r^{\rm N}_A\left(s\right) - \ve{\sigma} \cdot \ve{r}^{\rm N}_A\left(s\right)\right)
- \left(s \leftrightarrow s_0\right) \bigg],  
\label{logarithmic_term_4}
\\
\nonumber\\
\fl \ve{S}_5 = + 2\,\frac{m_A}{c^2}\bigg[
\ve{\sigma} \cdot \ve{a}_A\left(s\right)\,\ve{d}^{\rm N}_A\left(s\right)
\ln \left(r^{\rm N}_A\left(s\right)-\ve{\sigma}\cdot\ve{r}^{\rm N}_A\left(s\right)\right)  
- \left(s \leftrightarrow s_0\right) \bigg].  
\label{logarithmic_term_5}
\end{eqnarray}

\noindent
The argument of the logarithm in (\ref{logarithmic_term_1}) is dimensionless, but the   
arguments of the logarithms in (\ref{logarithmic_term_2}) - (\ref{logarithmic_term_5}) have the dimension of a length   
and in this sense they are meaningless expressions as they stand. A meaning to these ill-defined logarithms can be attributed  
by a series expansion in terms of inverse powers of the speed of light, 
\begin{eqnarray}
\ve{v}_A\left(s\right) &=& \ve{v}_A\left(s_0\right) + \ve{a}_A\left(s_0\right)\left(s - s_0\right) + {\cal O}\left(\dot{a}_A\right),
\label{series_expansion_1}
\\
\nonumber\\
\ve{a}_A\left(s\right) &=& \ve{a}_A\left(s_0\right) + {\cal O}\left(\dot{a}_A\right). 
\label{series_expansion_2}
\end{eqnarray}

\noindent
We will show how all improperly defined logarithms cancel each other. The proof is restricted to logarithms proportional to  
$\ve{v}_A$ and $\ve{a}_A$, but one may show that each ill-defined logarithms proportional to higher time-derivatives 
do cancel each other as well. Using (\ref{series_expansion_1}) - (\ref{series_expansion_2}) we obtain  
\begin{eqnarray}
\fl \ve{S}_2 = -\,2\,m_A\,\frac{\ve{v}_A\left(s_0\right)}{c}\,
\ln \frac{r^{\rm N}_A\left(s\right) - \ve{\sigma}\cdot\ve{r}^{\rm N}_A\left(s\right)}{r^{\rm N}_A\left(s_0\right)-\ve{\sigma}\cdot\ve{r}^{\rm N}_A\left(s_0\right)} 
\nonumber\\
\fl \hspace{0.9cm} -2\,m_A\,\frac{\ve{a}_A\left(s_0\right)}{c} \left(s - s_0\right) \ln \left(r^{\rm N}_A\left(s\right) 
- \ve{\sigma} \cdot \ve{r}^{\rm N}_A\left(s\right)\right) 
+ {\cal O}\left(\dot{a}_A\right),  
\label{logarithmic_term_6}
\\
\nonumber\\
\fl \ve{S}_3 = + m_A\,\ve{\sigma}\,\frac{v_A^2\left(s_0\right)}{c^2}\,\ln  
\frac{r^{\rm N}_A\left(s\right)-\ve{\sigma} \cdot \ve{r}^{\rm N}_A\left(s\right)}{r^{\rm N}_A\left(s_0\right)-\ve{\sigma}\cdot\ve{r}^{\rm N}_A\left(s_0\right)} 
+ {\cal O}\left(v_A\,a_A\right),  
\label{logarithmic_term_7}
\\
\nonumber\\ 
\fl \ve{S}_4 = - 2\,m_A \frac{\ve{a}_A\left(s_0\right)}{c^2}
\left(r^{\rm N}_A\left(s_0\right) - \ve{\sigma} \cdot \ve{r}^{\rm N}_A\left(s_0\right) \right)
\ln \frac{r^{\rm N}_A\left(s\right)-\ve{\sigma}\cdot\ve{r}^{\rm N}_A\left(s\right)}{r^{\rm N}_A\left(s_0\right)-\ve{\sigma}\cdot\ve{r}^{\rm N}_A\left(s_0\right)}
\nonumber\\
\nonumber\\
\fl \hspace{0.9cm} + 2\,m_A\,
\frac{\ve{a}_A\left(s_0\right)}{c} \left(s - s_0\right) \ln \left(r^{\rm N}_A\left(s\right) - \ve{\sigma} \cdot \ve{r}^{\rm N}_A\left(s\right)\right)
 + {\cal O}\left(a_A\,v_A\right) + {\cal O}\left(\dot{a}_A\right),  
\label{logarithmic_term_8}
\\
\nonumber\\
\fl \ve{S}_5 = + 2\,m_A \frac{\ve{\sigma} \cdot \ve{a}_A\left(s_0\right)}{c^2}\,\ve{d}^{\rm N}_A\left(s_0\right)
\,\ln 
\frac{r^{\rm N}_A\left(s\right)-\ve{\sigma}\cdot\ve{r}^{\rm N}_A\left(s\right)}{r^{\rm N}_A\left(s_0\right)-\ve{\sigma}\cdot\ve{r}^{\rm N}_A\left(s_0\right)} 
+ {\cal O}\left(v_A\,a_A\right).  
\label{logarithmic_term_9}
\end{eqnarray}

\noindent
In order to get (\ref{logarithmic_term_8}) we have used  
$\ve{\sigma} \cdot \ve{x}_A\left(s\right) - \ve{\sigma} \cdot \ve{x}_A\left(s_0\right) = {\cal O}\left(v_A\right)$ and the relation  
\begin{eqnarray}
\fl r^{\rm N}_A\left(s\right) - \ve{\sigma} \cdot \ve{r}^{\rm N}_A\left(s\right) = r^{\rm N}_A\left(s_0\right) - \ve{\sigma} \cdot \ve{r}^{\rm N}_A\left(s_0\right) 
- c\left(s - s_0\right) + \ve{\sigma} \cdot \ve{x}_A\left(s\right) - \ve{\sigma} \cdot \ve{x}_A\left(s_0\right),  
\label{logarithmic_term_10}
\end{eqnarray}

\noindent 
which is valid up to terms of the order ${\cal O}\left(c^{-2}\right)$ and which follows from (\ref{Retarded_Time_1}) and (\ref{Retarded_Time_0}) 
as well as (\ref{vector_C}) and (\ref{vector_D}), and which corresponds to Eq.~(96) in \cite{KopeikinSchaefer1999}.  
One recognizes that the arguments of the second logarithmic function in (\ref{logarithmic_term_6}) and (\ref{logarithmic_term_8}) have  
still the dimension of a length, but they do cancel each other. So there remain no improperly defined logarithms if one takes into account 
the acceleration terms correctly. In other words, the seemingly occurrence of ill-defined logarithms appears as an artifact of the 
approximation to neglect acceleration terms. Accordingly, by series expansion in inverse powers of the speed of light one recognizes that the 
solution for the light trajectory (\ref{Second_Integration}) contains, in fact, only well-defined logarithms. This observation allows to rearrange 
the solution for the light trajectory (\ref{Second_Integration}) in such a manner that it contains only properly-defined logarithms from  the very beginning,  
which is the subject of the next Section.

\subsection{The final expression for the 2PN light trajectory}

According to the results of the previous Section, the light trajectory in 2PN approximation in the field of one moving monopole can be 
rearranged in such a way that it contains only well-defined logarithmic functions:  
\begin{eqnarray}
\fl \ve{x}_{\rm 2PN}\left(t\right) = \ve{x}_0 + c \left(t - t_0\right) \ve{\sigma}
+ \,m_A\,\bigg( 
\ve{B}_1\left(\ve{r}^{\rm 1PN}_A\left(s\right)\right) - \ve{B}_1\left(\ve{r}^{\rm 1PN}_A\left(s_0\right)\right)
\bigg) 
\nonumber\\
\nonumber\\
\fl \hspace{1.7cm} + \,m_A\,\bigg(\ve{B}^A_2\left(\ve{r}^{\rm N}_A\left(s\right),\ve{v}_A\left(s_0\right)\right)
- \ve{B}^A_2\left(\ve{r}^{\rm N}_A\left(s_0\right),\ve{v}_A\left(s_0\right)\right)\bigg)
\nonumber\\
\nonumber\\
\fl \hspace{1.7cm} + \,m_A\,\bigg(\ve{B}^B_2\left(\ve{r}^{\rm N}_A\left(s\right),\ve{v}_A\left(s\right)\right)
- \ve{B}^B_2\left(\ve{r}^{\rm N}_A\left(s_0\right),\ve{v}_A\left(s_0\right)\right)\bigg)  
\nonumber\\
\nonumber\\
\fl \hspace{1.7cm} + \,m_A^2\,\bigg(\ve{B}_3\left(\ve{r}^{\rm N}_A\left(s\right)\right) - \ve{B}_3\left(\ve{r}^{\rm N}_A\left(s_0\right)\right) \bigg)
+ m_A\,\ve{\epsilon}_2\left(s,s_0\right),  
\label{Second_Integration_Rearranged}
\end{eqnarray}

\noindent
where the arguments are given by Eqs.~(\ref{Distance_Vector_N}) and (\ref{Distance_Vector_1PN}). It should be noticed that in  
Eq.~(\ref{Second_Integration_Rearranged}) one may formally replace these approximate arguments by the exact expression in (\ref{vector_C}) because  
of $\ve{r}_A\left(s\right) = \ve{r}^{\rm N}_A\left(s\right) + {\cal O}\left(c^{-2}\right)$ and 
$\ve{r}_A\left(s\right) = \ve{r}^{\rm 1PN}_A\left(s\right) + {\cal O}\left(c^{-3}\right)$  
such a replacement would cause an error of the order ${\cal O}\left(c^{-5}\right)$ which is beyond 2PN approximation.
The vectorial functions are given as follows:
\begin{eqnarray}  
\fl \ve{B}_1\left(\ve{x}\right) = -\,2\, \frac{\ve{\sigma} \times \left(\ve{x} \times \ve{\sigma}\right)}{x - \ve{\sigma} \cdot \ve{x}}
+\,2\,\ve{\sigma}\,\ln \left(x - \ve{\sigma} \cdot \ve{x}\right),
\label{Vectorial_Function_C1}
\\
\nonumber\\
\fl \ve{B}^A_2\left(\ve{x},\ve{v}\right) = - 2\,\frac{\ve{v}}{c}\,\ln \left(x - \ve{\sigma} \cdot \ve{x}\right),
\label{Vectorial_Function_C2_A}
\\
\nonumber\\
\fl \ve{B}^B_2\left(\ve{x},\ve{v}\right) =
+ 2\,\frac{\ve{\sigma} \times \left(\ve{x} \times \ve{\sigma}\right)}{x - \ve{\sigma} \cdot \ve{x}}\,\frac{\ve{\sigma} \cdot \ve{v}}{c} 
+ 2\,\frac{\ve{v}}{c}\,,  
\label{Vectorial_Function_C2_B}
\\
\nonumber\\
\fl \ve{B}_3\left(\ve{x}\right) = + 4\,\frac{\ve{\sigma}}{x - \ve{\sigma} \cdot \ve{x}}
+\,4\,\frac{\ve{\sigma} \times \left(\ve{x} \times \ve{\sigma}\right)}{\left(x - \ve{\sigma} \cdot \ve{x}\right)^2}
+\,\frac{1}{4}\,\frac{\ve{x}}{x^2}
-\,\frac{15}{4}\,\frac{\ve{\sigma}}{\left|\ve{\sigma} \times \ve{x}\right|} \,
\arctan \frac{\ve{\sigma} \cdot \ve{x}}{\left|\ve{\sigma} \times \ve{x}\right|}
\nonumber\\
\nonumber\\
\fl \hspace{1.55cm} -\,\frac{15}{4}\,\left(\ve{\sigma} \cdot \ve{x}\right) \frac{\ve{\sigma} \times \left(\ve{x} \times \ve{\sigma}\right)}
{\left|\ve{\sigma} \times \ve{x}\right|^3} \left(\arctan \frac{\ve{\sigma} \cdot \ve{x}}{\left|\ve{\sigma} \times \ve{x}\right|}
+ \frac{\pi}{2}\right). 
\label{Vectorial_Function_C3}
\end{eqnarray}

\noindent
We recall, that the second term in the vectorial function $\ve{B}^B_2$ would vanish in case of closed system of 
$N$ bodies moving under their mutual gravitational interaction owing to relation (\ref{Retarded_Light_Trajectory_M_3}).  
It is important to realize that the velocity in the vectorial functions $\ve{B}^A_2$ in (\ref{Second_Integration_Rearranged}) is taken at the very same instant  
of retarded time $s_0$, which ensures the logarithm in (\ref{Vectorial_Function_C2_A}) in combination with (\ref{Second_Integration_Rearranged}) to be well-defined.  
The explicit form of the vectorial function $\ve{\epsilon}_2$ with well-defined logarithms is separated into two pieces, proportional to $\ve{v}_A^2$ and  
$\ve{a}_A$, and reads  
\begin{eqnarray}
\fl \ve{\epsilon}_2\left(s,s_0\right) = \ve{\epsilon}^A_2\left(s,s_0\right) + \ve{\epsilon}^B_2\left(s,s_0\right),  
\label{epsilon_3}
\\
\nonumber\\
\fl \ve{\epsilon}^A_2\left(s,s_0\right) = - \frac{v_A^2\left(s\right)}{c^2}\,  
\frac{\ve{\sigma} \times \left(\ve{r}^{\rm N}_A\left(s\right) \times \ve{\sigma}\right)}{r^{\rm N}_A\left(s\right) - \ve{\sigma} \cdot \ve{r}^{\rm N}_A\left(s\right)}
+ \frac{v_A^2\left(s_0\right)}{c^2}\,
\frac{\ve{\sigma} \times \left(\ve{r}^{\rm N}_A\left(s_0\right) \times \ve{\sigma}\right)}{r^{\rm N}_A\left(s_0\right) - \ve{\sigma} \cdot \ve{r}^{\rm N}_A\left(s_0\right)} 
\nonumber\\
\nonumber\\
\fl \hspace{1.9cm} + \frac{v_A^2\left(s_0\right)}{c^2}\, \ve{\sigma}\,
\ln \frac{r^{\rm N}_A\left(s\right) - \ve{\sigma} \cdot \ve{r}^{\rm N}_A\left(s\right)}{r^{\rm N}_A\left(s_0\right) - \ve{\sigma} \cdot \ve{r}^{\rm N}_A\left(s_0\right)}\,, 
\label{epsilon_3a}
\\
\nonumber\\
\fl \ve{\epsilon}^B_2\left(s,s_0\right) = + 2\,\ve{d}^{\rm N}_A\left(s_0\right)\,\frac{\ve{\sigma} \cdot \ve{a}_A\left(s_0\right)}{c^2}\,
\,\ln \frac{r^{\rm N}_A\left(s\right) - \ve{\sigma} \cdot \ve{r}^{\rm N}_A\left(s\right)}{r^{\rm N}_A\left(s_0\right) - \ve{\sigma} \cdot \ve{r}^{\rm N}_A\left(s_0\right)} 
\nonumber\\
\nonumber\\
\fl \hspace{1.9cm} + 2\,\frac{\ve{a}_A\left(s_0\right)}{c^2}
\left[r^{\rm N}_A\left(s\right) - \ve{\sigma} \cdot \ve{r}^{\rm N}_A\left(s\right)
- r^{\rm N}_A\left(s_0\right) + \ve{\sigma} \cdot \ve{r}^{\rm N}_A\left(s_0\right) \right]
\nonumber\\
\nonumber\\
\fl \hspace{1.9cm} - 2\,\frac{\ve{a}_A\left(s_0\right)}{c^2}
\left(r^{\rm N}_A\left(s_0\right) - \ve{\sigma} \cdot \ve{r}^{\rm N}_A\left(s_0\right) \right)
\ln \frac{r^{\rm N}_A\left(s\right) - \ve{\sigma} \cdot \ve{r}^{\rm N}_A\left(s\right)}{r^{\rm N}_A\left(s_0\right) - \ve{\sigma} \cdot \ve{r}^{\rm N}_A\left(s_0\right)}\,. 
\label{epsilon_3b}
\end{eqnarray}

\noindent
In order to determine the relevance of $\ve{\epsilon}_2$ it is more meaningful to consider $\left|\ve{\sigma} \cdot \ve{\epsilon}_2\right|$ which determines 
its impact on the Shapiro time delay, rather than its absolute value $\left|\ve{\epsilon}_2\right|$. Complying  
to this proposition in anticipation of the Shapiro time delay (\ref{Shaprio_Time_Delay}), we consider the following upper limit  
(note that $\ve{\sigma} = \ve{k} + {\cal O}\left(c^{-2}\right)$):  
\begin{eqnarray}
\fl \left|\ve{\sigma} \cdot \ve{\epsilon}_2\left(s,s_0\right)\right| \le \left|\ve{\sigma} \cdot \ve{\epsilon}^A_2\left(s,s_0\right)\right| 
+ \left|\ve{\sigma} \cdot \ve{\epsilon}^B_2\left(s,s_0\right)\right|\,, 
\label{absolute_value_epsilon_3}
\\
\nonumber\\ 
\fl \left|\ve{\sigma} \cdot \ve{\epsilon}^A_2\left(s,s_0\right)\right| \le \frac{v_A^2\left(s_0\right)}{c^2}\,
\ln \frac{4\,r^{\rm N}_A\left(s_0\right)\,r^{\rm N}_A\left(s\right)}{\left(d^{\rm N}_A\left(s\right)\right)^2}\,,  
\label{absolute_value_epsilon_3a}  
\\
\nonumber\\
\fl \left|\ve{\sigma} \cdot \ve{\epsilon}^B_2\left(s,s_0\right)\right| \le 4\,\frac{a_A\left(s_0\right)}{c^2}\,r^{\rm N}_A\left(s_0\right)
\left(1 + \ln \frac{4\,r^{\rm N}_A\left(s_0\right)\,r^{\rm N}_A\left(s\right)}{\left(d^{\rm N}_A\left(s\right)\right)^2}\right),   
\label{absolute_value_epsilon_3b}  
\end{eqnarray}

\noindent
where an astrometric configuration with $r^{\rm N}_A\left(s_0\right) \simeq - \ve{\sigma} \cdot \ve{r}^{\rm N}_A\left(s_0\right)$  
and $r^{\rm N}_A\left(s\right) \simeq \ve{\sigma} \cdot \ve{r}^{\rm N}_A\left(s\right)$ has been assumed 
and we recall the constraint in (\ref{constraint_acceleration}) for the near-zone of the Solar System. Using the parameters in Table~\ref{Table1}, 
numerical values for the upper limit of the expression in (\ref{absolute_value_epsilon_3}) are given in Table~\ref{Table2}.  
\begin{table}
\caption{\label{Table1}The numerical parameters Schwarzschild radius $m_A$, equatorial radius $P_A$, orbital velocity $v_A$, and orbital acceleration
$a_A$ of Solar System bodies \cite{JPL}.  
The maximal distance $r^{\rm max}_A\left(s\right)$ between observer
and body at its retarded position is computed under the assumption that the observer
is located at Lagrange point $L_2$, i.e. $1.5 \times 10^9\,{\rm m}$ from the Earth's orbit.
For the maximal distance between light-source and body we assume $r_{A}\left(s_0\right) = 10^{13}\,{\rm m}$
so that the light-source is for sure located inside the near-zone of the Solar System, cf. Eq.~(\ref{near-zone_2}).}
\footnotesize
\begin{tabular}{@{}cccccc}
\br 
Object & $m_A\,[{\rm m}]$ & $P_A\,[{\rm 10^6\,m}]$ & $v_A/c$ & $a_A\,[{\rm 10^{-3}\,m/s^2}]$ & $r^{\rm max}_A\left(s\right)\,[{\rm 10^{12}\,m}]$ \\ 
\mr 
Sun & $1476$ & $ 696$ & $4.0 \times 10^{-8}$ & $ - $ & $ 0.147 $ \\ 
Mercury & $ 0.245 \times 10^{-3}$ & $ 2.440$ & $15.8 \times 10^{-5}$ & $ 38.73 $ & $0.208$ \\ 
Venus & $ 3.615 \times 10^{-3} $& $ 6.052$ & $11.7 \times 10^{-5}$ & $ 11.34 $ & $0.258$ \\ 
Earth & $ 4.438 \times 10^{-3}$ & $ 6.378$ & $9.9 \times 10^{-5}$ & $ 5.93$ & $0.0015$ \\ 
Mars & $ 0.477 \times 10^{-3}$ & $ 3.396$ & $8.0 \times 10^{-5}$ & $ 2.55$ & $0.399$ \\ 
Jupiter & $1.410$ & $ 71.49$ & $4.4 \times 10^{-5}$ & $0.21$ & $ 0.898$ \\ 
Saturn & $ 0.422 $ & $ 60.27$ & $3.2 \times 10^{-5}$ & $ 0.06$ & $ 1.646$ \\ 
Uranus & $ 0.064 $ & $ 25.56$ & $2.3 \times 10^{-5}$ & $ 0.016$ & $ 3.142$ \\ 
Neptune & $ 0.076 $ & $ 24.76$ & $1.8 \times 10^{-5}$ & $ 0.0065$ & $4.638$ \\ 
\br
\end{tabular}\\  
\end{table} 
\normalsize 

\noindent 
A comment should be in order about the solutions (\ref{First_Integration}) and (\ref{Second_Integration_Rearranged}).  
It is remarkable that these solutions depend on the position  
and velocity of the body taken at the retarded instant of time, while it is independent of the history of the motion of the body.  
Due to the fact that gravitational fields are long-ranged, one actually expects that the trajectory of a light signal will be influenced during the
entire time of propagation, i.e. from the time of emission until the time of reception. This picture is absolutely true and the above standing
solutions just state that the integral effect of gravitational fields on the entire light trajectory can approximately be written
in such a form that it depends only of the position and velocity of the body at its retarded position.

\Table{\label{Table2}The upper limits $m_A \left|\ve{\epsilon}_1\right|$ (i.e. grazing light rays) and $m_A \left|\ve{\sigma} \cdot \ve{\epsilon}_2\right|$
(i.e. grazing light rays and maximal values for $r_A\left(s_0\right)$ and $r_A\left(s\right)$) as determined according to
the estimates (\ref{estimate_epsilon_1}) and (\ref{absolute_value_epsilon_3}), respectively.
These extremely tiny numerical values for the upper limits clearly
indicate that the vectorial functions $\ve{\epsilon}_1$ in Eq.~(\ref{epsilon_1}) and $\ve{\epsilon}_2$ in Eq.~(\ref{epsilon_3})
are negligible even for extremely high-precision astrometry on nano-arcsecond and
time delay measurements on pico-second level. These numbers also do illustrate the fact that acceleration terms in the solution for the
light trajectory are only of relevance in order to get logarithms with dimensionless arguments.}
\footnotesize
\begin{tabular}{@{}ccc}
\br 
Object & $m_A \left| \ve{\epsilon}_1\right|\,[{\rm nas}]$ & $m_A \left|\ve{\sigma} \cdot \ve{\epsilon}_2\right|/c\;[{\rm ps}]$ \\
\mr 
Sun & $1.3 \times 10^{-5}$ & $0.1 \times 10^{-6}$ \\
Mercury & $0.9 \times 10^{-2}$ & $0.4 \times 10^{-3}$ \\
Venus & $0.4 \times 10^{-1}$ & $0.2 \times 10^{-2}$ \\
Earth & $0.2 \times 10^{-1}$ & $0.9 \times 10^{-3}$ \\
Mars & $0.4 \times 10^{-2}$ & $0.5 \times 10^{-4}$ \\
Jupiter & $1.4 \times 10^{-1}$ & $ 0.1 \times 10^{-1}$ \\
Saturn & $0.2 \times 10^{-1}$ & $0.9 \times 10^{-3}$ \\
Uranus & $0.5 \times 10^{-2}$ & $0.4 \times 10^{-1}$ \\
Neptune & $0.4 \times 10^{-2}$ & $0.2 \times 10^{-1}$ \\
\br 
\end{tabular}\\  
\endTable
\normalsize

\section{Observables}\label{Section6} 

If light signals travel in the gravitational field of one massive body, two important effects of general relativity occur, namely bending of the  
light trajectory and delay of the travel time of the light signal. These observable effects are of specific importance for astrometry and will be considered 
in this Section.

\subsection{Total light deflection} 

The total light deflection is defined as angle $\varphi$ between the tangent vectors along the light trajectory at past and future null infinity,  
\begin{eqnarray}
\ve{\sigma} &=& \lim_{t \rightarrow - \infty}\,\frac{\dot{\ve{x}}\left(t\right)}{c} \,, 
\label{Total_Light_Deflection_1} 
\\
\nonumber\\
\ve{\nu} &=& \lim_{t \rightarrow + \infty}\,\frac{\dot{\ve{x}}\left(t\right)}{c}\,.   
\label{Total_Light_Deflection_2}
\end{eqnarray}

\noindent
The photon's coordinate velocity $\dot{\ve{x}}\left(t\right)$ is only known up to terms of the order ${\cal O}\left(c^{-5}\right)$, hence  
for the angle of total light deflection we may use  
\begin{eqnarray}
\varphi &=& \left| \ve{\sigma} \times \ve{\nu} \right| + {\cal O}\left(c^{-6}\right).  
\label{Total_Light_Deflection_0}
\end{eqnarray}

\noindent  
While these tangent vectors in (\ref{Total_Light_Deflection_1}) and (\ref{Total_Light_Deflection_2}) are exact definitions, 
the approach presented is only valid in the near-zone of the gravitating bodies. 
It is therefore clear that the limits in (\ref{Total_Light_Deflection_1}) and (\ref{Total_Light_Deflection_2}) can only be determined approximately.  
By inserting the coordinate velocity of a light signal in 2PN approximation (\ref{First_Integration}) into  
(\ref{Total_Light_Deflection_2}) one obtains the expression for $\ve{\nu}$ as presented by Eq.~(47) in \cite{Zschocke3}. Such an expression   
depends on both the impact vector in 1PN approximation, $\ve{d}^{\rm 1PN}_A$, 
as well as in Newtonian approximation, $\ve{d}^{\rm N}_A$. These impact vectors 
are related to each other subject to Eq.~(\ref{Relation_Impact_Vectors_2}), which allows to express $\ve{\nu}$ fully in terms of $\ve{d}^{\rm N}_A\,$,  
\begin{eqnarray}
\fl \ve{\nu} = \ve{\sigma}
+ \lim_{t \rightarrow + \infty} \Bigg[- 4\,m_A\,\frac{\ve{d}^{\rm N}_A\left(s\right)}{\left(d^{\rm N}_A\left(s\right)\right)^2}
\left(1 - \frac{\ve{\sigma} \cdot \ve{v}_A\left(s\right)}{c}\right)
- 8\,m_A^2\,\frac{\ve{\sigma}}{\left(d^{\rm N}_A\left(s\right)\right)^2}
\nonumber\\
\nonumber\\
\fl \hspace{2.7cm} - \frac{15}{4}\,\pi\,m_A^2\,\frac{\ve{d}^{\rm N}_A\left(s\right)}{\left(d^{\rm N}_A\left(s\right)\right)^3}
- 8\,m_A^2\,\frac{\ve{d}^{\rm N}_A\left(s_0\right)}{\left(d^{\rm N}_A\left(s\right)\right)^2}
\frac{1}{r^{\rm N}_A\left(s_0\right) - \ve{\sigma} \cdot \ve{r}^{\rm N}_A\left(s_0\right)}   
\nonumber\\ 
\nonumber\\ 
\fl \hspace{2.7cm} + 16\,m_A^2\,\frac{\ve{d}^{\rm N}_A\left(s\right)}{\left(d^{\rm N}_A\left(s\right)\right)^4}\,  
\frac{\ve{d}_A^{\rm N}\left(s\right) \cdot \ve{d}_A^{\rm N}\left(s_0\right)} 
{r^{\rm N}_A\left(s_0\right) - \ve{\sigma} \cdot \ve{r}^{\rm N}_A\left(s_0\right)}\Bigg]  
+ {\cal O}\left(c^{-5}\right).  
\label{nu_1} 
\end{eqnarray}

\noindent
The limit in (\ref{nu_1}) is treated by means of   
\begin{eqnarray}
\fl \hspace{2.0cm} \lim_{t \rightarrow + \infty} \,s 
= s_0 + \frac{r^{\rm N}_A\left(s_0\right) - \ve{\sigma} \cdot \ve{r}^{\rm N}_A\left(s_0\right)}{c}
\left(1 + \frac{\ve{\sigma} \cdot \ve{v}_A\left(s_0\right)}{c}\right) + {\cal O}\left(c^{-3}\right),  
\label{Limit_1_Retarded_Time}
\end{eqnarray}

\noindent
which is finite for worldlines of bodies restricted in the spatial domain of the Solar System 
and follows from (\ref{Retarded_Time_1}) - (\ref{Retarded_Time_0}) by series expansion;
compare text below Eq.~(25) in \cite{KopeikinSchaefer1999} which is in accordance with (\ref{Limit_1_Retarded_Time}).
On the other side, let us note for completeness that the opposite limit  
\begin{eqnarray}
\lim_{t \rightarrow - \infty} \,s &=& - \infty  
\label{Limit_2_Retarded_Time}
\end{eqnarray}

\noindent
is infinite.  
According to Eqs.~(\ref{Total_Light_Deflection_2}) and (\ref{nu_1}), the following impact vector and its absolute value naturally arises,  
\begin{eqnarray}
\fl \ve{D}^{\rm N}_A\left(s_0\right) = \lim_{t \rightarrow + \infty} \ve{d}^{\rm N}_A\left(s\right) 
\nonumber\\
\fl \hspace{1.3cm} = \ve{d}^{\rm N}_A\left(s_0\right)  
- \ve{\sigma} \times \left(\ve{v}_A\left(s_0\right) \times \ve{\sigma}\right)
\frac{r^{\rm N}_A\left(s_0\right) - \ve{\sigma} \cdot \ve{r}^{\rm N}_A\left(s_0\right)}{c}  
+ {\cal O}\left(c^{-2}\right),  
\label{Limit_A}
\\
\nonumber\\
\fl D^{\rm N}_A\left(s_0\right) = \lim_{t \rightarrow + \infty} d^{\rm N}_A\left(s\right) 
\nonumber\\
\fl \hspace{1.3cm} = \bigg[\left(d^{\rm N}_A\left(s_0\right)\right)^2 - 2\,\frac{\ve{d}^{\rm N}_A\left(s_0\right) \cdot \ve{v}_A\left(s_0\right)}{c}
\left(r^{\rm N}_A\left(s_0\right) - \ve{\sigma} \cdot \ve{r}^{\rm N}_A\left(s_0\right)\right)
\nonumber\\
\fl \hspace{2.0cm} + \left(\frac{\ve{\sigma} \times \ve{v}_A\left(s_0\right)}{c}\right)^2\,
\left(r^{\rm N}_A\left(s_0\right) - \ve{\sigma} \cdot \ve{r}^{\rm N}_A\left(s_0\right)\right)^2\bigg]^{\case{1}{2}} 
+ {\cal O}\left(c^{-2}\right), 
\label{Limit_B}
\end{eqnarray}

\noindent
which are induced by series expansion with regard to relation (\ref{Limit_1_Retarded_Time}). 
As indicated, terms of the order ${\cal O}\left(c^{-2}\right)$ can safely be neglected in these relations for Solar System bodies; here 
we note that $\ve{\sigma} = \ve{\nu} + {\cal O}\left(c^{-2}\right)$.  
The second term on the r.h.s. in (\ref{Limit_A}) proportional to the orbital velocity of the body is not a tiny
correction but of the same order as the first term on the r.h.s. and must not be neglected in what follows.
This can already be concluded from the fact that $d^{\rm N}_A\left(s_0\right)$ could be small and even zero,
while $d^{\rm N}_A\left(s\right) \gg m_A$ for any moment of time; cf. Eq.~(\ref{weak_field_1}) and text below that equation.
For the same reason, all the terms in the square root can be of similar magnitude for Solar System objects, which prevents 
a series expansion of this expression in inverse powers of the speed of light.  
In particular, the condition (\ref{weak_field_2}) discussed in Appendix implies that $D^{\rm N}_A\left(s_0\right) \gg m_A$,
but actually $D^{\rm N}_A\left(s_0\right) \ge P_A$ which follows from the condition (\ref{grazing_ray_2}) for grazing light rays.
Furthermore, one needs the relation  
\begin{eqnarray}
\fl \hspace{2.0cm} \lim_{t \rightarrow + \infty} \ve{v}_A\left(s\right) = \ve{v}_A\left(s_0\right) 
+ \ve{a}_A\left(s_0\right) \frac{r^{\rm N}_A\left(s_0\right) - \ve{\sigma} \cdot \ve{r}^{\rm N}_A\left(s_0\right)}{c} 
+ {\cal O}\left(c^{-2}\right),   
\label{Limit_C}
\end{eqnarray} 

\noindent 
where the second term on the r.h.s. in (\ref{Limit_C}) proportional to the orbital acceleration
of the body is actually negligible on the nas-level of accuracy. 
By inserting (\ref{Limit_A}) and (\ref{Limit_C}) into (\ref{nu_1}) one obtains  
\begin{eqnarray}
\fl \ve{\nu} = \ve{\sigma} - 4\,m_A\,\frac{\ve{D}^{\rm N}_A\left(s_0\right)}{\left(D^{\rm N}_A\left(s_0\right)\right)^2} 
\left(1 - \frac{\ve{\sigma} \cdot \ve{v}_A\left(s_0\right)}{c} \right) 
- \frac{15}{4}\,\pi\,m_A^2\, \frac{\ve{D}^{\rm N}_A\left(s_0\right)}{\left(D^{\rm N}_A\left(s_0\right)\right)^3} 
\nonumber\\ 
\nonumber\\ 
\fl \hspace{1.0cm} +\,8\,m_A^2\,\frac{\ve{D}^{\rm N}_A\left(s_0\right)}{\left(D^{\rm N}_A\left(s_0\right)\right)^2}
\,\frac{1}{r^{\rm N}_A\left(s_0\right) - \ve{\sigma} \cdot \ve{r}^{\rm N}_A\left(s_0\right)} 
 - 8\,m_A^2\,\frac{\ve{\sigma}}{\left(D^{\rm N}_A\left(s_0\right)\right)^2} + {\cal O}\left(c^{-5}\right).   
\label{nu_2}
\end{eqnarray}

\noindent 
In case of motionless body (\ref{nu_2}) exactly coincides with Eq.~(64) in \cite{Article_Zschocke1} and 
with Eq.~(3.2.43) in \cite{Brumberg1991}.  
Using the fact that $|\ve{\sigma} \times \ve{D}^{\rm N}_A\left(s_0\right)| = D^{\rm N}_A\left(s_0\right)$ 
one obtains from (\ref{nu_2}) the following expression for the total light deflection, 
\begin{eqnarray}
\fl \left| \ve{\sigma} \times \ve{\nu} \right| = 4\,\frac{m_A}{D^{\rm N}_A\left(s_0\right)} \left(1 - \frac{\ve{\sigma} \cdot \ve{v}_A\left(s_0\right)}{c} \right) 
\nonumber\\ 
\nonumber\\ 
\fl \hspace{1.45cm} + \frac{15}{4}\,\pi\,\frac{m_A^2}{\left(D^{\rm N}_A\left(s_0\right)\right)^2} 
- 8\,\frac{m_A^2}{D^{\rm N}_A\left(s_0\right)}\,\frac{1}{r^{\rm N}_A\left(s_0\right) - \ve{\sigma} \cdot \ve{r}^{\rm N}_A\left(s_0\right)} 
+ {\cal O}\left(c^{-5}\right),  
\label{Total_Light_Deflection_3}
\end{eqnarray}

\noindent
which in case of body at rest coincides with Eq.~(65) in \cite{Article_Zschocke1} and also with the expression as given by  
Eq.~(3.2.44) in \cite{Brumberg1991}.
For astrometry on the nano-arcsecond level each of the terms  
in (\ref{Total_Light_Deflection_3}) are relevant and cannot be neglected. The total light deflection in (\ref{Total_Light_Deflection_3}) is a function  
of the initial values (\ref{Introduction_6}) of the light trajectory, $\ve{x}_0$ and $\ve{\sigma}$, as well as of the retarded position and velocity  
of the massive body, $\ve{x}_A\left(s_0\right)$ and $\ve{v}_A\left(s_0\right)$. This fact is not surprising, because in the approximation made there 
are no acceleration terms and, therefore, the motion of the body at any moment of time is determined by its position and velocity at the retarded time.  

A final comment should be in order. In case of light propagation in the field of one monopole at rest (Schwarzschild metric) there 
exists an integral of motion of the geodesic equations, denoted as three-vector $\ve{D}$ in Eq.~(59) in \cite{Article_Zschocke1}. The absolute value   
of it is nothing else but Chandrasekhar's impact parameter $D$ (defined by Eq.~(215) in $3^{\rm th}$ Section in \cite{Chandrasekhar1983}).  
This integral of motion is related to coordinate-independent impact vectors, $\ve{d}^{\prime}$ and $\ve{d}^{\prime\prime}$, which 
were defined by Eqs.~(57) and (58) in \cite{Article_Zschocke1}. In particular, their absolute values agree with each other, 
$D = d^{\prime} = d^{\prime\prime}$. Therefore, it is meaningful to rewrite the expression for the total light deflection in terms of 
such impact parameters for light propagation in the field of one monopole at rest.  
However, for light propagation in the field of one moving monopole there exists no integral of motion for the null geodesics.  
For instance, in line with the investigations in \cite{Article_Zschocke1} one might want to introduce the impact vectors
\begin{eqnarray}
\ve{D}^{\prime}_A\left(s_0\right) &=& \lim_{t \rightarrow - \infty} \; \ve{\sigma} \times \left(\ve{r}_A\left(s\right) \times \ve{\sigma}\right),
\label{Total_Light_Deflection_4a}
\\
\ve{D}^{\prime\prime}_A\left(s_0\right) &=& \lim_{t \rightarrow + \infty} \; \ve{\nu} \times \left(\ve{r}_A\left(s\right) \times \ve{\nu}\right),
\label{Total_Light_Deflection_4b}
\end{eqnarray}

\noindent
where $\ve{r}_A\left(s\right)$ was introduced by Eq.~(\ref{vector_C}) with the photon's trajectory $\ve{x}\left(t\right)$. These impact vectors  
(\ref{Total_Light_Deflection_4a}) and (\ref{Total_Light_Deflection_4b}) correspond to Eqs.~(57) and (58) in \cite{Article_Zschocke1}. But their absolute values  
are not related to Chandrasekhar's impact parameter $D$ and they also differ among each other: $D \neq D^{\prime}_A \neq D^{\prime\prime}_A$. In other words, 
a physical meaning cannot be attributed to such kind of impact vectors. For these reasons it appears not meaningful to rewrite (\ref{Total_Light_Deflection_3})  
in terms of impact vectors like (\ref{Total_Light_Deflection_4a}) or (\ref{Total_Light_Deflection_4b}). Furthermore,   
practical astrometric measurements are processed within concrete reference systems, so the expression for the total light deflection in
(\ref{Total_Light_Deflection_3}) represents the appropriate form for real astrometric data reduction; see also text below Eq.~(60) in \cite{Article_Zschocke1}.

In order to get an idea about the impact of the individual terms in (\ref{Total_Light_Deflection_3}) we use the following estimates for grazing light rays,   
\begin{eqnarray}
\left| \ve{\sigma} \times \ve{\nu} \right| &=& \varphi_{\rm 1PN} + \varphi_{\rm 1.5PN} + \varphi_{\rm 2PN} + {\cal O}\left(c^{-5}\right),  
\label{Total_Light_Deflection_5}
\\
\quad \left| \varphi_{\rm 1PN} \right| &\le& 4\,\frac{m_A}{P_A}\,,
\label{Total_Light_Deflection_5a}
\\
\left| \varphi_{\rm 1.5PN} \right| &\le & 4\,\frac{m_A}{P_A}\,\frac{v_A\left(s_0\right)}{c}\,,
\label{Total_Light_Deflection_5b}
\\
\left| \varphi_{\rm 2PN} \right| &\le& \frac{15}{4}\,\pi\,\frac{m_A^2}{P_A^2} 
+ 8\,\frac{m_A^2}{P_A\,r_A^{\rm N}\left(s_0\right)}\,,   
\label{Total_Light_Deflection_5c}
\end{eqnarray}

\noindent
where in the last term in (\ref{Total_Light_Deflection_5c}) we have used that  
$ - r_A^{\rm N}\left(s_0\right) \le \ve{\sigma} \cdot \ve{r}^{\rm N}_A\left(s_0\right) \le 0$. 
Numerical values of these expressions are presented in Table \ref{Table3}; for the numerical magnitude of these light deflection terms 
see also \cite{Klioner2003b} where the light deflection in the field of one uniformly moving monopole in 1PN and 1.5PN approximation  
and in the field of one monopole at rest in 2PN approximation has been determined.  
\Table{\label{Table3}The upper limits of total light deflection as given by Eqs.~(\ref{Total_Light_Deflection_5a}) - (\ref{Total_Light_Deflection_5c}).
Besides the smallness of 2PN terms in the total light deflection it has to be emphasized that not all 2PN effects are negligible in \muas-astrometry
or nas-astrometry; see also a comment in the main text.}
\footnotesize
\begin{tabular}{@{}cccc}
\br 
Object & $\left|\varphi_{\rm 1PN}\right|\,[\muas]$ & $\left|\varphi_{\rm 1.5PN}\right|\,[\muas]$ & $\left|\varphi_{\rm 2PN}\right|\,[\muas]$ \\
\mr 
Sun & $1.75 \times 10^{6}$ & $0.07$ & $10.9$ \\
Mercury & $82.8$ & $0.013$ & $ 2 \times 10^{-8} $ \\
Venus & $492.8$ & $0.058$ & $ 9 \times 10^{-7} $\\
Earth & $574.1$ & $0.057$ & $ 1 \times 10^{-6} $ \\
Mars & $115.9$ & $0.009$ & $5 \times 10^{-8}$ \\
Jupiter & $16.272 \times 10^{3}$ & $ 0.72$ & $0.9 \times 10^{-3} $\\
Saturn & $5.776 \times 10^{3}$ & $0.18$ & $0.1 \times 10^{-3} $ \\
Uranus & $2.066 \times 10^{3}$ & $0.047$ & $2 \times 10^{-5}$ \\
Neptune & $2.532 \times 10^{3}$ & $0.046$ & $2 \times 10^{-5}$ \\
\br 
\end{tabular}\\ 
\endTable
\normalsize

\noindent 
Here we should briefly raise another subject for avoiding incorrect conclusions about the impact of 2PN terms. 
Namely, in view of the small numerical magnitude of the 2PN terms in Table \ref{Table3} one might be led to believe that 2PN effects are negligible for 
nas-astrometry. Such a conclusion is, however, absolutely wrong. While the 2PN terms in (\ref{Total_Light_Deflection_5c}) are negligible, detailed  
investigations in \cite{Article_Zschocke1,Teyssandier,AshbyBertotti2010} have recovered that there are also enhanced 2PN effects which are of decisive importance
already for astrometry on the \muas-level of accuracy. In order to recognize that highly important fact about the existence of enhanced 2PN terms 
one has to consider the boundary-value problem  
(e.g. Section 6 in \cite{Article_Zschocke1}) which, however, will not be on the scope of the present investigation, which mainly aims at a correct 
description for the light trajectory in 2PN approximation.  
In the next Section we merely touch the topic of the boundary-value problem in order to determine the Shapiro effect of time delay.

\subsection{Shapiro time delay} 

According to the theory of gravity, the speed of a light signal depends on the strength of the gravitational fields along its curvilinear trajectory. 
The corresponding effect of time delay of a light signal which propagates in the field of one monopole at rest is the so-called 
classical Shapiro effect. Since the pioneering works of Shapiro \cite{Shapiro1,Shapiro2,Shapiro3}, measurements of the time delay  
of light signals propagating through the Solar System become one of the four classical tests of general relativity.  

For the Shapiro effect one has to consider the travel time of a light signal which is supposed to be emitted  
at some space-time point $\left(\ve{x}_0,t_0\right)$ and is received at some space-time point $\left(\ve{x}_1,t_1\right)$.  
Accordingly, one has to consider the boundary value problem where a unique solution of geodesic equation (\ref{Geodesic_Equation3}) is   
defined by the boundary conditions \cite{Brumberg1991,Kopeikin_Efroimsky_Kaplan,KlionerKopeikin1992,Klioner2003a,Klioner2003b,Book_Clifford_Will}  
\begin{eqnarray}
\ve{x}\left(t_0\right) = \ve{x}_0\,, \quad \ve{x}\left(t_1\right) = \ve{x}_1\,,  
\label{Boundary}
\end{eqnarray}
 
\noindent
rather than the initial-boundary problem (\ref{Introduction_6}). Furthermore, we introduce the unit-vector which  
points from the source to the observer,  
\begin{eqnarray}
\ve{k} &=& \frac{\ve{R}}{R} \quad {\rm with} \quad \ve{R} = \ve{x}_1 - \ve{x}_0\,, 
\label{Boundary_3}
\end{eqnarray}

\noindent 
where the light source at $\ve{x}_0$ and observer at $\ve{x}_1$ both are assumed to be at rest with respect to the global coordinate system. 
Then, the Shapiro time delay is defined as the time interval $\Delta t = \left(t_1 - t_0\right) - R/c$.  
In order to find the expression for the total time of propagation of the light signal, $t_1 - t_0$, one needs the relation between 
the unit-vectors $\ve{\sigma}$ and $\ve{k}$. Such a relation follows immediately from (\ref{Second_Integration_Rearranged}) and reads  
\begin{eqnarray}
\fl \ve{\sigma} = \ve{k} + \frac{m_A}{R} \,
\Bigg(\ve{k} \times \bigg[ \ve{k} \times
\left(\ve{B}_1 \left(\ve{r}_A\left(s_1\right)\right) - \ve{B}_1 \left(\ve{r}_A\left(s_0\right)\right)\right) \bigg] \Bigg)
+ {\cal O}\left(c^{-3}\right)
\label{iteration_10}
\\
\nonumber\\
\fl \hspace{0.4cm} = \ve{k} + 2\,\frac{m_A}{R}
\left(\frac{\ve{k} \times \left(\ve{r}_A\left(s_1\right) \times \ve{k}\right)}{r_A\left(s_1\right) - \ve{k} \cdot \ve{r}_A\left(s_1\right)}
- \frac{\ve{k} \times \left(\ve{r}_A\left(s_0\right) \times \ve{k}\right)}{r_A\left(s_0\right) - \ve{k} \cdot \ve{r}_A\left(s_0\right)}\right)
+ {\cal O}\left(c^{-3}\right),
\label{iteration_15}
\end{eqnarray}

\noindent
where we have accounted for $\ve{x}_1 = \ve{x}_{\rm 1PN}\left(t_1\right) + {\cal O}\left(c^{-3}\right)$,  
$\ve{r}_A\left(s\right) = \ve{r}^{\rm N}_A\left(s\right) + {\cal O}\left(c^{-2}\right)$, and 
$\ve{r}_A\left(s\right) = \ve{r}^{\rm 1PN}_A\left(s\right) + {\cal O}\left(c^{-3}\right)$. It is important to realize  
that in (\ref{iteration_10}) and (\ref{iteration_15}) the vector $\ve{\sigma}$ has been replaced by vector $\ve{k}$ in the arguments of the  
vectorial function $\ve{B}_1$ subject to $\ve{\sigma} = \ve{k} + {\cal O}\left(c^{-2}\right)$. Now we are in the position to determine the Shapiro effect.  
With the aid of (\ref{iteration_10}) one obtains from (\ref{Second_Integration_Rearranged}) the Shapiro effect,  
\begin{eqnarray}
\fl c\left(t_1 - t_0\right)  = R - m_A\,\ve{k} \cdot
\bigg[\ve{B}_1\left(\ve{r}_A\left(s_1\right)\right) - \ve{B}_1 \left(\ve{r}_A\left(s_0\right)\right)\bigg]  
\nonumber\\
\nonumber\\
\fl \hspace{2.25cm} - m_A\,\ve{k} \cdot
\bigg[\ve{B}^A_2\left(\ve{r}_A\left(s_1\right),\ve{v}_A\left(s_0\right)\right) - \ve{B}^A_2(\ve{r}_A\left(s_0\right),\ve{v}_A\left(s_0\right)\bigg]
\nonumber\\
\nonumber\\
\fl \hspace{2.25cm} - m_A\,\ve{k} \cdot
\bigg[\ve{B}^B_2\left(\ve{r}_A\left(s_1\right),\ve{v}_A\left(s_1\right)\right) - \ve{B}^B_2(\ve{r}_A\left(s_0\right),\ve{v}_A\left(s_0\right)\bigg]
\nonumber\\
\nonumber\\
\fl \hspace{2.25cm} - m_A^2\,\ve{k} \cdot \bigg[ \ve{B}_3\left(\ve{r}_A\left(s_1\right)\right) - \ve{B}_3 \left(\ve{r}_A\left(s_0\right)\right)\bigg]
- m_A\,\ve{k} \cdot \ve{\epsilon}_2\left(s_1,s_0\right) 
\nonumber\\
\nonumber\\
\fl \hspace{2.25cm} + \frac{m_A^2}{2\,R}\,
\bigg|\ve{k} \times \left( \ve{B}_1 \left(\ve{r}_A\left(s_1\right)\right) - \ve{B}_1 \left(\ve{r}_A\left(s_0\right)\right) \right)
\bigg|^2 + {\cal O}\left(c^{-5}\right),    
\label{iteration_5}
\end{eqnarray}

\noindent
where we also have used that 
\begin{eqnarray}
\fl \hspace{1.5cm} \ve{\sigma} \cdot \ve{k} = 1 - \frac{1}{2}\,\frac{m_A^2}{R^2}\,
\bigg|\ve{k} \times \left( \ve{B}_1 \left(\ve{r}_A\left(s_1\right)\right) - \ve{B}_1 \left(\ve{r}_A\left(s_0\right)\right) \right)\bigg|^2  
+ {\cal O}\left(c^{-5}\right).  
\label{iteration_6}
\end{eqnarray}

\noindent
The vectorial functions in (\ref{iteration_5}) still depend on the unit-vector $\ve{\sigma}$ and, therefore, the expression for the time delay is still  
implicit. According to relation (\ref{iteration_10}) we have $\ve{\sigma} = \ve{k} + {\cal O}\left(c^{-2}\right)$, that means one may immediately replace  
$\ve{\sigma}$ by vector $\ve{k}$ in the arguments of the vectorial functions of the second until the fifth line in (\ref{iteration_5}).  
Then, the insertion of (\ref{iteration_15}) in the argument of the vectorial functions $\ve{B}_1$ in the first line in (\ref{iteration_5})  
completes the assignment to find an expression for the Shapiro effect fully in terms of the boundary values (\ref{Boundary}),   
\begin{eqnarray}
\fl c \left(t_1 - t_0\right) = R - 2\,m_A\,\,
\ln \frac{r_A\left(s_1\right) - \ve{k} \cdot \ve{r}_A\left(s_1\right)}{r_A\left(s_0\right) - \ve{k} \cdot \ve{r}_A\left(s_0\right)} 
\left(1 - \frac{\ve{k} \cdot \ve{v}_A\left(s_0\right)}{c}\right)  
\nonumber\\ 
\nonumber\\ 
\fl \hspace{0.5cm} - 2\,m_A\left(\frac{\ve{k} \cdot \ve{v}_A\left(s_1\right)}{c} - \frac{\ve{k} \cdot \ve{v}_A\left(s_0\right)}{c}\right)  
 - \frac{1}{4}\,m_A^2
\left(\frac{\ve{k} \cdot \ve{r}_A\left(s_1\right)}{r^2_A\left(s_1\right)} - \frac{\ve{k} \cdot \ve{r}_A\left(s_0\right)}{r^2_A\left(s_0\right)}\right)
\nonumber\\ 
\nonumber\\ 
\fl \hspace{0.5cm} - 4\,m_A^2 
\left(\frac{1}{r_A\left(s_1\right) - \ve{k} \cdot \ve{r}_A\left(s_1\right)} - \frac{1}{r_A\left(s_0\right) - \ve{k} \cdot \ve{r}_A\left(s_0\right)}\right) 
- m_A\,\ve{k} \cdot \ve{\epsilon}_2\left(s_1,s_0\right) 
\nonumber\\ 
\nonumber\\ 
\fl \hspace{0.5cm} + \frac{15}{4}\,m_A^2 
\left(\frac{1}{\left|\ve{k} \times \ve{r}_A\left(s_1\right)\right|}\,
\arctan \frac{\ve{k} \cdot \ve{r}_A\left(s_1\right)}{\left|\ve{k} \times \ve{r}_A\left(s_1\right)\right|} 
- \frac{1}{\left|\ve{k} \times \ve{r}_A\left(s_0\right)\right|}\,
\arctan \frac{\ve{k} \cdot \ve{r}_A\left(s_0\right)}{\left|\ve{k} \times \ve{r}_A\left(s_0\right)\right|}\right) 
\nonumber\\ 
\nonumber\\ 
\fl \hspace{0.5cm} + 2\,\frac{m_A^2}{R} 
\left|\frac{\ve{k} \times \ve{r}_A\left(s_1\right)}{r_A\left(s_1\right) - \ve{k} \cdot \ve{r}_A\left(s_1\right)} 
- \frac{\ve{k} \times \ve{r}_A\left(s_0\right)}{r_A\left(s_0\right) - \ve{k} \cdot \ve{r}_A\left(s_0\right)}\right|^2  
+ {\cal O}\left(c^{-5}\right). 
\label{Shaprio_Time_Delay}
\end{eqnarray}

\noindent
The last term of the first line vanishes in case of $N$ bodies due to relation (\ref{Retarded_Light_Trajectory_M_3}) as mentioned above.  
This expression for the Shapiro time delay in 2PN approximation in the field of one moving body with monopole structure generalizes the 
expression for the Shapiro time delay in 2PN approximation in the field of one body at rest with monopole structure as provided by 
Eq.~(69) in \cite{Article_Zschocke1}. In particular, one may demonstrate that in the limit of monopole at rest the expression 
in Eq.~(\ref{Shaprio_Time_Delay}) would coincide with the expression in Eq.~(69) in \cite{Article_Zschocke1}.  

Let us recall that the term $m_A\,\ve{k} \cdot \ve{\epsilon}_2$ is negligible for time delay measurements on the pico-second level 
according to the tiny numerical magnitude of this term given in Table~\ref{Table2} which corresponds to the estimation in 
Eq.~(\ref{absolute_value_epsilon_3}). In order to get an idea about the impact of the individual terms in (\ref{Shaprio_Time_Delay}) for the 
Shapiro time-delay, we use the following estimates for grazing light rays,  
\begin{eqnarray}
\Delta t &=& \Delta t_{\rm 1PN} + \Delta t_{\rm 1.5PN} + \Delta t_{\rm 2PN} + {\cal O}\left(c^{-5}\right), 
\label{Shaprio_Time_Delay_Estimate}
\\
\nonumber\\
\left|\Delta t_{\rm 1PN}\right| &\le& 2\,\frac{m_A}{c} \, \ln \frac{4\,r_A\left(s_0\right)\,r_A\left(s_1\right)}{P_A^2}\,, 
\label{Shaprio_Time_Delay_Estimate1}
\\
\nonumber\\
\left|\Delta t_{\rm 1.5PN}\right| &\le& 2\,\frac{m_A}{c} \,\frac{v_A\left(s_0\right)}{c}\,\ln \frac{4\,r_A\left(s_0\right)\,r_A\left(s_1\right)}{P_A^2}\,,  
\label{Shaprio_Time_Delay_Estimate2}
\\
\nonumber\\
\left|\Delta t_{\rm 2PN}\right| &\le& 2\,\frac{m_A^2}{c}\,\frac{R}{P_A^2} + \frac{15}{4}\,\pi\,\frac{m_A^2}{c}\,\frac{1}{P_A}\,, 
\label{Shaprio_Time_Delay_Estimate3}
\end{eqnarray}

\noindent
where for the estimate (\ref{Shaprio_Time_Delay_Estimate3}) we have used results of our previous investigations \cite{Estimate_Shapiro1,Estimate_Shapiro2}.  
Numerical values of these expressions are presented in Table \ref{Table4}.
According to the magnitude of the time-delay as given in Table~\ref{Table4}, the Shapiro delay becomes detectable in nearest future
for all of the planets of the Solar System, while higher order effects beyond 1PN approximation are measurable only for the Sun.
\Table{\label{Table4}The upper limits of Shapiro time-delay as given by Eqs.~(\ref{Shaprio_Time_Delay_Estimate1}) - (\ref{Shaprio_Time_Delay_Estimate3}).
The given numerical values for the time delay should be compared with
aimed accuracies of astrometry missions proposed to ESA like
ASTROD \cite{Astrod1,Astrod2}, LATOR \cite{Lator1,Lator2}, ODYSSEY \cite{Odyssey}, SAGAS \cite{Sagas}, or TIPO \cite{TIPO}, which have aimed at an
accuracy in time delay measurements better than about $\Delta t \sim 10^{-1}\,{\rm ns}$. Accordingly, 1.5PN effects
in time delay will surely not be detectable even within the very next generation of high-precision space-based astrometry missions, while
2PN effects on time-delay become detectable for the Sun.}
\footnotesize
\begin{tabular}{@{}cccc}
\br 
Object & $\left|\Delta t_{\rm 1PN}\right|\,[{\rm ns}]$ & $\left|\Delta t_{\rm 1.5PN}\right|\,{[\rm ns}]$ & $\left|\Delta t_{\rm 2PN}\right|\,[{\rm ns}]$ \\
\mr  
Sun & $160 \times 10^{3}$ & $ 6 \times 10^{-3}$ & $304$ \\
Mercury & $0.05$ & $7 \times 10^{-6}$ & $7 \times 10^{-7}$ \\
Venus & $0.64$ & $7 \times 10^{-5}$ & $2 \times 10^{-5}$ \\
Earth & $0.68$ & $7 \times 10^{-5}$ & $3\times 10^{-5}$ \\
Mars & $0.09$ & $7 \times 10^{-6}$ & $1 \times 10^{-6}$ \\
Jupiter & $213$ & $9 \times 10^{-3}$ & $3 \times 10^{-2}$ \\
Saturn & $66$ & $2 \times 10^{-3}$ & $4 \times 10^{-3}$ \\
Uranus & $11$ & $2 \times 10^{-4}$ & $5 \times 10^{-4}$ \\
Neptune & $13$ & $2 \times 10^{-4}$ & $9 \times 10^{-4}$ \\
\br 
\end{tabular}\\ 
\endTable
\normalsize

\section{Summary}\label{Section7} 

Todays astrometric accuracy has reached a level of several micro-arcseconds in angular measurements \cite{GAIA,GAIA1}.  
It is clear that in foreseeable future the sub-micro-arcsecond level will be attained by prospective astrometry  
and several space-based missions have already been proposed to ESA aiming at such unprecedented accuracies \cite{Theia,NEAT1,NEAT2,Gaia_NIR}.  
One fundamental problem in relativistic astrometry concerns the accurate modeling of light rays emitted by the celestial light sources 
which propagate through the gravitational field of the Solar System and which finally arrive at the observer. It is well-known that   
astrometry on the sub-micro-arcsecond or nano-arcsecond level will not be feasible without accounting for several 
post-post-Newtonian effects in the theory of light propagation  
\cite{Xu_Wu,Xu_Gong_Wu_Soffel_Klioner,Minazzoli1,Deng_Xie,2PN_Light_PropagationA,Deng_2015,Xie_Huang,Minazzoli2,Conference_Cambridge}. However, so far  
the light trajectory in 2PN approximation has mainly been considered in the field of one motionless body with monopole structure  
\cite{Brumberg1991,KlionerKopeikin1992,Deng_Xie,Deng_2015,Brumberg1987,Article_Zschocke1,Teyssandier,Hees_Bertone_Poncin_Lafitte_2014b,AshbyBertotti2010},  
while investigations in the field of moving bodies are extremely rare. In fact, up to now the problem of  
2PN light propagation in the field of moving monopoles has only been considered in \cite{Bruegmann2005}, where additional approximations  
were imposed because that investigation was not intended for astrometry within the Solar System. 
Furthermore, the problem of time-delay of a light signal in the field of a Kerr-Newman black hole in uniform motion has been considered 
in \cite{Moving_Kerr_Black_Hole1,Moving_Kerr_Black_Hole2}, which also does not aim for astrometry in the Solar System. 
Recently, the coordinate velocity and light trajectory in 2PN approximation in the field of one moving monopole 
has been determined at the first time in \cite{Zschocke3}.  
The reinvestigation of that problem was mainly triggered by two peculiarities:  
First, it has been recognized that the solution for the light trajectory contains logarithmic functions which are improperly defined 
because their arguments carry the dimension of a length.  
Second, it has been found that the 2PN solution for the trajectory contains terms proportional to the acceleration of the body.  
Both these aspects are related to each other, because  
it has turned out that some of these acceleration terms are indispensable for getting well-defined logarithmic functions in the light trajectory,  
while all the other acceleration terms are negligible on the pico-second level of accuracy in time delay measurements.  
In respect thereof, the main results of this investigation can be summarized as follows:  
\begin{enumerate} 
\item[1.] The coordinate velocity of a light signal in 2PN approximation is given by Eq.~(\ref{First_Integration}) and confirms previous results obtained  
in \cite{Zschocke3}. 
\item[2.] Terms proportional to $v^2_A/c^2$ in Eq.~(\ref{First_Integration}) 
are negligible for astrometry on nano-arcsecond level in light deflection measurements.  
\item[3.] The preliminary solution for light trajectory in 2PN approximation (\ref{Second_Integration})   
can be rewritten in such a way that it contains only well-defined logarithms and is given by Eq.~(\ref{Second_Integration_Rearranged}).  
\item[4.] It has been found that the light trajectory, both the preliminary expression (\ref{Second_Integration}) as well as 
the final expression (\ref{Second_Integration_Rearranged}), depend in the acceleration $a_A$ of the body.  
\item[5.] Terms proportional to $v^2_A/c^2$ as well as the acceleration terms proportional to $a_A$ in  
Eq.~(\ref{Second_Integration_Rearranged}) are negligible for time-delay measurements on the pico-second level.  
\item[6.] An effort has been made which allows to compare some of our results with the post-Minkowskian approach up to terms of the order 
${\cal O}\left(c^{-4}\right)$ so that no space is left for any kind of incorrectness in the primary results presented by Eq.~(\ref{First_Integration})  
and Eq.~(\ref{Second_Integration_Rearranged}). The only difference of our results with the 1PM solution 
(second term in Eq.~(\ref{Vectorial_Function_C2_B})) has its origin in the consideration of only one moving body while for a closed system  
(i.e. no energy or momentum escapes the system at infinity) of $N$ bodies this term vanishes; cf. Eq.~(\ref{Retarded_Light_Trajectory_M_3}).  
\item[7.] The total light deflection in 2PN approximation is given by Eq.~(\ref{Total_Light_Deflection_3}).  
\item[8.] It has been shown that the 2PN terms (\ref{Total_Light_Deflection_5c}) in the total light deflection (\ref{Total_Light_Deflection_3})  
are negligible for nas-astrometry. But in order to avoid incorrect conclusions, it was emphasized that the consideration of the boundary value problem  
\cite{Zschocke5} will recover the existence of so-called enhanced 2PN terms which are much larger than the 2PN terms in (\ref{Total_Light_Deflection_5c}). 
That means 2PN effects are not negligible, neither for \muas-astrometry nor for nas-astrometry.  
\item[9.] The Shapiro time delay is given by Eq.~(\ref{Shaprio_Time_Delay}).  
\end{enumerate} 

\noindent
Finally, it should be mentioned that in the astrometrical science not only the light deflection by Solar System bodies is of interest, but also 
gravitational lensing caused by extra-solar massive bodies. In order to describe gravitational lensing one needs a solution for the light trajectory which is  
valid in the near-zone as well as in the far-zone of the lens (gravitating body).  
In \cite{Zschocke4} an approach has been described about how the light trajectory in 2PN approximation can be used in order  
to derive a generalized lens equation valid for light deflection in the field of one monopole at rest. For the case of monopole at rest  
the second post-Newtonian (2PN) solution coincides with second post-Minkowskian (2PM) solution, because the gravitational field is time-independent.  
Hence, for monopoles at rest the 2PN solution is valid in the near-zone as well as in the far-zone of the gravitating system.  
However, for arbitrarily moving bodies, i.e. time-dependent gravitational fields, the validity of the 2PN solution is restricted to the near-zone 
of the gravitating system, while the 2PM solution is valid in the near-zone as well as in the far-zone of the gravitating system.  
That means, in order to obtain a generalized lens equation in the field of arbitrarily moving monopoles  
one has to solve the light trajectory in the second post-Minkowskian approximation, which is out of the scope of the present investigation.  
Of course, a prerequisite for solving the light trajectory in 2PM approximation is the determination of 
the metric perturbation in 2PM approximation (\ref{metric_perturbation_1PM_5}). While the 1PM terms $h^{\rm 1PM}_{\alpha \beta}$ are known 
for $N$ arbitrarily moving bodies (see Eq.~(\ref{Arbitrarily_Moving_Body_4})), the 2PM terms $h^{\rm 2PM}_{\alpha \beta}$ might be determined within 
the Multipolar Post-Minkowskian (MPM) formalism \cite{Blanchet_Damour1}.  
As soon as the 2PM terms of the metric tensor are in reach, the light trajectory in 2PM approximation seems to touch the realms of possibility in view  
of advanced mathematical techniques for integrating the equations of null geodesics as developed  
in \cite{Kopeikin_Efroimsky_Kaplan,KopeikinSchaefer1999,KopeikinSchaefer1999_Gwinn_Eubanks}.  
For several reasons such 2PM solution would be of great value in the astrometrical science. For instance, if  
such a 2PM solution has been achieved, then it might be that the approach described in \cite{Zschocke4} can be used
in order to derive a generalized lens equation for light deflection in the field of one arbitrarily moving monopole.

In summary, by this investigation the initial value problem of light propagation in the near-zone of the gravitational field 
generated by one arbitrarily moving monopole has been determined in the 2PN approximation. Initiated by very general considerations about the
finite speed at which gravitational action travels, the solutions were given in terms of the retarded position of
the massive body. But this fact does not mean they would be in coincidence with a possible solution in second post-Minkowskian approximation.
Such an agreement could only be established up to terms of the order ${\cal O}\left(c^{-5}\right)$, as soon as such
a solution in second post-Minkowskian approximation might be obtained in future. In a prospective investigation \cite{Zschocke5} the 
corresponding boundary value problem of light propagation will be determined which is necessary for practical astrometry in the Solar System.  
Accordingly, for the moment being we consider this investigation as a further step towards a general relativistic light propagation model
aiming at high-precision astrometry on the sub-micro-arcsecond-level of accuracy.

\section{Acknowledgment}

The author would like to express his sincere gratitude to Prof. S.A. Klioner and Prof. M.H. Soffel  
for kind encouragement and permanent support. He also wish to thank Dr. A.G. Butkevich for enlightening  
remarks about modern approaches in the field of space-based astrometrical science. Furthermore,  
Prof. R. Sch\"utzhold, Prof. L.P. Csernai, Priv.-Doz. Dr. G. Plunien, and Prof. B. K\"ampfer are greatfully 
acknowledged for inspiring discussions in general theory of relativity during recent years.  
This work was funded by the German Research Foundation (Deutsche Forschungsgemeinschaft DFG) under grant number 263799048.  

\newpage

\appendix

\section{Notation}\label{Appendix0}

Throughout the investigation the following notation is in use.

\begin{itemize}
\item $G$ is the Newtonian constant of gravitation.
\item $c$ is the vacuum speed of light in Minkowskian space-time.
\item $\eta_{\alpha\beta} = \eta^{\alpha \beta} = {\rm diag}\left(-1,+1,+1,+1\right)$ is the metric tensor of flat space-time.  
\item $g^{\alpha\beta}$ and $g_{\alpha\beta}$ are the contravariant and covariant components of the metric tensor with signature $\left(-,+,+,+\right)$.  
\item $g = {\rm det}\left(g_{\mu \nu}\right)$ is the determinant of metric tensor. 
\item $\displaystyle {f}_{\,,\,\mu} = \partial_{\mu}\,f = \frac{\partial f}{\partial x^{\mu}}$ is partial derivative of function $f$. 
\item $A^{\alpha}_{\,;\,\mu} = A^{\alpha}_{\,,\,\mu} + \Gamma^{\alpha}_{\mu\nu}\,A^{\nu}$ is covariant derivative of first rank tensor.  
\item $B^{\alpha\beta}_{\;\;\;\;;\,\mu} = B^{\alpha\beta}_{\;\;\;\;,\,\mu} + \Gamma^{\alpha}_{\mu\nu}\,B^{\nu\beta} + \Gamma^{\beta}_{\mu\nu}\,B^{\alpha\nu}$ 
is covariant derivative of second rank tensor.  
\item $M_A$ denotes the rest mass of the body.  
\item $m_A = G\,M_A/c^2$ is the Schwarzschild radius of the body.
\item $P_A$ denotes the equatorial radius of the body.
\item $v_A$ denotes the orbital velocity of the body.
\item $a_A$ denotes the orbital acceleration of the body.
\item $\displaystyle 1\,{\rm mas}\; ({\rm milli-arcsecond}) = \frac{\pi}{180 \times 60 \times 60}\,10^{-3}\,{\rm rad} \simeq 4.85 \times 10^{-9}\,{\rm rad}$.
\item $\displaystyle 1\,\muas\; ({\rm micro-arcsecond}) = \frac{\pi}{180 \times 60 \times 60}\,10^{-6}\,{\rm rad} \simeq 4.85 \times 10^{-12}\,{\rm rad}$.
\item $\displaystyle 1\,{\rm nas}\; ({\rm nano-arcsecond}) = \frac{\pi}{180 \times 60 \times 60}\,10^{-9}\,{\rm rad} \simeq 4.85 \times 10^{-15}\,{\rm rad}$.
\item $\displaystyle 1\,{\rm ns}\; ({\rm nano-second}) = 10^{-9}\,{\rm seconds}$.
\item $\displaystyle 1\,{\rm ps}\; ({\rm pico-second}) = 10^{-12}\,{\rm seconds}$.
\item Lower case Latin indices take values 1,2,3.
\item Lower case Greek indices take values 0,1,2,3.
\item The three-dimensional coordinate quantities (three-vectors) referred to
the spatial axes of the reference system are in boldface: $\ve{a}$.
\item The contravariant components of three-vectors: $a^{i} = \left(a^1,a^2,a^3\right)$.
\item The contravariant components of four-vectors: $a^{\mu} = \left(a^0,a^1,a^2,a^3\right)$.
\item The absolute value of a three-vector:  
$a = |\ve{a}| = \sqrt{a^1\,a^1+a^2\,a^2+a^3\,a^3}$.
\item The scalar product of two three-vectors:  
$\ve{a}\,\cdot\,\ve{b}=\delta_{ij}\,a^i\,b^j=a^i\,b^i$ with Kronecker delta $\delta_{ij}$.
\item The vector product of two three-vectors reads  
$\left(\ve{a}\times\ve{b}\right)^i=\varepsilon_{ijk}\,a^j\,b^k$
with Levi-Civita symbol $\varepsilon_{ijk}$.
\end{itemize}

\section{Impact vectors}\label{Appendix2}

In case of moving bodies it is appropriate to introduce a set of several impact vectors as they naturally arise in the practical calculations 
about light propagation. It is a tacitly important issue to have a clear definition of them, because the expressions for the light trajectory depend  
sensitively on these impact vectors.  
First of all, we introduce the impact vectors with respect to the unperturbed light trajectory,  
\begin{eqnarray}
\ve{d}^{\rm N}_A\left(t\right) &=& \ve{\sigma} \times \left(\ve{r}^{\rm N}_A\left(t\right) \times \ve{\sigma} \right),  
\label{Impact_Vector_1}
\\ 
\nonumber\\ 
\ve{d}^{\rm N}_A\left(s\right) &=& \ve{\sigma} \times \left(\ve{r}^{\rm N}_A\left(s\right) \times \ve{\sigma} \right),  
\label{Impact_Vector_2}
\end{eqnarray}

\noindent
where $\ve{r}^{\rm N}_A\left(t\right)$ and $\ve{r}^{\rm N}_A\left(s\right)$ are given by 
Eqs.~(\ref{Example_20}) and (\ref{Distance_Vector_N}), respectively. For an illustration of these impact vectors see Figure \ref{Diagram}.  
The impact vector are time-dependent. In particular, we need the time-derivative of (\ref{Impact_Vector_1}) which reads 
\begin{eqnarray}
\frac{d}{d ct}\,\ve{d}^{\rm N}_A\left(t\right) = \ve{\sigma} \times \left(\ve{\sigma} \times \frac{\ve{v}_A\left(t\right)}{c}\right) 
= {\cal O}\left(\frac{v_A}{c}\right). 
\label{Time_Derivative_Impact_Vector_1}
\end{eqnarray}

\noindent  
Furthermore, we define the impact vectors with respect to the light trajectory in first post-Newtonian approximation,  
\begin{eqnarray}
\ve{d}^{\rm 1PN}_A\left(t\right) &=& \ve{\sigma} \times \left(\ve{r}^{\rm 1PN}_A\left(t\right) \times \ve{\sigma} \right),
\label{Impact_Vector_3}
\\
\nonumber\\
\ve{d}^{\rm 1PN}_A\left(s\right) &=& \ve{\sigma} \times \left(\ve{r}^{\rm 1PN}_A\left(s\right) \times \ve{\sigma} \right),
\label{Impact_Vector_4}
\end{eqnarray}

\noindent
where $\ve{r}^{\rm 1PN}_A\left(t\right)$ and $\ve{r}^{\rm 1PN}_A\left(s\right)$ are given by
Eqs.~(\ref{Example_55}) and (\ref{Distance_Vector_1PN}), respectively.

These impact vectors (\ref{Impact_Vector_1}) - (\ref{Impact_Vector_2}) and (\ref{Impact_Vector_3}) - (\ref{Impact_Vector_4})  
are uniquely related to each other and one may switch from one impact vector to another
whenever it is necessary. It is not a matter of taste which concrete impact vector the most appropriate one is, but solely depends
on the expression under consideration. What we need to have is the relation between the impact vector (\ref{Impact_Vector_1}) and (\ref{Impact_Vector_3}) 
as well as between (\ref{Impact_Vector_2}) and (\ref{Impact_Vector_4}), which read   
\begin{eqnarray}
\fl \hspace{1.0cm} \ve{d}^{\rm 1PN}_A\left(t\right) = \ve{d}^{\rm N}_A\left(t\right)
- 2\,m_A \left(\frac{\ve{d}^{\rm N}_A\left(t\right)}{r^{\rm N}_A\left(t\right) - \ve{\sigma} \cdot \ve{r}^{\rm N}_A\left(t\right)}
- \frac{\ve{d}^{\rm N}_A\left(t_0\right)}{r^{\rm N}_A\left(t_0\right) - \ve{\sigma} \cdot \ve{r}^{\rm N}_A\left(t_0\right)}\right), 
\label{Relation_Impact_Vectors_1}
\\
\nonumber\\
\fl \hspace{1.0cm} \ve{d}^{\rm 1PN}_A\left(s\right) = \ve{d}^{\rm N}_A\left(s\right)
- 2\,m_A \left(\frac{\ve{d}^{\rm N}_A\left(s\right)}{r^{\rm N}_A\left(s\right) - \ve{\sigma} \cdot \ve{r}^{\rm N}_A\left(s\right)}
- \frac{\ve{d}^{\rm N}_A\left(s_0\right)}{r^{\rm N}_A\left(s_0\right) - \ve{\sigma} \cdot \ve{r}^{\rm N}_A\left(s_0\right)}\right), 
\label{Relation_Impact_Vectors_2}
\end{eqnarray}

\noindent
where $\ve{r}^{\rm N}_A\left(t_0\right) = \ve{x}_0 - \ve{x}_A\left(t_0\right)$ and
$\ve{r}^{\rm N}_A\left(s_0\right) = \ve{x}_0 - \ve{x}_A\left(s_0\right)$. 
Furthermore, we need the corresponding impact vectors with regard of the exact light trajectory, namely  
\begin{eqnarray}
\ve{d}_A\left(t\right) &=& \ve{\sigma} \times \left(\ve{r}_A\left(t\right) \times \ve{\sigma} \right),  
\label{Impact_Vector_5}
\\ 
\nonumber\\ 
\ve{d}_A\left(s\right) &=& \ve{\sigma} \times \left(\ve{r}_A\left(s\right) \times \ve{\sigma} \right),  
\label{Impact_Vector_6}
\end{eqnarray}

\noindent
where $\ve{r}_A\left(t\right)$ and $\ve{r}_A\left(s\right)$ are defined by Eqs.~(\ref{vector_B}) and (\ref{vector_C}), respectively.  
The difference between these exact impact vectors and the impact vectors in Newtonian and post-Newtonian approximation is of the order  
\begin{eqnarray}
\ve{d}_A\left(t\right) &=& \ve{d}^{\rm N}_A\left(t\right) + {\cal O}\left(c^{-2}\right), \quad 
\hspace{0.25cm} \ve{d}_A\left(s\right) = \ve{d}^{\rm N}_A\left(s\right) + {\cal O}\left(c^{-2}\right),  
\label{Relation_1}
\\
\nonumber\\
\ve{d}_A\left(t\right) &=& \ve{d}^{\rm 1PN}_A\left(t\right) + {\cal O}\left(c^{-3}\right), \quad   
\ve{d}_A\left(s\right) = \ve{d}^{\rm 1PN}_A\left(s\right) + {\cal O}\left(c^{-3}\right).  
\label{Relation_2}
\end{eqnarray}

\noindent 
Finally, we emphasize that the term 'weak gravitational field' implies a constraint for the impact vector of the exact light ray  
(cf. Eq.~(34) in \cite{Zschocke3}),
\begin{eqnarray}
d_A\left(s\right) \gg m_A\,,
\label{weak_field_1}
\end{eqnarray}

\noindent
and because of $d^{\rm N}_A\left(s\right) > d_A\left(s\right)$ the same condition is valid for the impact vector of the unperturbed light ray. 
On the other side, impact vectors at retarded time $s_0$ can be small and even be zero in some astrometric configurations, so that  
$d_A\left(s_0\right) = 0$ or $d^{\rm N}_A\left(s_0\right) = 0$ is possible. In such astrometric configurations the unit-direction $\ve{\sigma}$ 
of the light ray is anti-parallel to $\ve{r}^{\rm N}_A\left(s_0\right)$. During the time of propagation of the light signal the body moves with  
velocity $\ve{v}_A$ out of the line of sight between light source and body, so that the smallest distance between body and light signal has to be  
much larger than the Schwarzschild radius of the body, which implies the following constraint for the orbital speed,  
\begin{eqnarray}
\left|\frac{\ve{\sigma} \times \ve{v}_A\left(s_0\right)}{c}\right| 
\left(r^{\rm N}_A\left(s_0\right) - \ve{\sigma} \cdot \ve{r}^{\rm N}_A\left(s_0\right)\right) \gg m_A\,,    
\label{weak_field_2}
\end{eqnarray}

\noindent
which is not an additional condition but follows from (\ref{weak_field_1}) and (\ref{Limit_A}). It states that even in case $d^{\rm N}_A\left(s_0\right) = 0$  
the minimal distance between light ray and body will be larger than the Schwarzschild radius of that body. It also implies that  
configurations with $d^{\rm N}_A\left(s_0\right) = 0$ are only possible if the body is in motion.  

Since in reality a light ray in the optical band cannot propagate inside the body, the condition (\ref{weak_field_1}) can be replaced by  
a stronger restriction which is valid for grazing light rays, where the impact vector $d_A\left(s\right)$ equals the equatorial radius of the body,  
while in general it will be larger, so we get  
\begin{eqnarray}
d_A\left(s\right) &\ge& P_A\,, 
\label{grazing_ray_1}
\end{eqnarray}

\noindent
which implies $d^{\rm N}_A\left(s\right) \ge P_A$. 
This condition states that even in case $d_A\left(s_0\right) = 0$ the minimal distance between light ray and body will be larger than the 
body's equatorial radius. From (\ref{grazing_ray_1}) and (\ref{Limit_A}) is follows that  
\begin{eqnarray}
\left|\frac{\ve{\sigma} \times \ve{v}_A\left(s_0\right)}{c}\right|\left(r_A\left(s_0\right) - \ve{\sigma} \cdot \ve{r}_A\left(s_0\right)\right) 
&\ge& P_A\,,
\label{grazing_ray_2}
\end{eqnarray}
 
\noindent
for practical astrometry.  

\section{Light propagation in 1.5PN approximation}\label{Appendix_15PN}

The coordinate velocity and trajectory of a light signal in 1.5PN approximation in the field of arbitrarily shaped bodies in arbitrary motion  
has been determined in \cite{Zschocke2}. Here we need these results for the much simpler case of one moving body with monopole structure.  

\subsection{The coordinate velocity of a light signal in 1.5PN approximation} 

The coordinate velocity of a light signal in the gravitational field of one arbitrarily moving monopole reads 
(Eqs.~(110) - (111) in \cite{Zschocke2}):  
\begin{eqnarray}
\frac{\dot{\ve{x}}\left(t\right)}{c} &=& \ve{\sigma} + \frac{\Delta \dot{\ve{x}}_{\rm 1PN}\left(t\right)}{c}
+ \frac{\Delta \dot{\ve{x}}_{\rm 1.5PN}\left(t\right)}{c} + {\cal O}\left(c^{-4}\right),
\label{Light_Deflection_M_0}
\end{eqnarray}

\noindent
with 
\begin{eqnarray}
\frac{\Delta \dot{\ve{x}}_{\rm 1PN}\left(t\right)}{c} = - \frac{2\,m_A}{r^{\rm N}_A\left(t\right)}
\left(\frac{\ve{d}^{\rm N}_A\left(t\right)}{r^{\rm N}_A\left(t\right) - \ve{\sigma}\cdot\ve{r}^{\rm N}_A\left(t\right)} + \ve{\sigma}\right),
\label{Light_Deflection_M_1}
\\
\nonumber\\
\fl \frac{\Delta \dot{\ve{x}}_{\rm 1.5PN}\left(t\right)}{c} = + \frac{2\,m_A}{r^{\rm N}_A\left(t\right)}
\frac{\ve{\sigma}\cdot\ve{v}_A\left(t\right)}{c}\,\frac{\ve{d}^{\rm N}_A\left(t\right)}{r^{\rm N}_A\left(t\right) - \ve{\sigma}\cdot\ve{r}^{\rm N}_A\left(t\right)}
+ \frac{4\,m_A}{r^{\rm N}_A\left(t\right)}\,\frac{\ve{v}_A\left(t\right)}{c}
\nonumber\\
\nonumber\\
\fl - \frac{2\,m_A}{r^{\rm N}_A\left(t\right) - \ve{\sigma}\cdot\ve{r}^{\rm N}_A\left(t\right)}\,
\frac{\ve{\sigma}\times\left(\ve{v}_A\left(t\right) \times \ve{\sigma}\right)}{c}
+ \frac{2\,m_A}{\left(r^{\rm N}_A\left(t\right) - \ve{\sigma}\cdot\ve{r}^{\rm N}_A\left(t\right)\right)^2}\,
\frac{\ve{d}^{\rm N}_A\left(t\right)\cdot\ve{v}_A\left(t\right)}{c}\,\frac{\ve{d}^{\rm N}_A\left(t\right)}{r^{\rm N}_A\left(t\right)}\,. 
\nonumber\\
\label{Light_Deflection_M_2}
\end{eqnarray}

\noindent 
Using relations (\ref{retarded_time_B}) - (\ref{retarded_time_D}) one may rewrite the expressions
in (\ref{Light_Deflection_M_0}) - (\ref{Light_Deflection_M_2}) in terms of retarded time (cf. Eq.~(177) in \cite{Zschocke2}):  
\begin{eqnarray}
\frac{\dot{\ve{x}}\left(t\right)}{c} &=& \ve{\sigma} + \frac{\Delta \dot{\ve{x}}_{\rm 1PN}\left(s\right)}{c}
+ \frac{\Delta \dot{\ve{x}}_{\rm 1.5PN}\left(s\right)}{c} + {\cal O}\left(c^{-4}\right),
\label{Light_Deflection_M_6}
\end{eqnarray}

\noindent
with
\begin{eqnarray}
\frac{\Delta \dot{\ve{x}}_{\rm 1PN}\left(s\right)}{c} = - \frac{2\,m_A}{r^{\rm N}_A\left(s\right)}
\left(\frac{\ve{d}^{\rm N}_A\left(s\right)}{r^{\rm N}_A\left(s\right) - \ve{\sigma}\cdot\ve{r}^{\rm N}_A\left(s\right)} + \ve{\sigma}\right),
\label{Light_Deflection_M_7}
\\
\nonumber\\
\nonumber\\
\fl \hspace{1.5cm} \frac{\Delta \dot{\ve{x}}_{\rm 1.5PN}\left(s\right)}{c} = + \frac{2\,m_A}{r^{\rm N}_A\left(s\right)}
\frac{\ve{\sigma}\cdot\ve{v}_A\left(s\right)}{c}\,
\frac{\ve{d}^{\rm N}_A\left(s\right)}{r^{\rm N}_A\left(s\right) - \ve{\sigma}\cdot\ve{r}^{\rm N}_A\left(s\right)}
+ \frac{4\,m_A}{r^{\rm N}_A\left(s\right)}\,\frac{\ve{v}_A\left(s\right)}{c}
\nonumber\\
\nonumber\\
\fl \hspace{1.5cm} + \frac{2\,m_A}{\left(r^{\rm N}_A\left(s\right)\right)^2}\,\frac{\ve{\sigma}\cdot\ve{v}_A\left(s\right)}{c}\,\ve{d}^{\rm N}_A\left(s\right)
- \frac{2\,m_A}{\left(r^{\rm N}_A\left(s\right)\right)^2}\,\frac{\ve{v}_A\left(s\right) \cdot \ve{d}^{\rm N}_A\left(s\right)}{c}\,
\frac{\ve{d}^{\rm N}_A\left(s\right)}{r^{\rm N}_A\left(s\right) - \ve{\sigma}\cdot\ve{r}^{\rm N}_A\left(s\right)}
\nonumber\\
\nonumber\\
\fl \hspace{1.5cm} - \frac{2\,m_A}{\left(r^{\rm N}_A\left(s\right)\right)^2}\,\frac{\ve{r}^{\rm N}_A\left(s\right) \cdot \ve{v}_A\left(s\right)}{c}\,\ve{\sigma}\,, 
\label{Light_Deflection_M_8}
\end{eqnarray}

\noindent 
which up to terms of the order $\displaystyle {\cal O}\left(c^{-4}\right)$ agrees with the post-Minkowskian solution 
in \cite{KopeikinSchaefer1999}; cf. Eq.~(\ref{Kopeikin_Schaefer_15}). 

\subsection{The trajectory of a light signal in 1.5PN approximation} 

The trajectory of a light signal in the gravitational field of one arbitrarily moving monopole reads (Eqs.~(118) - (119) in \cite{Zschocke2}):  
\begin{eqnarray}
\fl \ve{x}\left(t\right) = \ve{x}_0 + c \left(t-t_0\right) \ve{\sigma}
+ \Delta\ve{x}_{\rm 1PN}\left(t,t_0\right) 
+ \Delta\ve{x}_{\rm 1.5PN}\left(t,t_0\right) + {\cal O}\left(c^{-4}\right),
\label{Light_Trajectory_M_0}
\\
\nonumber\\
\Delta\ve{x}_{\rm 1PN}\left(t,t_0\right) = 
\Delta\ve{x}_{\rm 1PN}\left(t\right) - \Delta\ve{x}_{\rm 1PN}\left(t_0\right),  
\label{Delta_t1}
\\
\nonumber\\
\Delta\ve{x}_{\rm 1.5PN}\left(t,t_0\right) =  
\Delta\ve{x}_{\rm 1.5PN}\left(t\right) - \Delta\ve{x}_{\rm 1.5PN}\left(t_0\right),   
\label{Delta_t2}
\end{eqnarray}

\noindent
with   
\begin{eqnarray}
\fl \Delta\ve{x}_{\rm 1PN}\left(t\right) = 
- 2\,m_A\,\frac{\ve{d}^{\rm N}_A\left(t\right)}{r^{\rm N}_A\left(t\right) - \ve{\sigma} \cdot \ve{r}^{\rm N}_A\left(t\right)}
+ 2\,m_A\,\ve{\sigma}\;\ln\left(r^{\rm N}_A\left(t\right) - \ve{\sigma} \cdot \ve{r}^{\rm N}_A\left(t\right)\right),
\label{Light_Trajectory_M_1}
\\
\nonumber\\
\fl \Delta \ve{x}_{\rm 1.5PN}\left(t\right) =
+ 2\,m_A\,\frac{\ve{v}_A\left(t\right) \cdot \ve{d}^{\rm N}_A\left(t\right)}{c}\,
\frac{\ve{d}^{\rm N}_A\left(t\right)}{\left(r^{\rm N}_A\left(t\right) - \ve{\sigma} \cdot \ve{r}^{\rm N}_A\left(t\right)\right)^2}
\nonumber\\
\nonumber\\
\fl \hspace{0.75cm} - 2\,m_A\,\frac{\ve{v}_A\left(t\right)}{c}\,\ln\left(r^{\rm N}_A\left(t\right) - \ve{\sigma} \cdot \ve{r}^{\rm N}_A\left(t\right)\right)
- 2\,m_A\,\frac{\ve{\sigma} \times \left(\ve{v}_A\left(t\right) \times \ve{\sigma}\right)}{c}\,
\frac{\ve{\sigma} \cdot \ve{r}^{\rm N}_A\left(t\right)}{r^{\rm N}_A\left(t\right) - \ve{\sigma} \cdot \ve{r}^{\rm N}_A\left(t\right)}  
\nonumber\\
\nonumber\\
\fl \hspace{0.75cm} + 2\,m_A\,\frac{\ve{v}_A\left(t\right)\cdot\ve{d}^{\rm N}_A\left(t\right)}{c}\,
\frac{\ve{\sigma}}{r^{\rm N}_A\left(t\right) - \ve{\sigma} \cdot \ve{r}^{\rm N}_A\left(t\right)}\,. 
\label{Light_Trajectory_M_2}
\end{eqnarray}

\noindent
One may check that the total time-derivative of (\ref{Light_Trajectory_M_0}) yields (\ref{Light_Deflection_M_0}). 
Using relations (\ref{retarded_time_B}) - (\ref{retarded_time_D}) one may rewrite the expressions  
in (\ref{Light_Trajectory_M_0}) - (\ref{Light_Trajectory_M_2}) in terms of retarded time:
\begin{eqnarray}
\fl \ve{x}\left(t\right) = \ve{x}_0 + c \left(t-t_0\right) \ve{\sigma}
+ \Delta \ve{x}_{\rm 1PN}\left(s,s_0\right) 
+ \Delta \ve{x}_{\rm 1.5PN}\left(s,s_0\right) + {\cal O}\left(c^{-4}\right),
\label{Retared_Light_Trajectory_M_0}
\\
\nonumber\\ 
\Delta \ve{x}_{\rm 1PN}\left(s,s_0\right) = \Delta \ve{x}_{\rm 1PN}\left(s\right) 
- \Delta \ve{x}_{\rm 1PN}\left(s_0\right), 
\label{Delta_s1}
\\ 
\nonumber\\ 
\Delta \ve{x}_{\rm 1.5PN}\left(s,s_0\right) = \Delta \ve{x}_{\rm 1.5PN}\left(s\right) 
- \Delta \ve{x}_{\rm 1.5PN}\left(s_0\right),  
\label{Delta_s2}
\end{eqnarray}

\noindent
with 
\begin{eqnarray}
\fl \Delta\ve{x}_{\rm 1PN}\left(s\right) = 
- 2\,m_A\,\frac{\ve{d}^{\rm N}_A\left(s\right)}{r^{\rm N}_A\left(s\right) - \ve{\sigma} \cdot \ve{r}^{\rm N}_A\left(s\right)}
+ 2\,m_A\,\ve{\sigma}\;\ln\left(r^{\rm N}_A\left(s\right) - \ve{\sigma} \cdot \ve{r}^{\rm N}_A\left(s\right)\right),
\label{Retarded_Light_Trajectory_M_1}
\\
\nonumber\\
\fl \Delta \ve{x}_{\rm 1.5PN}\left(s\right) = + 2\,m_A\,\frac{\ve{\sigma}\cdot\ve{v}_A\left(s\right)}{c}\,
\frac{\ve{d}^{\rm N}_A\left(s\right)}{r^{\rm N}_A\left(s\right) - \ve{\sigma} \cdot \ve{r}^{\rm N}_A\left(s\right)}
+ 2\,m_A \frac{\ve{v}_A\left(s\right)}{c} 
\nonumber\\
\nonumber\\
\fl \hspace{2.3cm} - 2\,m_A \frac{\ve{v}_A\left(s\right)}{c}\,\ln\left(r^{\rm N}_A\left(s\right) - \ve{\sigma} \cdot \ve{r}^{\rm N}_A\left(s\right)\right), 
\label{Retarded_Light_Trajectory_M_2}
\end{eqnarray}

\noindent
which up to terms of the order $\displaystyle {\cal O}\left(c^{-4}\right)$ agrees with the post-Minkowskian solution 
in \cite{KopeikinSchaefer1999}; cf. Eqs.~(\ref{Kopeikin_Schaefer_40}) - (\ref{Kopeikin_Schaefer_45}). 
The logarithm in (\ref{Retarded_Light_Trajectory_M_1}) is well-defined in view of (\ref{Delta_s1}) which results in dimensionless arguments. 
But besides of (\ref{Delta_s2}) the logarithm in (\ref{Retarded_Light_Trajectory_M_2}) remains ill-defined.  
As outlined in the main text, a meaning of this function is attributed by the series expansion (\ref{series_expansion_1})  
up to terms beyond 1.5PN approximation.  

A comment should be in order about the last term in the first line of (\ref{Retarded_Light_Trajectory_M_2}).
It has much long been known \cite{Landau_Lifschitz} that the vanishing of covariant derivative of stress-energy tensor of matter $T^{\alpha \beta}$ implies  
\cite{Kopeikin_Efroimsky_Kaplan,MTW,Landau_Lifschitz,Poisson_Lecture_Notes}  
\begin{eqnarray}
T^{\alpha \beta}_{\;\;\;\;\;;\,\beta} = 0 \quad \Longrightarrow \quad  
\left[\left(- g \right) \left(T^{\alpha \beta} + t_{\rm LL}^{\alpha \beta}\right)\right]_{\,,\,\beta} = 0\;, 
\label{Momentum_Conservation} 
\end{eqnarray}

\noindent  
where the Landau-Liftschitz pseudotensor of the gravitational fields $t_{\rm LL}^{\alpha \beta}$ has already been encountered in the field equations
of gravity (\ref{Field_Equations_10}). This local conservation equation admits the formulation of global conservation laws.  
Especially, a global four-momentum for isolated gravitational systems can be defined  
(e.g. Eqs.~(20.23a) and (20.23c) in \cite{MTW} or Eqs.~(1.1.7) and (1.2.1) in \cite{Poisson_Lecture_Notes}),
\begin{eqnarray}
\fl \hspace{1.0cm} P^{\alpha} = \frac{1}{c} \int d^3 x\,\left(-g\left(t,\ve{x}\right)\right)
\left(T^{\alpha 0}\left(t,\ve{x}\right) + t_{\rm LL}^{\alpha 0}\left(t,\ve{x}\right)\right) \quad {\rm with} \quad 
\frac{d P^{\alpha}}{d t} = 0\,.   
\label{Total_Momentum}
\end{eqnarray}

\noindent
The four-momentum (\ref{Total_Momentum}) is coordinate-independent and convergent for isolated systems, hence is well-defined 
and a physical meaning can be attributed.   
Moreover, the second equation in (\ref{Total_Momentum}) states that the four-momentum is strictly conserved 
for such systems. This fact can be proven by means of Gauss theorem with boundary at infinity (e.g. Eq.~(20.25) in \cite{MTW})  
because the flux through the surface of the volume vanishes at infinity due to Eqs.~(\ref{Asymptotic_1}) and (\ref{Asymptotic_2}). 
From (\ref{Total_Momentum}) we readily find  
(for the stress-energy tensor of $N$ pointlike bodies moving under their gravitational interaction 
we refer to Eq.~(2.363) in \cite{Kopeikin_Efroimsky_Kaplan} or Eqs.~(1) - (2) in \cite{KopeikinSchaefer1999}, while the Landau-Lifschitz pseudotensor  
is neglected due to $t_{\rm LL}^{\alpha \beta} = {\cal O}\left(c^{-2}\right)$)  
that the total three-momentum of $N$ arbitrarily moving pointlike bodies is strictly conserved to order ${\cal O}\left(c^{-2}\right)$  
\footnote{Likewise, the total three-momentum in post-Newtonian approximation $\ve{P}_{\rm 1PN}$ of a isolated gravitating system
of $N$ arbitrarily moving bodies (cf. Eq.~(5.4.21) in \cite{Poisson_Lecture_Notes}) is strictly conserved to order ${\cal O}\left(c^{-4}\right)$,
that means $\displaystyle \frac{d {\ve{P}}_{\rm 1PN}}{d t} = {\cal O}\left(c^{-4}\right)$ (cf. Eq.~(5.4.24) in \cite{Poisson_Lecture_Notes}), 
but this fact is beyond the approximations of the investigation. We also note that the gravitational potentials among the bodies are taken into account by  
the Landau-Lifschitz pseudotensor $t_{\rm LL}^{\alpha \beta} = {\cal O}\left(c^{-2}\right)$ hence they do not contribute  
in the Newtonian approximation $\ve{P}_{\rm N}$ in Eq.~(\ref{Total_Newtonian_Momentum}) but appear in the  
post-Newtonian approximation $\ve{P}_{\rm 1PN}$ (cf. Eq.~(5.4.25) in \cite{Poisson_Lecture_Notes}).},  
\begin{eqnarray}
\ve{P}_{\rm N} = \sum \limits_{A=1}^N M_A\,\ve{v}_A\left(t\right) \quad {\rm with} \quad  
\frac{d \ve{P}_{\rm N}}{d t} = {\cal O}\left(c^{-2}\right),   
\label{Total_Newtonian_Momentum}
\end{eqnarray}

\noindent
where the index ${\rm N}$ refers to the Newtonian approximation. 
The relation (\ref{Total_Newtonian_Momentum}) implies that (cf. Eqs.~(7.213) in \cite{Kopeikin_Efroimsky_Kaplan},  
Eqs.~(155) in \cite{KopeikinSchaefer1999} or Eq.~(5.4.25) in \cite{Poisson_Lecture_Notes}) 
\begin{eqnarray}
\sum\limits_{A=1}^N m_A \left(\frac{\ve{v}_A\left(s\right)}{c} - \frac{\ve{v}_A\left(s_0\right)}{c} \right) &=& {\cal O}\left(c^{-5}\right)\,,   
\label{Retarded_Light_Trajectory_M_3}
\end{eqnarray}

\noindent
where we recall   
$\ve{v}_A\left(s\right) = \ve{v}_A\left(t\right) + {\cal O}\left(c^{-1}\right)$ and 
$\ve{v}_A\left(s_0\right) = \ve{v}_A\left(t_0\right) + {\cal O}\left(c^{-1}\right)$.  
It clearly shows that the last term in the first line of (\ref{Retarded_Light_Trajectory_M_2}) is solely caused by the model  
of only one arbitrarily moving monopole which necessarily is an open system, while this term vanishes in case of a closed system  
(i.e. no energy or momentum escapes the system) of $N$ bodies.  

\section{Some relations for integration by parts}\label{Appendix_Integration_by_Parts}

Some relations of total time derivatives are listed which are useful for integrating the geodesic equation by parts.  
Time-arguments are omitted to simplify the notation. 

\subsection{Relations valid to order ${\cal O}\left(c^{-1}\right)$}

\begin{eqnarray}
\fl \frac{1}{r_A^{\rm N}} =
- \frac{d}{d ct}\,\ln \left(r_A^{\rm N} - \ve{\sigma} \cdot\ve{r}_A^{\rm N}\right) + {\cal O}\left(\frac{v_A}{c}\right).
\label{Appendix_Time_Derivative_1}
\end{eqnarray}

\vspace{0.5cm} 
\begin{eqnarray}
\fl \frac{1}{\left(r_A^{\rm N}\right)^2}
= \frac{d}{d ct}\,\frac{1}{d^{\rm N}_A}\,\arctan \frac{\ve{\sigma}\cdot\ve{r}_A^{\rm N}}{d^{\rm N}_A}
+ {\cal O}\left(\frac{v_A}{c}\right).
\label{Appendix_Time_Derivative_2}
\end{eqnarray}

\vspace{0.5cm} 
\begin{eqnarray}
\fl \frac{1}{\left(r_A^{\rm N}\right)^3}
= \frac{d}{d ct}\,\frac{1}{r_A^{\rm N}}\,\frac{1}{r_A^{\rm N} - \ve{\sigma} \cdot\ve{r}_A^{\rm N}}
+ {\cal O}\left(\frac{v_A}{c}\right).  
\label{Appendix_Time_Derivative_3}
\end{eqnarray}

\vspace{0.5cm} 
\begin{eqnarray}
\fl \frac{1}{\left(r_A^{\rm N}\right)^4} = \frac{1}{2}\,
\frac{d}{d ct}\,\left(\frac{\ve{\sigma}\cdot\ve{r}_A^{\rm N}}{\left(d_A^{\rm N}\right)^2\,\left(r_A^{\rm N}\right)^2}
+ \frac{1}{\left(d_A^{\rm N}\right)^3}\,\arctan \frac{\ve{\sigma}\cdot\ve{r}_A^{\rm N}}{d_A^{\rm N}}\right)
+ {\cal O}\left(\frac{v_A}{c}\right).
\label{Appendix_Time_Derivative_4}
\end{eqnarray}

\vspace{0.5cm} 
\begin{eqnarray}
\fl \frac{1}{\left(r_A^{\rm N}\right)^5} = \frac{d}{d ct}\,
\left[\frac{2}{3}\,\frac{\ve{\sigma}\cdot\ve{r}_A^{\rm N}}{r_A^{\rm N}}\,\frac{1}{\left(d_A^{\rm N}\right)^4}
+ \frac{1}{3}\,\frac{\ve{\sigma}\cdot\ve{r}_A^{\rm N}}{\left(r_A^{\rm N}\right)^3}\,\frac{1}{\left(d_A^{\rm N}\right)^2}\right]
+ {\cal O}\left(\frac{v_A}{c}\right).
\label{Appendix_Time_Derivative_5}
\end{eqnarray}

\vspace{0.5cm} 
\begin{eqnarray}
\fl \frac{\ve{\sigma}\cdot\ve{r}_A^{\rm N}}{r_A^{\rm N}} =
\frac{d}{d ct}\,r_A^{\rm N} + {\cal O}\left(\frac{v_A}{c}\right).
\label{Appendix_Time_Derivative_6A}
\end{eqnarray}

\vspace{0.5cm} 
\begin{eqnarray}
\fl \frac{\ve{\sigma} \cdot \ve{r}_A^{\rm N}}{\left(r_A^{\rm N}\right)^2} =
\frac{d}{d ct}\,\ln \,r_A^{\rm N} + {\cal O}\left(\frac{v_A}{c}\right).
\label{Appendix_Time_Derivative_6B}
\end{eqnarray}

\vspace{0.5cm} 
\begin{eqnarray}
\fl \frac{\ve{\sigma}\cdot\ve{r}_A^{\rm N}}{\left(r_A^{\rm N}\right)^n} = - \frac{1}{n-2}\,
\frac{d}{d ct}\,\frac{1}{\left(r_A^{\rm N}\right)^{n-2}} + {\cal O}\left(\frac{v_A}{c}\right) \quad {\rm for} \quad n \ge 3\,.
\label{Appendix_Time_Derivative_7}
\end{eqnarray}

\vspace{0.5cm} 
\begin{eqnarray}
\fl \frac{\left(\ve{\sigma}\cdot\ve{r}_A^{\rm N}\right)^2}{\left(r_A^{\rm N}\right)^3} =
- \frac{d}{d ct} \left[\frac{\ve{\sigma}\cdot \ve{r}_A^{\rm N}}{r_A^{\rm N}}\,
+ \ln \left(r_A^{\rm N} - \ve{\sigma}\cdot \ve{r}_A^{\rm N}\right)\right] + {\cal O}\left(\frac{v_A}{c}\right).
\label{Appendix_Time_Derivative_8}
\end{eqnarray}

\vspace{0.5cm} 
\begin{eqnarray}
\fl \frac{\left(\ve{\sigma}\cdot\ve{r}_A^{\rm N}\right)^2}{\left(r_A^{\rm N}\right)^4}
= + \frac{1}{2}\,\frac{d}{d ct} \left[\arctan \frac{\ve{\sigma}\cdot\ve{r}_A^{\rm N}}{d^{\rm N}_A}
- \frac{d^{\rm N}_A\,\left(\ve{\sigma}\cdot\ve{r}_A^{\rm N}\right)}{\left(r_A^{\rm N}\right)^2} \right]
+ {\cal O}\left(\frac{v_A}{c}\right).
\label{Appendix_Time_Derivative_9}
\end{eqnarray}

\vspace{0.5cm} 
\begin{eqnarray}
\fl \frac{\left(\ve{\sigma}\cdot\ve{r}_A^{\rm N}\right)^2}{\left(r_A^{\rm N}\right)^5} =
\frac{1}{3}\,\frac{d}{d ct}\,\frac{\left(\ve{\sigma}\cdot\ve{r}_A^{\rm N}\right)^3}{\left(r_A^{\rm N}\right)^3}\,
\frac{1}{\left(d_A^{\rm N}\right)^2} + {\cal O}\left(\frac{v_A}{c}\right).
\label{Appendix_Time_Derivative_10}
\end{eqnarray}

\vspace{0.5cm} 
\begin{eqnarray}
\fl \frac{\left(\ve{\sigma}\cdot\ve{r}_A^{\rm N}\right)^3}{\left(r_A^{\rm N}\right)^5} =
- \frac{d}{d ct} \left[\frac{2}{3}\,\frac{1}{r_A^{\rm N}}
+ \frac{1}{3}\,\frac{\left(\ve{\sigma}\cdot\ve{r}_A^{\rm N}\right)^2}{\left(r_A^{\rm N}\right)^3} \right]
+ {\cal O}\left(\frac{v_A}{c}\right).
\label{Appendix_Time_Derivative_11}
\end{eqnarray}

\vspace{0.5cm} 
\begin{eqnarray}
\fl \frac{\left(\ve{\sigma}\cdot\ve{r}_A^{\rm N}\right)^2}{\left(r_A^{\rm N}\right)^6} =
\frac{1}{4}\,\frac{d}{d ct} \left[\frac{1}{2}\,
\frac{\ve{\sigma}\cdot\ve{r}_A^{\rm N}}{\left(d_A^{\rm N}\right)^2\,\left(r_A^{\rm N}\right)^2}
- \frac{\ve{\sigma}\cdot\ve{r}_A^{\rm N}}{\left(r_A^{\rm N}\right)^4}
+ \frac{1}{2}\,\frac{1}{\left(d^{\rm N}_A\right)^3}\,\arctan \frac{\ve{\sigma}\cdot\ve{r}_A^{\rm N}}{d^{\rm N}_A} \right]
+ {\cal O}\left(\frac{v_A}{c}\right).
\nonumber\\ 
\label{Appendix_Time_Derivative_12}
\end{eqnarray}

\vspace{0.5cm} 
\begin{eqnarray}
\fl \frac{\left(\ve{\sigma}\cdot\ve{r}_A^{\rm N}\right)^3}{\left(r_A^{\rm N}\right)^6}
= \frac{1}{2}\,\frac{d}{d ct} \left[\frac{1}{2}\,\frac{\left(d_A^{\rm N}\right)^2}{\left(r_A^{\rm N}\right)^4}
- \frac{1}{\left(r_A^{\rm N}\right)^2} \right] + {\cal O}\left(\frac{v_A}{c}\right).
\label{Appendix_Time_Derivative_13}
\end{eqnarray}

\vspace{0.5cm} 
\begin{eqnarray}
\fl \frac{1}{r_A^{\rm N}}\,\frac{1}{r_A^{\rm N} - \ve{\sigma} \cdot\ve{r}_A^{\rm N}} =
\frac{d}{d ct}\,\frac{1}{r_A^{\rm N} - \ve{\sigma} \cdot\ve{r}_A^{\rm N}} + {\cal O}\left(\frac{v_A}{c}\right).
\label{Appendix_Time_Derivative_14}
\end{eqnarray}

\vspace{0.5cm} 
\begin{eqnarray}
\fl \frac{1}{r_A^{\rm N}}\,\frac{1}{\left(r_A^{\rm N} - \ve{\sigma} \cdot\ve{r}_A^{\rm N}\right)^2} =
\frac{1}{2}\,\frac{d}{d ct}\,\frac{1}{\left(r_A^{\rm N} - \ve{\sigma} \cdot\ve{r}_A^{\rm N}\right)^2}
+ {\cal O}\left(\frac{v_A}{c}\right).
\label{Appendix_Time_Derivative_15}
\end{eqnarray}

\vspace{0.5cm} 
\begin{eqnarray}
\fl \frac{1}{r_A^{\rm N} - \ve{\sigma} \cdot\ve{r}_A^{\rm N}} =
\frac{1}{2}\,\frac{d}{d ct}\,\frac{r_A^{\rm N}}{r_A^{\rm N} - \ve{\sigma} \cdot\ve{r}_A^{\rm N}}
- \frac{1}{2}\,\frac{d}{d ct}\,\ln \left(r_A^{\rm N} - \ve{\sigma} \cdot\ve{r}_A^{\rm N}\right) + {\cal O}\left(\frac{v_A}{c}\right).
\label{Appendix_Time_Derivative_16}
\end{eqnarray}

\vspace{0.5cm} 
\begin{eqnarray}
\fl \arctan \frac{\ve{\sigma} \cdot\ve{r}_A^{\rm N}}{d^{\rm N}_A} = \frac{d}{d ct}\,
\left[\ve{\sigma} \cdot\ve{r}_A^{\rm N}\,\arctan \frac{\ve{\sigma} \cdot\ve{r}_A^{\rm N}}{d^{\rm N}_A}
- d^{\rm N}_A\, \ln r_A^{\rm N}\right] + {\cal O}\left(\frac{v_A}{c}\right).
\label{Appendix_Time_Derivative_17}
\end{eqnarray}

\vspace{0.5cm} 
\begin{eqnarray}
\fl \ln \left(r_A^{\rm N} - \ve{\sigma} \cdot \ve{r}_A^{\rm N}\right) = + \frac{d}{d ct}
\left[r_A^{\rm N} + \ve{\sigma} \cdot \ve{r}_A^{\rm N}
\ln \left(r_A^{\rm N} - \ve{\sigma} \cdot \ve{r}_A^{\rm N}\right)\right] + {\cal O}\left(\frac{v_A}{c}\right).  
\label{Appendix_Time_Derivative_18}
\end{eqnarray}

\subsection{Exact relations}

\begin{eqnarray}
\fl \frac{1}{r_A^{\rm N}} =
- \frac{d}{d ct}\,\ln \left(r_A^{\rm N} - \ve{\sigma} \cdot\ve{r}_A^{\rm N}\right)
+ \frac{1}{r_A^{\rm N}}\,\frac{\ve{\sigma}\cdot\ve{v}_A}{c}
- \frac{1}{r_A^{\rm N} - \ve{\sigma} \cdot\ve{r}_A^{\rm N}}
\,\frac{\ve{d}_A^{\rm N} \cdot \ve{v}_A}{c\;r_A^{\rm N}}\,.
\label{Exact_Time_Derivative_1}
\end{eqnarray}

\vspace{0.5cm} 
\begin{eqnarray}
\fl \frac{\ve{\sigma} \cdot\ve{r}_A^{\rm N}}{r_A^{\rm N}} =
+ \frac{d}{d ct}\,r_A^{\rm N} + \frac{1}{r_A^{\rm N}}\,
\frac{\ve{r}_A^{\rm N} \cdot \ve{v}_A}{c}\,.
\label{Exact_Time_Derivative_2a}
\end{eqnarray}

\vspace{0.5cm} 
\begin{eqnarray}
\fl \frac{\ve{\sigma} \cdot\ve{r}_A^{\rm N}}{\left(r_A^{\rm N}\right)^3} =
- \frac{d}{d ct}\,\frac{1}{r_A^{\rm N}} + \frac{1}{\left(r_A^{\rm N}\right)^3}\,
\frac{\ve{r}_A^{\rm N} \cdot \ve{v}_A}{c}\,.
\label{Exact_Time_Derivative_2b}
\end{eqnarray}

\vspace{0.5cm} 
\begin{eqnarray}
\fl \frac{1}{\left(r_A^{\rm N}\right)^3} = + \frac{d}{d ct}\,\frac{1}{r_A^{\rm N}}\,
\frac{1}{r_A^{\rm N} - \ve{\sigma} \cdot \ve{r}_A^{\rm N}}
+ \frac{1}{\left(r_A^{\rm N}\right)^3}\,
\frac{\ve{\sigma}\cdot\ve{v}_A}{c}
- \frac{1}{\left(r_A^{\rm N}\right)^3}\,\frac{1}{r_A^{\rm N} - \ve{\sigma} \cdot \ve{r}_A^{\rm N}}\,
\frac{\ve{d}_A^{\rm N} \cdot \ve{v}_A}{c}
\nonumber\\
\nonumber\\
\fl \hspace{1.5cm} - \frac{1}{\left(r_A^{\rm N}\right)^2}\,\frac{1}{\left(r_A^{\rm N} - \ve{\sigma} \cdot \ve{r}_A^{\rm N}\right)^2}\,
\frac{\ve{d}_A^{\rm N} \cdot \ve{v}_A}{c}\,.
\label{Exact_Time_Derivative_3}
\end{eqnarray}

\vspace{0.5cm} 
\begin{eqnarray}
\fl \frac{1}{r_A^{\rm N}}\,\frac{1}{r_A^{\rm N} - \ve{\sigma}\cdot\ve{r}_A^{\rm N}}
= + \frac{d}{d c t}\,\frac{1}{r_A^{\rm N} - \ve{\sigma}\cdot\ve{r}_A^{\rm N}}
- \frac{1}{\left(r_A^{\rm N} - \ve{\sigma} \cdot \ve{r}_A^{\rm N}\right)^2}\,
\frac{\ve{d}_A^{\rm N} \cdot \ve{v}_A}{c\;r_A^{\rm N}}
\nonumber\\ 
\nonumber\\ 
\fl \hspace{2.9cm} + \frac{1}{r_A^{\rm N}}\,
\frac{1}{r_A^{\rm N} - \ve{\sigma}\cdot\ve{r}_A^{\rm N}}\,\frac{\ve{\sigma}\cdot\ve{v}_A}{c}\,.
\label{Exact_Time_Derivative_4}
\end{eqnarray}

\vspace{0.5cm} 
\begin{eqnarray}
\fl \frac{1}{\left(r_A^{\rm N}\right)^2}
= + \frac{d}{d ct}\,\frac{1}{d^{\rm N}_A}\,\arctan \frac{\ve{\sigma}\cdot\ve{r}_A^{\rm N}}{d^{\rm N}_A}
- \frac{1}{\left(d_A^{\rm N}\right)^3}\,\frac{\ve{d}_A^{\rm N} \cdot \ve{v}_A}{c}\,
\arctan \frac{\ve{\sigma}\cdot\ve{r}_A^{\rm N}}{d^{\rm N}_A}
+ \frac{1}{\left(r_A^{\rm N}\right)^2}\,\frac{\ve{\sigma} \cdot \ve{v}_A}{c}
\nonumber\\ 
\nonumber\\ 
\fl \hspace{1.5cm} - \frac{1}{\left(d_A^{\rm N}\right)^2}\,\frac{\ve{\sigma}\cdot\ve{r}_A^{\rm N}}{\left(r_A^{\rm N}\right)^2}\,
\frac{1}{\left(d_A^{\rm N}\right)^2}\,\frac{\ve{d}_A^{\rm N} \cdot \ve{v}_A}{c}\,.
\label{Exact_Time_Derivative_5}
\end{eqnarray}

\vspace{0.5cm} 
\begin{eqnarray}
\fl \ln \left(r_A^{\rm N} - \ve{\sigma} \cdot \ve{r}_A^{\rm N}\right) = + \frac{d}{d ct} 
\left[r_A^{\rm N} + \ve{\sigma} \cdot \ve{r}_A^{\rm N} 
\ln \left(r_A^{\rm N} - \ve{\sigma} \cdot \ve{r}_A^{\rm N}\right)\right] 
+ \frac{\ve{\sigma} \cdot \ve{v}_A}{c} \ln \left(r_A^{\rm N} - \ve{\sigma} \cdot \ve{r}_A^{\rm N}\right) 
\nonumber\\ 
\fl \hspace{2.9cm} + \frac{\ve{d}_A^{\rm N} \cdot \ve{v}_A}{c} \frac{1}{r_A^{\rm N} - \ve{\sigma} \cdot \ve{r}_A^{\rm N}}\,. 
\label{Exact_Time_Derivative_6}
\end{eqnarray}

\vspace{0.5cm} 
\begin{eqnarray}
\fl \frac{1}{r_A^{\rm N} - \ve{\sigma} \cdot\ve{r}_A^{\rm N}} =
+ \frac{1}{2}\,\frac{d}{d ct}\,\frac{r_A^{\rm N}}{r_A^{\rm N} - \ve{\sigma} \cdot\ve{r}_A^{\rm N}}
- \frac{1}{2}\,\frac{d}{d ct}\,\ln \left(r_A^{\rm N} - \ve{\sigma} \cdot\ve{r}_A^{\rm N}\right) 
+ \frac{1}{r_A^{\rm N} - \ve{\sigma} \cdot\ve{r}_A^{\rm N}} \frac{\ve{\sigma} \cdot\ve{v}_A}{c} 
\nonumber\\ 
\fl \hspace{2.25cm} - \frac{1}{2} \frac{1}{\left(r_A^{\rm N} - \ve{\sigma} \cdot \ve{r}_A^{\rm N}\right)^2} \frac{\ve{d}_A^{\rm N} \cdot \ve{v}_A}{c}\,.  
\label{Exact_Time_Derivative_7} 
\end{eqnarray}

\section{Light propagation in first post-Minkowskian approximation}\label{Appendix3}

In view of the emphasized importance of the acceleration terms for clear definition of the logarithms it is 
advisable to have an independent check of the results obtained. Such a check is possible, at least up to terms of the order ${\cal O}\left(c^{-4}\right)$, 
by a comparison with results in the post-Minkowskian approach.  
The light trajectory in the gravitational field of $N$ arbitrarily moving monopoles in first post-Minkowskian approximation, 
i.e. exact to the first order in the gravitational constant and to all orders in the velocities of the bodies,  
was determined in \cite{KopeikinSchaefer1999}. We shall briefly summarize these results for comparison with our solution.   
Here we keep the notation of \cite{KopeikinSchaefer1999} except for their $\ve{k}$ (defined by Eq.~(12) in \cite{KopeikinSchaefer1999}) 
which is just our $\ve{\sigma}$ (defined by Eq.~(\ref{Introduction_6})).  

\subsection{The metric and geodesic equation in 1PM approximation}

In the post-Minkowskian expansion one assumes the gravitational fields to be weak while the speed of matter is not restricted   
and could even be ultra-relativistic, hence this expansion is performed in terms of the gravitational constant, while the speed of matter is taken into  
account to any order. The corresponding series expansion of the metric in powers of the gravitational constant reads  
\begin{eqnarray}
g_{\alpha\beta}\left(t,\ve{x}\right) &=& \eta_{\alpha\beta}
+ h^{\rm 1PM}_{\alpha\beta}\left(t,\ve{x}\right) + h^{\rm 2PM}_{\alpha \beta} \left(t,\ve{x}\right) + {\cal O} \left(G^3\right),  
\label{metric_perturbation_1PM_5}
\end{eqnarray}

\noindent
where $h^{\rm 1PM}_{\alpha \beta} = {\cal O}\left(G^1\right)$ and $h^{\rm 2PM}_{\alpha \beta} = {\cal O}\left(G^2\right)$  
are small metric perturbations which describe deviations from the flat space-time.  
The expansion in (\ref{metric_perturbation_1PM_5}) is exact up to terms of the third order in the gravitational constant. At the present time,  
for the case of bodies at rest the 1PM metric perturbation $h^{\rm 1PM}_{\alpha \beta}$ as well as the 2PM metric perturbations $h^{\rm 2PM}_{\alpha \beta}$  
have been determined within the Multipolar Post-Minkowskian (MPM) formalism in \cite{2PN_Metric1,2PN_Metric2}, while for the general case of arbitrarily  
moving bodies of finite size they are far out of reach. But in case of moving pointlike bodies they are known in the first post-Minkowskian approximation  
\cite{KopeikinSchaefer1999,KopeikinMashhoon2002,Zschocke_Soffel},  
\begin{eqnarray}
\fl h_{\alpha \beta}^{\rm 1PM} \left(t,\ve{x}\right) =
\frac{4\,m_A}{\gamma_A\left(s\right)\,\left(r_A\left(s\right) -
\frac{\displaystyle \ve{v}_A\left(s\right) \cdot \ve{r}_A\left(s\right)}{\displaystyle c}\right)}\,
\left(\frac{u^A_{\alpha}\left(s\right)}{c}\,\frac{u^A_{\beta}\left(s\right)}{c} + \frac{\eta_{\alpha\beta}}{2}\right),
\label{Arbitrarily_Moving_Body_4}
\end{eqnarray}

\noindent
where $\gamma_A\left(s\right) = \left(1 - v^2_A\left(s\right)/c^2\right)^{- \case{1}{2}}$ is the Lorentz factor.
The covariant components of the four-velocity of the body are
$u^A_{\alpha}\left(s\right) = \gamma_A\left(s\right)\left(- c, \ve{v}_A\left(s\right)\right)$,
and $\ve{v}_A\left(s\right)$ is the three-velocity of the body in the global system.
The vector pointing from the retarded position $\ve{x}_A\left(s\right)$ of the body $A$ towards the field-point $\ve{x}$ reads 
$\ve{r}_A\left(s\right) = \ve{x} - \ve{x}_A\left(s\right)$ (cf. Eq.~(\ref{vector_A})),  
where the retarded time $s$ is related to the coordinate time $t$ via the implicit relation (\ref{Retarded_Time_1}).  
Inserting (\ref{metric_perturbation_1PM_5}) into (\ref{Geodetic_Equation2}) yields up to terms of the order ${\cal O}\left(G^2\right)$ 
\cite{KopeikinSchaefer1999_Gwinn_Eubanks}:  
\begin{eqnarray}
\fl \frac{\ddot{x}^i \left(t\right)}{c^2} = + \frac{1}{2}\,h_{00,i}^{\rm 1PM}
- h_{00,j}^{\rm 1PM} \frac{\dot{x}^i\left(t\right)}{c}\frac{\dot{x}^j\left(t\right)}{c}
- h_{ij,k}^{\rm 1PM}\,\frac{\dot{x}^j\left(t\right)}{c}\frac{\dot{x}^k\left(t\right)}{c}
+ \frac{1}{2}\,h_{jk,i}^{\rm 1PM}\,\frac{\dot{x}^j\left(t\right)}{c}\frac{\dot{x}^k\left(t\right)}{c}
\nonumber\\
\nonumber\\
\fl \hspace{1.25cm} - \frac{1}{2}\,h_{00,0}^{\rm 1PM}\,\frac{\dot{x}^i\left(t\right)}{c}
- h_{ij,0}^{\rm 1PM} \frac{\dot{x}^j\left(t\right)}{c}
+ \frac{1}{2}\,h_{jk,0}^{\rm 1PM} \frac{\dot{x}^i\left(t\right)}{c}
\frac{\dot{x}^j\left(t\right)}{c}\frac{\dot{x}^k\left(t\right)}{c}
- h_{0i,j}^{\rm 1PM} \frac{\dot{x}^j\left(t\right)}{c}
\nonumber\\
\nonumber\\
\fl \hspace{1.25cm} + h_{0j,i}^{\rm 1PM} \frac{\dot{x}^j\left(t\right)}{c}
- h_{0j,k}^{\rm 1PM}\frac{\dot{x}^i\left(t\right)}{c}\frac{\dot{x}^j\left(t\right)}{c}\frac{\dot{x}^k\left(t\right)}{c}
- h_{0i,0}^{\rm 1PM} 
+ {\cal O}\left(G^2\right).   
\label{geodesic_equation_1PM}
\end{eqnarray}

\noindent 
Let us consider the solution of the geodesic equation (\ref{geodesic_equation_1PM})  
as it has been provided in \cite{KopeikinSchaefer1999}, that means excluding terms of the order ${\cal O}\left(G^2\right)$.

\subsection{The coordinate velocity of a light signal} 

The coordinate velocity of a light signal is given by Eqs.~(32) and (34) in \cite{KopeikinSchaefer1999}.   
Using $h^{\alpha \beta}$ and $\hat{\partial}_i B^{\alpha \beta}$ as given by Eqs.~(10) and (30) in \cite{KopeikinSchaefer1999} it reads  
\begin{eqnarray}
\fl \frac{\dot{\ve{x}}\left(t\right)}{c} = \ve{\sigma} - 2\,m_A \gamma_A\left(s\right)
\frac{\ve{d}_A\left(s\right)}{r_A\left(s\right) - \ve{\sigma} \cdot \ve{r}_A\left(s\right)}
\left(1 - \frac{\ve{\sigma}\cdot\ve{v}_A\left(s\right)}{c}\right)^2
\frac{1}{r_A\left(s\right) - {\displaystyle \frac{\ve{v}_A\left(s\right) \cdot \ve{r}_A\left(s\right)}{c}}}
\nonumber\\
\nonumber\\
\fl \hspace{1.15cm} + 4\,m_A\,\gamma_A\left(s\right) \frac{\ve{v}_A\left(s\right)}{c}\,
\frac{1}{r_A\left(s\right) - {\displaystyle \frac{\ve{v}_A\left(s\right) \cdot \ve{r}_A\left(s\right)}{c}}}
\left(1 - \frac{\ve{\sigma}\cdot\ve{v}_A\left(s\right)}{c}\right)
\nonumber\\
\nonumber\\
\fl \hspace{1.15cm} - 2\,m_A\,\gamma_A\left(s\right) \ve{\sigma}
\left(1 - \left(\frac{\ve{\sigma}\cdot\ve{v}_A\left(s\right)}{c}\right)^2\right)
\frac{1}{r_A\left(s\right) - {\displaystyle \frac{\ve{v}_A\left(s\right) \cdot \ve{r}_A\left(s\right)}{c}}} + {\cal O}\left(G^2\right), 
\label{Kopeikin_Schaefer_15}
\end{eqnarray}

\noindent
with $\ve{r}_A\left(s\right) = \ve{x}\left(t\right) - \ve{x}_A\left(s\right)$ (cf. Eq.~(\ref{vector_C})),  
where $\ve{x}\left(t\right)$ is the exact spatial coordinate of the photon rather than the field point $\ve{x}$ in (\ref{vector_A}). 
The solution in (\ref{Kopeikin_Schaefer_15}) confirms the expression (C.4) in \cite{Klioner2003a}. 
By series expansion in inverse powers of the speed of light one may show that  
(\ref{Kopeikin_Schaefer_15}) agrees, up to terms of the order ${\cal O}\left(c^{-3}\right)$, 
with our 1.5PN solution provided by Eqs.~(\ref{Light_Deflection_M_6}) - (\ref{Light_Deflection_M_8}) and 
it agrees, up to terms of the order ${\cal O}\left(c^{-4}\right)$, with our 2PN solution in Eq.~(\ref{First_Integration}).

\subsection{The trajectory of a light signal} 

The trajectory of a light signal is defined by Eqs.~(33) and (35) in \cite{KopeikinSchaefer1999} and reads, 
\begin{eqnarray}
\ve{x}\left(t\right) = \ve{x}_0 + c \left(t-t_0\right) \ve{\sigma} + \ve{\Xi}\left(s\right) - \ve{\Xi}\left(s_0\right) + {\cal O}\left(G^2\right),
\label{Kopeikin_Schaefer_40}
\\
\nonumber\\
\fl \ve{\Xi}\left(s\right) = - 2\,m_A\,\gamma_A\left(s\right) \left(1 - \frac{\ve{\sigma}\cdot\ve{v}_A\left(s\right)}{c}\right)
\frac{\ve{d}_A\left(s\right)}{r_A\left(s\right) - \ve{\sigma} \cdot \ve{r}_A\left(s\right)}
\nonumber\\
\nonumber\\
\fl \hspace{1.25cm} + 2\,m_A\,\gamma_A\left(s\right)\,
\left(\ve{\sigma} - \frac{\ve{v}_A\left(s\right)}{c}\right) \ln \left(r_A\left(s\right) - \ve{\sigma} \cdot \ve{r}_A\left(s\right)\right)
+ \ve{{\cal I}}\left(s\right) + \ve{{\cal K}}\left(s\right), 
\label{Kopeikin_Schaefer_45}
\end{eqnarray}

\noindent
where $\ve{r}_A\left(s\right)$ is defined by Eq.~(\ref{vector_C}). The solution in (\ref{Kopeikin_Schaefer_45}) confirms the expressions  
for the light trajectory as given by Eqs.~(C.2) - (C.3) in \cite{Klioner2003a}. In order to get  
(\ref{Kopeikin_Schaefer_40}) - (\ref{Kopeikin_Schaefer_45}) we have used the expressions $B^{\alpha \beta}$ and $\hat{\partial}_i D^{\alpha \beta}$  
as defined by Eqs.~(26) and (31) in \cite{KopeikinSchaefer1999} and performed an integration by parts which yields 
(using (28) and (48) in \cite{KopeikinSchaefer1999}):  
\begin{eqnarray}
\fl B^{\alpha \beta}\left(s\right) = 
- 4\,\left(\hat{T}^{\alpha \beta}\left(s\right) - \frac{1}{2}\,\eta^{\alpha \beta}\,\hat{T}^{\lambda}_{\lambda}\left(s\right)\right) 
\frac{\ln \left(r_A\left(s\right) - \ve{\sigma} \cdot \ve{r}_A\left(s\right)\right)}{1 - {\displaystyle \frac{\ve{\sigma}\cdot\ve{v}_A\left(s\right)}{c}}} 
+ {\cal I}^{\alpha \beta}\left(s\right),  
\label{Kopeikin_Schaefer_30}
\\
\nonumber\\
\fl \hat{\partial}_i D^{\alpha \beta} \left(s\right) = 
- 4\,d_A^i\left(s\right)\,\left(\hat{T}^{\alpha \beta}\left(s\right) - \frac{1}{2}\,\eta^{\alpha \beta}\,\hat{T}^{\lambda}_{\lambda}\left(s\right)\right)
\frac{1}{1 - {\displaystyle \frac{\ve{\sigma}\cdot\ve{v}_A\left(s\right)}{c}}}\,\frac{1}{r_A\left(s\right) - \ve{\sigma} \cdot \ve{r}_A\left(s\right)} 
\nonumber\\ 
\nonumber\\ 
\fl \hspace{0.75cm} +\,4\,P_{ij}\,\frac{v_A^{j}\left(s\right)}{c}
\left(\hat{T}^{\alpha \beta}\left(s\right)-\frac{1}{2}\,\eta^{\alpha \beta}\,\hat{T}^{\lambda}_{\lambda}\left(s\right)\right)
\frac{\ln \left(r_A\left(s\right) - \ve{\sigma} \cdot \ve{r}_A\left(s\right)\right)}
{\left(1 - {\displaystyle \frac{\ve{\sigma}\cdot\ve{v}_A\left(s\right)}{c}}\right)^2}
+ {\cal K}_i^{\alpha \beta}\left(s\right),  
\label{Kopeikin_Schaefer_35}
\end{eqnarray}  

\noindent
where the components of the stress-energy tensor are given by 
$\hat{T}^{\alpha \beta} = M_A\,\gamma_A^{-1}\,u_A^{\alpha}\,u_A^{\beta}$ (cf. Eq.~(2) in \cite{KopeikinSchaefer1999}) where  
$u_A^{\alpha} = \gamma_A \left(c , \ve{v}_A\right) $ and  
the functions ${\cal I}^{\alpha\beta}$ and ${\cal K}_i^{\alpha\beta}$ are  
\begin{eqnarray}
\fl {\cal I}^{\alpha\beta}\left(s\right) = 
4 \int\limits_{-\infty}^{s} d \zeta\;\ln \left(r_A - \ve{\sigma} \cdot \ve{r}_A\right)
\frac{d}{d\zeta}\;\left[ \left(\hat{T}^{\alpha \beta} - \frac{1}{2}\,\eta^{\alpha \beta}\,\hat{T}^{\lambda}_{\lambda}\right)
\left(1 - \frac{\ve{\sigma}\cdot\ve{v}_A}{c}\right)^{-1}\right],  
\label{Remnant_Integrals_5}
\\
\nonumber\\
\fl {\cal K}_i^{\alpha\beta}\left(s\right) = 4 \int \limits_{-\infty}^{s} d\zeta\,\frac{d_A^i}{r_A - \ve{\sigma} \cdot \ve{r}_A}
\frac{d}{d\zeta}\;\left[ \left(\hat{T}^{\alpha \beta} - \frac{1}{2}\,\eta^{\alpha \beta}\,\hat{T}^{\lambda}_{\lambda}\right)
\left(1 - \frac{\ve{\sigma}\cdot\ve{v}_A}{c}\right)^{-1}\right]
\nonumber\\
\nonumber\\
\fl - 4 \int \limits_{-\infty}^{s} d\zeta\,\ln \left(r_A - \ve{\sigma} \cdot \ve{r}_A\right)
\frac{d}{d\zeta} \left[\left(\hat{T}^{\alpha \beta} - \frac{1}{2}\,\eta^{\alpha \beta}\,\hat{T}^{\lambda}_{\lambda}\right)
\left(1 - \frac{\ve{\sigma}\cdot\ve{v}_A}{c}\right)^{-2} P^{ij} \frac{v_A^j}{c}\right], 
\label{Remnant_Integrals_10}
\end{eqnarray}

\noindent 
where the arguments are omitted for simpler notation: $\ve{d}_A = \ve{d}_A\left(\zeta\right)$, 
$\hat{T}^{\alpha \beta} = \hat{T}^{\alpha \beta}\left(\zeta\right)$, $\ve{v}_A = \ve{v}_A\left(\zeta\right)$, and 
$\ve{r}_A = \ve{x}_0 + c \left(t-t_0\right) \ve{\sigma} - \ve{x}_A\left(\zeta\right)$.  
The vectorial functions $\ve{{\cal I}}$ and $\ve{{\cal K}}$ in (\ref{Kopeikin_Schaefer_45}) are defined by
\begin{eqnarray}
{\cal I}^i\left(s\right) &=& - \sigma_{\alpha} {\cal I}^{\alpha i}\left(s\right)
- \frac{1}{2}\,\sigma^i {\cal I}^{00}\left(s\right) + \frac{1}{2}\,\sigma^i \sigma_p \sigma_q {\cal I}^{pq}\left(s\right),
\label{Function_I}
\\
\nonumber\\
{\cal K}^i\left(s\right) &=& \frac{1}{2}\,\sigma_{\alpha} \sigma_{\beta}\,{\cal K}_i^{\alpha\beta}\left(s\right).
\label{Function_K}
\end{eqnarray}

\noindent
One obtains the following expression for the function (\ref{Function_I}) and (\ref{Function_K}):  
\begin{eqnarray}
\fl \ve{\cal I}\left(s\right) = + 2\,m_A \int \limits_{-\infty}^{s} d\zeta \ln \left(r_A - \ve{\sigma} \cdot \ve{r}_A\right)  
\gamma_A 
\nonumber\\ 
\nonumber\\ 
\fl \hspace{2.0cm} \times \left[2\,\frac{\ve{a}_A}{c} - \ve{\sigma}\left(\frac{\ve{\sigma}\cdot\ve{a}_A}{c}\right) + \gamma^2_A  
\left(2\,\frac{\ve{v}_A}{c} - \ve{\sigma} - \ve{\sigma}\,\frac{\ve{\sigma}\cdot\ve{v}_A}{c}\right) 
\left(\frac{\ve{v}_A \cdot \ve{a}_A}{c^2}\right)\right],  
\label{Remnant_Integrals_35}
\\ 
\nonumber\\ 
\nonumber\\ 
\fl \ve{\cal K}\left(s\right) = 
+ 2\,m_A \int \limits_{-\infty}^{s} d\zeta\,\frac{\ve{d}_A}{r_A - \ve{\sigma} \cdot \ve{r}_A}
\gamma_A \left[\gamma_A^2 \left(\frac{\ve{v}_A \cdot \ve{a}_A}{c^2}\right)
\left(1 - \frac{\ve{\sigma}\cdot\ve{v}_A}{c}\right) - \left(\frac{\ve{\sigma}\cdot\ve{a}_A}{c}\right) \right]  
\nonumber\\
\nonumber\\
\fl \hspace{0.5cm} - 2\,m_A \int \limits_{-\infty}^{s} d\zeta \ln \left(r_A - \ve{\sigma} \cdot \ve{r}_A\right)
\gamma_A \left[\gamma_A^2 \left(\frac{\ve{v}_A \cdot \ve{a}_A}{c^2}\right)
\frac{\ve{\sigma} \times \left(\ve{v}_A \times \ve{\sigma}\right)}{c} + \frac{\ve{\sigma} \times \left(\ve{a}_A\times \ve{\sigma}\right)}{c}\right]. 
\nonumber\\
\label{Remnant_Integrals_40}
\end{eqnarray}

\noindent 
Using Eqs.~(25) and (29) in \cite{KopeikinSchaefer1999}) one may show that the expressions in (\ref{Remnant_Integrals_35}) and (\ref{Remnant_Integrals_40})  
are in agreement with the vectorial function $\ve{g}\left(t_0,t\right)$ defined by Eq.~(C.2) in \cite{Klioner2003a}, that means   
$-2 \,m_A\, \ve{g}\left(t_0,t\right) = \ve{\cal I}\left(s\right) - \ve{\cal I}\left(s_0\right) + \ve{\cal K}\left(s\right) - \ve{\cal K}\left(s_0\right)$.  
For comparison with our results in (\ref{Second_Integration}) we consider a series expansion of the expressions 
in (\ref{Remnant_Integrals_35}) - (\ref{Remnant_Integrals_40}) in inverse powers of the speed of light.  
According to (\ref{Kopeikin_Schaefer_40}) and (\ref{Kopeikin_Schaefer_45}) we consider  
\begin{eqnarray}
\fl \ve{\cal I}\left(s\right) - \ve{\cal I}\left(s_0\right) + \ve{\cal K}\left(s\right) - \ve{\cal K}\left(s_0\right) = 
+ 2\,m_A \int \limits_{s_0}^{s} d\zeta \ln \left(r_A - \ve{\sigma} \cdot \ve{r}_A\right) \frac{\ve{a}_A}{c}  
\nonumber\\ 
\nonumber\\ 
\fl \hspace{5.1cm} - 2\,m_A \int \limits_{s_0}^{s} d\zeta\,\frac{\ve{d}_A}{r_A - \ve{\sigma} \cdot \ve{r}_A}\,\frac{\ve{\sigma} \cdot \ve{a}_A}{c}  
+ {\cal O}\left(c^{-5}\right),  
\label{Remnant_Integrals_65}
\end{eqnarray}

\noindent
where it has been taken into account that integrals having an integrand $\ve{v}_A \cdot \ve{a}_A$ turn out to be of the order ${\cal O}\left(c^{-5}\right)$.  
Integration by parts, using relations (28) and (48) in \cite{KopeikinSchaefer1999}, yields   
\begin{eqnarray}
\fl \ve{\cal I}\left(s\right) - \ve{\cal I}\left(s_0\right) + \ve{\cal K}\left(s\right) - \ve{\cal K}\left(s_0\right) = 
+ 2\,m_A \left(\frac{\ve{\sigma}\cdot\ve{a}_A\left(s\right)}{c^2}\right) \ve{d}_A\left(s\right)\,
\ln \left(r_A\left(s\right) - \ve{\sigma} \cdot \ve{r}_A\left(s\right)\right)
\nonumber\\
\nonumber\\
\fl - 2\,m_A \left(\frac{\ve{\sigma}\cdot\ve{a}_A\left(s_0\right)}{c^2}\right)\ve{d}_A\left(s_0\right)\,
\ln \left(r_A\left(s_0\right) - \ve{\sigma} \cdot \ve{r}_A\left(s_0\right)\right)
+ {\cal O}\left(\dot{\ve{a}}_A\right) 
\nonumber\\ 
\nonumber\\ 
\fl + 2\,m_A\, \frac{\ve{a}_A\left(s\right)}{c^2} \left(r_A\left(s\right) - \ve{\sigma} \cdot \ve{r}_A\left(s\right)\right)
\left[1 - \ln \left(r_A\left(s\right) - \ve{\sigma} \cdot \ve{r}_A\left(s\right)\right)\right]
\nonumber\\
\nonumber\\
\fl - 2\,m_A \frac{\ve{a}_A\left(s_0\right)}{c^2} \left(r_A\left(s_0\right) - \ve{\sigma} \cdot \ve{r}_A\left(s_0\right)\right)
\left[1 - \ln \left(r_A\left(s_0\right) - \ve{\sigma} \cdot \ve{r}_A\left(s_0\right)\right)\right]  
+ {\cal O}\left(c^{-5}\right),   
\label{Remnant_Integrals_70}
\end{eqnarray}

\noindent
which might also be compared with solutions for the integrals given by the Eqs.~(49) - (50) and (202) in \cite{KopeikinSchaefer1999}. 
This expression in (\ref{Remnant_Integrals_70}) is in coincidence with  
$\ve{\tilde \epsilon}^B_2\left(\ve{r}_A\left(s\right), \ve{a}\left(s\right)\right) - \ve{\tilde \epsilon}^B_2\left(\ve{r}_A\left(s_0\right), \ve{a}\left(s_0\right)\right)$   
where the vectorial function $\ve{\tilde \epsilon}^B_2$ has been defined by Eq.~(\ref{epsilon_2_a}), and we recall 
that $\ve{r}_A = \ve{r}^{\rm N}_A + {\cal O}\left(c^{-2}\right)$.


\end{document}